\documentclass{bmcart}

\usepackage{amsthm,amsmath}
\usepackage[utf8]{inputenc} 

\usepackage{graphicx}
\usepackage{booktabs}
\usepackage[authoryear,longnamesfirst]{natbib}
\usepackage{amsmath,amssymb,amsfonts}
\usepackage{algorithmic}
\usepackage{textcomp}
\usepackage{xcolor}
\usepackage{lipsum,lineno}
\usepackage[a4paper, margin=1in]{geometry}

\usepackage{todonotes}
\usepackage{svg}
\usepackage{array}\newcolumntype{P}[1]{>{\centering\arraybackslash}p{#1}}
\usepackage{ifthen}
\usepackage{xcolor}
\usepackage{verbatim}
\usepackage{placeins}
\usepackage{lscape}
\usepackage{multirow}
\usepackage{graphicx}
\usepackage{float}
\usepackage{url}
\usepackage{lipsum}
\usepackage{siunitx} 
\usepackage{changepage} 
\usepackage{enumitem}

\usepackage[final, commandnameprefix=ifneeded, xcolor=black]{changes}

\makeatletter
\newcommand\notsotiny{\@setfontsize\notsotiny{6.31415}{7.1828}}
\makeatother

\definecolor{lavenderblush}{rgb}{0.93, 0.88, 0.90}

\colorlet{shadecolor}{lavenderblush}

\startlocaldefs

\newcommand{\mytitle}{
How are Primary School Computer Science Curricular Reforms Contributing to Equity? 
Impact on Student Learning, Perception of the Discipline, and Gender Gaps
}

\newboolean{ISanonymous}
\setboolean{ISanonymous}{false}
\ifthenelse{\boolean{ISanonymous}}{\newcommand*{\Anonymous}{}
\newcommand*{\IsAnonymous}{}
}{}

\newcommand{\KWA}{Computer Science}
\newcommand{\KWB}{Curricular Reform}
\newcommand{\KWC}{Elementary Education}
\newcommand{\KWD}{Learning Achievement}
\newcommand{\KWE}{Perception Survey}
\newcommand{\KWF}{Equity}

\newcommand{\EmailLH}{laila.elhamamsy@epfl.ch}
\newcommand{\EmailBB}{barbara.bruno@epfl.ch}
\newcommand{\EmailFM}{francesco.mondada@epfl.ch}
\newcommand{\EmailJDZ}{jessica.dehlerzufferey@epfl.ch}

\newcommand{\EmailSA}{sunny.avry@epfl.ch}

\newcommand{\EmailMC}{morgane.chevalier@hepl.ch}
\newcommand{\EmailCA}{catherine.audrin@hepl.ch}

\newcommand{\IJSTEMEduc}{
}
\endlocaldefs

\begin{document}

\begin{frontmatter}

\begin{fmbox}
\dochead{Research}


\title{\mytitle}

\ifdefined\IsAnonymous

\else

\author[
  addressref={aff1, aff2},                   
  corref={aff1},                       
  email={\EmailLH}   
]
{\inits{L.E.-H.}\fnm{Laila} \snm{El-Hamamsy}} 
\author[
  addressref={aff3},
  email={\EmailBB}
]{\inits{B.B.}\fnm{Barbara} \snm{Bruno}}
\author[
  addressref={aff4},
  email={\EmailCA}
]{\inits{C.A.}\fnm{Catherine} \snm{Audrin}}
\author[
  addressref={aff4},
  email={\EmailMC}
]{\inits{M.C.}\fnm{Morgane} \snm{Chevalier}}
\author[
  addressref={aff2},
  email={\EmailSA}
]{\inits{S.A.}\fnm{Sunny} \snm{Avry}}
\author[
  addressref={aff2},
  email={\EmailJDZ}
]{\inits{J.D.Z.}\fnm{Jessica} \snm{Dehler Zufferey}}
\author[
  addressref={aff1, aff2},
  email={\EmailFM}
]{\inits{F.M.}\fnm{Francesco} \snm{Mondada}}


\address[id=aff1]{
  \orgdiv{MOBOTS Group}, 
  \orgname{EPFL, Ecole Polytechnique Fédérale de Lausanne},          
  \city{Lausanne},
  \cny{Switzerland} 
}
\address[id=aff2]{%
  \orgdiv{LEARN, Center for Learning Sciences},             
  \orgname{EPFL, Ecole Polytechnique Fédérale de Lausanne},          
  \city{Lausanne},                              
  \cny{Switzerland}
}

\address[id=aff3]{%
  \orgdiv{Computer Human Interaction in Learning and Instruction (CHILI) Laboratory},             
  \orgname{EPFL, Ecole Polytechnique Fédérale de Lausanne},          
  \city{Lausanne},                              
  \cny{Switzerland}
}

\address[id=aff4]{%
  \orgname{University of Teacher Education (Haute Ecole Pédagogique) Vaud},          
  \city{Lausanne},                              
  \cny{Switzerland}
}

\fi


\end{fmbox}


\begin{abstractbox}

\begin{abstract} 

\parttitle{Background} Early exposure to Computer Science (CS) and Computational Thinking (CT) for all is critical to broaden participation and promote equity in the field. But how does the introduction of CS \& CT into primary school curricula impact learning, perception, and gaps between groups of students?

\parttitle{Methodology} We investigate a CS-curricular reform and teacher Professional Development (PD) program from an equity standpoint by applying hierarchical regression and structural equation modelling on student learning and perception data from three studies with respectively 1384, 2433 and 1644 grade 3-6 students (ages 7-11) and their 83, 142 and 95 teachers.

\parttitle{Results} 

Regarding learning, exposure to CS instruction appears to contribute to closing the performance gap between low-achieving and high-achieving students, as well as pre-existing gender gaps. Despite a lack of direct influence of what was taught on student learning, there is no impact of teachers' demographics or motivation on student learning, with teachers’ perception of the CS-PD positively influencing learning. 

Regarding perception, students perceive CS and its teaching tools (robotics, tablets) positively, and even more so when they perceive a role model close to them as doing CS.  Nonetheless gender differences exist all around with boys perceiving CS more positively than girls despite access to CS education. However, access to CS-education affects boys and girls differently: larger gender gaps are closing (namely those related to robotics), while smaller gaps are increasing (namely those related to CS and tablets).

\parttitle{Conclusion} This article highlights how a CS curricular reform impacts learning, perception, and equity and supports the importance of i) early introductions to CS for all, ii) preparing teachers to teach CS all the while removing the influence of teacher demographics and motivation on student outcomes, and iii) having developmentally appropriate activities that signal to all groups of students.

\end{abstract}


\begin{keyword}
\kwd{\KWA}
\kwd{\KWB}
\kwd{\KWC}
\kwd{\KWD}
\kwd{Computational Thinking}
\kwd{\KWE}
\kwd{\KWF}
\kwd{Gender Gaps}
\end{keyword}


\end{abstractbox}
%

\end{frontmatter}




\section{Introduction and Related Work}
\label{sec:introduction}

\subsection{Introducing Computer Science and Computational Thinking for all from an equity perspective}

The past decades have seen a growing international consensus regarding the importance of teaching Computer Science (CS) and Computational Thinking (CT) to ensure that students are digitally literate in today's societies \citep{webb_computer_2017}. \added{Indeed, Computer Science is more and more often considered as a subset of STEM education which must be rendered as available to students as mathematics or science education given that computing is increasingly ubiquitous in today's societies \citep{guzdial_growing_2016}.}
\added{Introducing CS into formal education is also considered to foster Computational Thinking (CT),}\deleted{ While CT is traditionally fostered through CS and is considered} an essential skill for everyone in the 21st century \citep{jiang_exploring_2022} which is as important as reading, writing, and arithmetics 
\citep{wing_computational_2006}\added{. Teaching CT is not only considered by researchers to benefit STEM related disciplines \citep{swaid_bringing_2015}, but is also considered transversal with benefits that extend} \deleted{an increasing number of researchers advocate that CT is transversal and that the benefit of teaching CT extends} beyond CS or mathematics ~\citep{mannila_computational_2014, denning_computational_2021, weintrop_defining_2016, weintrop_assessing_2021, li_computational_2020}, providing an additional lever to introduce both CS and CT to all. 
Although studies on CS education and CT have increased significantly in recent years \citep{hsu_how_2018, bers_state_2022}, introducing CS \& CT into curricula has been a challenge internationally. \citet{ottenbreit-leftwich_computational_2022} recently expressed the importance of a ``system-wide implementation of CT'' from an equity perspective to ensure that all students are introduced to CT, and not just those of a select number of teachers who choose to teach CT. This is echoed by \citet{bers_state_2022} who advocate that exposure to CS \& CT should happen in early foundational years (ages 3-8) ``from a social equity perspective to prevent stereotypes and ensure [that] all young children receive equal opportunities to develop their digital literacy'', thus increasing the likelihood that a more diverse and inclusive set of people persist in these fields.
Two key points emerge from this discourse and must be addressed to broaden participation and promote equity in these fields: 

\begin{itemize}
  \item Structural barriers are access-related and limit (early) CS \& CT experiences for all, but can be addressed through curricular reforms \citep{ottenbreit-leftwich_computational_2022} 
 
  \item Social barriers, often stereotype (and therefore gender) related, arise despite equal access and regardless of socioeconomic status \citep{wang_diversity_2017}, but can be addressed through early exposure to mitigate the effects of existing stereotypes \citep{bers_state_2022}.
\end{itemize} 

The consequence is that disparities are present at multiple levels, including performance (i.e. learning) and attitudes towards CS (i.e. perception), which ultimately contribute to having under-represented groups in CS \& CT related fields. While the main focus of the article is on the impact of a CS-curricular reform generally, and its impact and on gender-equity (i.e. reducing significant differences between boys' and girls' \added{perception and performance}) \citep{cheryan_why_2017, jiang_exploring_2022}, performance-equity (i.e. reducing significant differences between initially low and high performers), and self-efficacy equity (i.e. reducing significant differences between students who have low or high self-efficacy), one must not neglect the importance of equity in terms of socioeconomic status \citep{wang_diversity_2017, vandenberg_interaction_2021}. 

\subsection{The influence of social \& structural barriers on learning-related equity}

Several studies have shown that unequal access to (high-quality) CS education \citep{wang_diversity_2017, bers_state_2022} contributes to performance gaps. A recent large-scale analysis of performance (46,000 students from 14 countries) by \citet{karpinski_computational_2021} found that socioeconomic background was related to persistent gaps in CT performance. In particular, students from ``less advantaged backgrounds had lower levels of computer skills [...], especially in CT'' \citep{karpinski_computational_2021}. Unfortunately, regardless of access, several studies have found that boys perform better than girls \ifdefined\Anonymous  (\citealp{roman-gonzalez_which_2017, polat_comprehensive_2021, kong_validating_2022}, \added{Anonymous authors - details removed for peer review -, 202X)} \else \citep{roman-gonzalez_which_2017, polat_comprehensive_2021, kong_validating_2022, el-hamamsy_competent_2022} \fi, even in kindergarten \citep{sullivan_girls_2016}, due to the existence of stereotypes 
(see section \ref{sec:barriers_on_perception}). 
Although access to developmentally appropriate CS \& CT education can increase students' skills from a young age \citep{bers_computational_2014, bers_coding_nodate, relkin_learning_2021, bers_state_2022}, several studies suggest that perception of the discipline can also influence performance. \citet{rachmatullah_toward_2022} found that the gender-performance gap was more prevalent in countries where the ``socio-cultural context'' tends to promote such stereotypes and ``influenc[e] gender diversity in the CS field''. This is corroborated by \citet{hinckle_relationship_2020} who found that student learning was not directly influenced by prior experience, but was mediated by their perception of CS. Numerous studies in higher education have also found that motivational and affective factors influence performance and participation in the field \citep{lishinski_self-efficacy_2022}, and that they are influenced by gender and ethnicity \citep{lishinski_self-efficacy_2022, warner_gender_2022}. These studies confirm the importance of:

\begin{itemize}
    \item developing CS and CT initiatives that broaden participation to all students,
    \item considering their impact on performance and perception to verify whether the gaps between different groups of participants are decreasing.
\end{itemize}

\subsection{The influence of social \& structural barriers on equity related to the perception of the discipline}
\label{sec:barriers_on_perception}

Perception-related biases are considered to contribute to disparities and under-representation in CS for women \citep{wang_diversity_2017, rachmatullah_toward_2022}, and more generally for under-represented minorities \citep{lishinski_self-efficacy_2022, warner_gender_2022}, due to stereotype threat (i.e., conforming to / inducing a stereotype simply because you know it exists). 
Unfortunately, the developmental literature has found that basic stereotypes develop in children as young as 2-3 years old \citep{bers_state_2022}. Multiple studies identified CS-related stereotypes in young children (e.g., starting at 6 years old as shown by \citealt{master_gender_2021}, and even kindergarten as shown by \citealt{sullivan_girls_2016}). 
Being exposed to negative CS-stereotypes, students in the stereotyped group (here girls) tend to endorse those beliefs \citep{plante_gender_2013, vandenberg_interaction_2021} which negatively impacts their performance, motivation, and career intentions \citep{master_cultural_2020, plante_gender_2013, vandenberg_interaction_2021}. For instance, 
\citet{cheryan_stereotypical_2013} found that women who were presented non-stereotypical views on computer scientists were more likely to express an interest in majoring in CS. 
Therefore, students may make early career decisions informed by such stereotypes, contributing to an early gender gap \citep{wang_diversity_2017}, and long-term disparities in the fields of CS and engineering \citep{master_gender_2021}.

As gender-related stereotypes are prevalent, it is not surprising that numerous studies find that girls perceive CS more negatively than boys \ifdefined\Anonymous (\citealp{witherspoon_gender_2016, kong_study_2018, vandenberg_interaction_2021}, \added{ Anonymous authors - details removed for peer review - g, 202X)} \else \citep{witherspoon_gender_2016, el-hamamsy_C3PP_2022, kong_study_2018, vandenberg_interaction_2021}\fi, contributing to a lower sense of belonging \citep{cheryan_stereotypical_2013, cheryan_why_2017, vandenberg_interaction_2021, opps_who_2022}, self-efficacy \citep{kong_study_2018, vandenberg_interaction_2021, beyer_why_2014}, and interest \citep{master_gender_2021, beyer_why_2014}. Provided the importance of such factors for academic achievement and career decisions \citep{bandura_perceived_1993, beyer_why_2014, olivier_student_2019, howard_student_2021}, the consequence is that CS ``suffers from the lowest participation of girls than other science, technology, engineering, and mathematics (STEM) subjects \citep{cheryan_why_2017}'' \citep{jiang_exploring_2022, hinckle_relationship_2020}. 
As prior experience may positively affect attitudes toward CS \citep{hinckle_relationship_2020}, researchers have suggested that engaging early in CS-related activities that ``signal equally to both girls and boys that they belong and can succeed'' \citep{cheryan_why_2017} in CS, may increase girls' interest, and ultimately contribute to addressing gender equity in the field \citep{cheryan_why_2017, hinckle_relationship_2020, jiang_exploring_2022}.
Therefore, in the rest of the article we refer to perception equity as the reduction of the influence of stereotypes around CS \& CT that lead to biases between groups of people (namely gender) and may influence their motivation, engagement, participation and persistence in these fields.

\subsection{How are CS \& CT curricular reforms having an impact and contributing to equity in these fields?}

Early CS and CT opportunities for all students are essential to address structural and social barriers, broaden CS participation, and promote equity in the field. An increasing number of initiatives have therefore sought to include CS and CT in compulsory K-12 worldwide \citep{voogt_computational_2015, hubwieser_global_2015, balanskat_computer_2015, webb_computer_2017, european_union_digital_2019, bocconi_reviewing_2022, bers_state_2022}. In this context, it is essential to establish how such initiatives affect students \citep{guskey_professional_2002}. This should extend beyond learning to include perception, and investigate how these dimensions interrelate \citep{hinckle_relationship_2020} to ensure that expanding CS to K-12 ``neither exacerbates existing equity gaps in education nor hinders efforts to diversify the field of CS'' \citep{wang_diversity_2017}. 
The student-level impact of widespread CS and CT curricular reforms, and professional development (PD) programs, is however seldom evaluated. As a pre-requisite to achieving equity is that the reform has an impact, this means that there is little insight into whether these reforms are contributing to equity and reducing learning and perception gaps between different groups of students. Indeed, ''studies that relate student's learning achievement and teachers' capacity building are still rare in the literature of CT \citep{mason_preparing_2019}'' \citep{kong_effects_2022}. 
This is likely due to the difficulties countries face implementing CS \& CT reforms, including adequately training a sufficient number of teachers to teach the new concepts \citep{bocconi_reviewing_2022, el-hamamsy_computer_2021}. Difficulties of assessing teachers' mastery of Computational Pedagogical Content Knowledge \citep{hickmott_assess_2018}, and what is implemented after PD programs \ifdefined\Anonymous \added{(Anonymous authors - details removed for peer review - d, 202X)} \else \citep{el-hamamsy_tacs_2022} \fi also exist, despite their direct influence on student learning \citep{kong_effects_2022}. To the best of our knowledge, only \citet{kong_effects_2022} linked 81 teachers' content knowledge with 3226 students' achievement in their evaluation of a PD program. However, these teachers chose to participate in the PD program and were required to teach a year-long curriculum. This differs significantly from mandatory curricular reform contexts, where the PD program is imposed on all teachers, resulting in teachers who implement the pedagogical content to varying degrees, if at all. 

Since a ``K-12 curriculum is a zero-sum game, where adding a subject means [removing] something'' \citep{ottenbreit-leftwich_computational_2022}, it is essential to establish the effectiveness of implementing CS \& CT curricula in formal education, notably given i) the need to improve corresponding PD programs and curricula \citep{hickmott_assess_2018}, ii) the objective of sustaining the reform in teachers' practices \citep{hubers_paving_2020}, and iii) the importance of alleviating concerns of funding agencies and government bodies regarding the impact of the reform and PD program on teachers \citep{hickmott_assess_2018} and students. Such studies are pressing since teachers are not necessarily convinced that their students are learning as a result of teaching these novel curricula \ifdefined\Anonymous  (\citealp{toh_leading_2016}, \added{Anonymous authors - details removed for peer review - e, 202X)} \else \citep{toh_leading_2016, el-hamamsy_sustainability_2022} \fi. Indeed, establishing the benefits at the student-level is not only necessary to have a complete evaluation of reforms \ifdefined\Anonymous  (\citealp{guskey_evaluating_2000}, \added{Anonymous authors - details removed for peer review - f, 202X, Anonymous authors - details removed for peer review - d, 202X)} \else \citep{guskey_evaluating_2000, avry_evaluating_2022, el-hamamsy_sustainability_2022}\fi, but is also a key factor found to affect teachers' decision to continue to implement a new practice in the long term \citep{klingner_examining_2001, howard_designing_2021}.  \\

\subsection{Problem Statement and Research Questions}
\label{sec:present_study}

The present study therefore looks to contribute to understanding the influence of CS curricular reforms on student learning and perception and determining to what extent they contribute to equity with respect to gender, performance, and self-efficacy. 
We propose to address this overarching question in two steps: first investigating whether and how the reform significantly influences perception and learning (impact), and then how the results differ according to student populations (equity).
To that effect, we investigate the impact of a mandatory CS curricular reform and teacher PD program (see section \ref{sec:EduNum}) to understand \emph{whether and how the primary school Computer Science curricular reform is contributing to reaching equity goals} (i.e. broadening participation in the field to a larger number and a more diverse set of people). 
Therefore, we consider the following research questions:

\begin{enumerate}[label=\textbf{(RQ\arabic*)}, leftmargin=2cm]
  \item How does teaching CS pedagogical content impact student learning? And how does it impact learning-related gender- and performance-equity?
  \item How does teaching CS pedagogical content impact students' perception of CS \added{and the tools used to teach it (i.e. robots and tablets)}? And how does it impact perception-related self-efficacy and gender-equity?
\end{enumerate}

To answer these questions we employ data collected between January 2021 and June 2022 in the context of a mandatory primary school CS-curricular reform that is presently being deployed to all grade 1-6 teachers in the region after a piloting phase. The data stems from three studies (see Table \ref{tab:studies_recap}), the first on student learning (RQ1), the second on perception of the discipline and performance (RQ1, RQ2), and the third on perception of the discipline (RQ2). These studies involved respectively $n_1=1384$, $n_2=2433$ and $n_3=1644$ grade 3-6 students (ages 7-11) and their $n_1=83$, $n_2=142$ and $n_3=95$ teachers. The data is analysed through hierarchical linear modelling for student learning, and Structural Equation Modelling for perception, to establish the link between teaching CS and these key outcome variables.

\begin{table*}[h]
\centering
\caption{Synthesis of the three studies evaluating the impact of the CS-curricular reform at the student-level}
\label{tab:studies_recap}
\footnotesize
\begin{tabular}{lcP{3cm}P{5cm}}
\toprule
 & \textbf{Study 1 - Learning} & \textbf{Study 2 - Perception \& Performance} & \textbf{Study 3 - Perception}  \\ \midrule
\textbf{Date} & January and June 2021 & November 2021 & June 2022  \\ 
\textbf{Grades} & 3-4 (ages 7-9) & 3-6 (ages 7-11) & 3-6 (ages 7-11)  \\ 
\textbf{Number of schools} & 7 CS-schools & 7 CS-schools & 3 CS-schools \& 2 non CS-schools \\ 
\textbf{Number of teachers} & $83$ & $142$ & $95$  \\ 
\textbf{Number of students} & $1384$ & $2433$ & $1644$ \\
\textbf{Student - CT-Concepts} & x & x &   \\ 
\textbf{Student - Perception of CS}   & & x & x  \\ 
\textbf{Teacher - Perception of CS} & x &   &  \\ 
\textbf{Teacher - Activities taught}  & & x &   \\ 
\bottomrule
\end{tabular}
\end{table*}

\section{Context: A Computer Science Curricular Reform for all to Promote Equity starting Primary School}
\label{sec:EduNum}

The research is part of a \added{large scale project seeking to introduce Digital Education (also referred to as Computing Education) as a new discipline} \deleted{mandatory K-12 Digital Education curricular reform} for all students in the \ifdefined\Anonymous Anonymous region (\added{Anonymous authors - details removed for peer review - a, 202X)} \else the Canton of Vaud in Switzerland \citep{el-hamamsy_computer_2021}\fi. The \added{curricular} reform relies on the collaboration between four institutions in the region (the department of education, the university of teacher education, a higher education university and the technical university) within a research practice partnership to develop the curriculum  and corresponding \added{mandatory teacher-}PD program for CS, Information and Communication Technology and Digital Citizenship. To ensure the sustainability and scalability of the reform, the project began with a piloting phase with 10 representative schools from the region (hereby referred to as CS-schools) before large-scale deployment. The \added{CS-curriculum and teacher PD-program} was piloted for the first time and iteratively adjusted for grades 1-4 in 2018-2019, and for grades 5-6 in 2019-2020, with all the teachers from the 10 CS-schools (approximately $n_{grades1-4}=350$, and $n_{grades5-6}=180$)
\footnote{The up-to-date Computer Science curriculum can be accessed at \url{https://www.plandetudes.ch/web/guest/education-numerique}}. 
This resulted in a reference manual \footnote{The 2021-2022 version of the pedagogical content can be accessed at
\url{https://www.vd.ch/fileadmin/user\_upload/accueil/Communique\_presse/decodage.pdf}} 
containing pedagogical activities (for CS, $n_{grades1-4}=13$, $n_{grades5-6}=12$) that the teachers can choose from to achieve the curricular objectives \added{(in terms of algorithms and languages, machines and networks, information and data, and the impact of CS on society)}. The teachers were trained to teach these activities \added{during a mandatory CS-PD that they participated in prior to the present study} and were encouraged to teach the novel \added{discipline}\deleted{CS curriculum} which is now part of the regional study plan\added{. They were however} not required to do so. Given that in primary school there is no dedicated hour in the grid for digital education, and that the discipline is not evaluated, this leads to a large variability in both what and how much is taught. This therefore required analysing the student-level impact of the curricular reform, and the influence being taught specific pedagogical content by teachers (which we refer to as adoption). 
While the initial focus was on student learning (see study 1 in section \ref{sec:study1_methodology}), a parallel pilot study in grade 9 (ages 13-14) in Spring 2021 indicated that there were already significant perception-related gender gaps \ifdefined\Anonymous \added{ (Anonymous authors - details removed for peer review - g, 202X)} \else \citep{el-hamamsy_C3PP_2022}\fi. This lead to the introduction of a student perception survey in Fall 2021 (see study 2 and study 3 in sections \ref{sec:study2}, and \ref{sec:study3_methodology}) to determine when gender gaps appear and whether teaching CS contributes to closing these gaps.

\section{Study 1 - Student Learning and the link with what teachers from the CS-schools implemented}
\label{sec:study1_methodology}

\subsection{Methodology}

\subsubsection{Participants and Data Collection}
\label{sec:study1_participants_data}

\begin{figure*}[htbp!]
 \centering \includegraphics[width=0.48\textwidth]{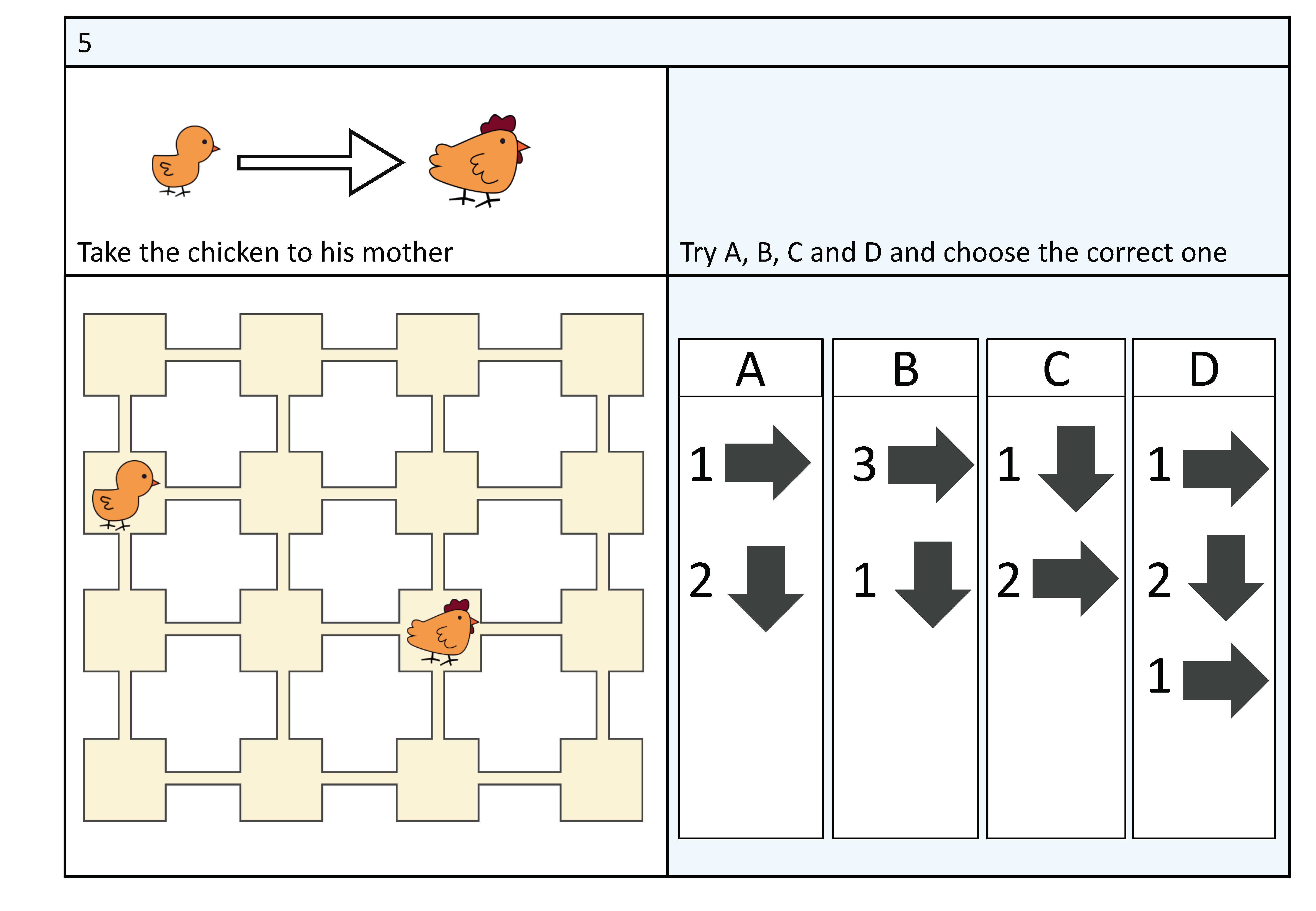} \includegraphics[width=0.48\textwidth]{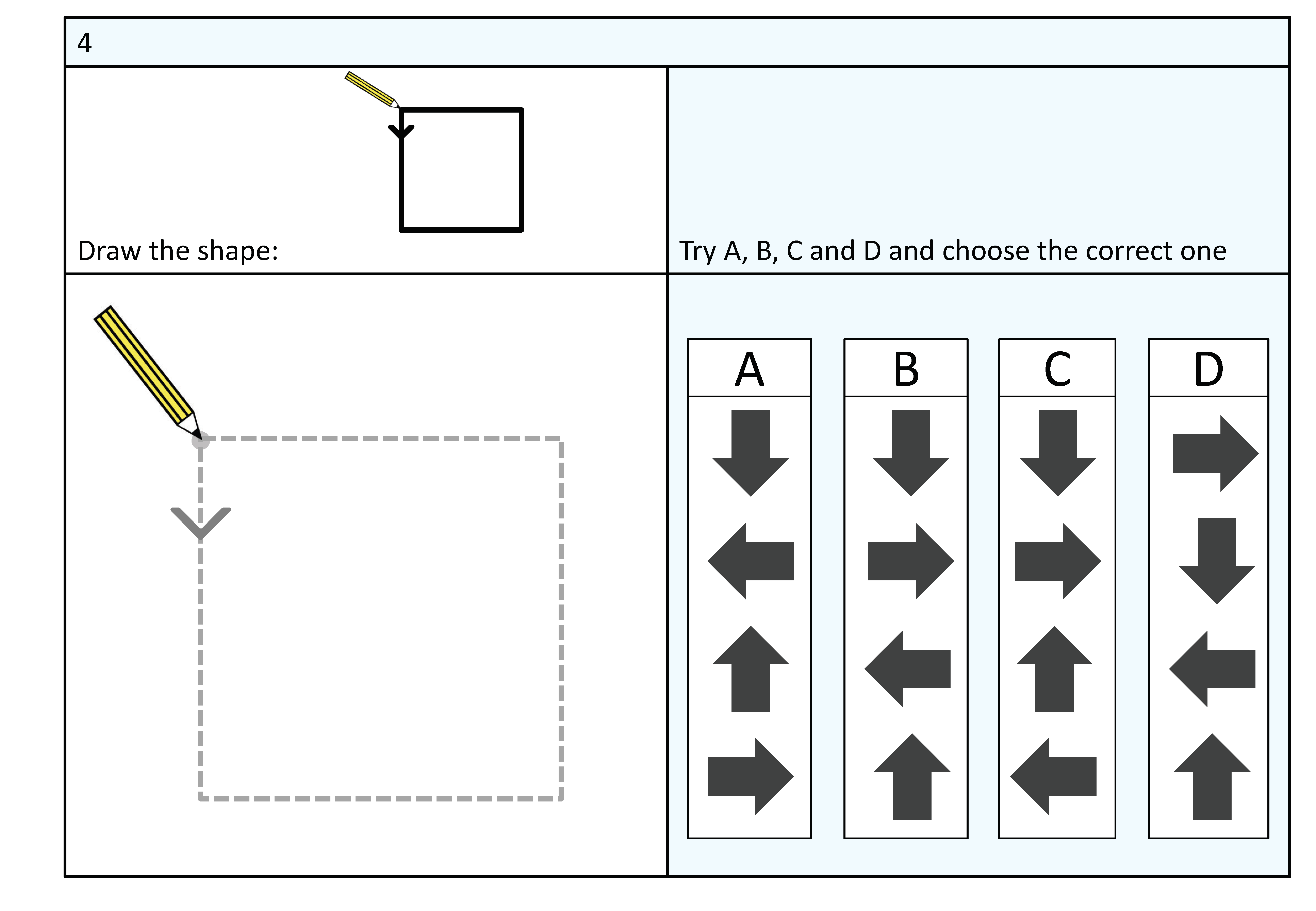}
 \vspace{-20pt}
 \caption{Two cCTt question formats: grid (left) and canvas (right) (Figure taken from~\citet{el-hamamsy_competent_2022}).}
 \label{fig:cCT_format}
\end{figure*}

The first study follows all the grade 3-4 students from 7 CS-schools over 6 months to evaluate learning in a pre- post-test design. These students were all introduced to CS for the first time during the 2018-2019 academic year and therefore had approximately 2 years of prior CS experience. The objective of the study was therefore to see to what extent these students progressed \added{over that time period in relation to what they were taught. Given the scale of the study, the objective was to focus on a subset of the learning objectives that could be measured in a valid and reliable way, and at a large scale, in grades 3-6. We therefore chose to focus on the} \deleted{in terms of} CT-concepts \deleted{that are aligned} \added{(as defined by} \citealt{brennan_new_2012}
\footnote{\citet{brennan_new_2012}\added{'s operational definition of CT decomposes CT into CT concepts (i.e. the concepts that computer scientists engage with), practices (i.e. the processes they employ to resolve computational problems) and perspectives (i.e. their perception of CT). Please note that at the time of the study there were no valid, reliable and scalable instruments to measure CT-practices and perspectives. }}\added{)}
which align with the \added{region's} CS curricular objectives (sequences, loops, conditionals, and while statements), all the while considering what the teachers taught between the pre- and post-tests. To that effect, we employed the competent Computational Thinking test (cCTt, \citealp{el-hamamsy_competent_2022}), a 25-item CT-concepts' assessment (see example questions in Fig. \ref{fig:cCT_format}) originally developed and validated for grades 3-4 that evaluates CS concepts of sequences, loops, if-else statements and while statements. \added{This instrument was later validated for grades 3-6, including a Differential Item Functioning analysis which demonstrates that the cCTt is not biased towards genders (i.e. it is gender fair) and can therefore be used to measure significant differences between boys' and girls' responses \ifdefined\Anonymous \added{ (Anonymous authors - details removed for peer review - h, 202X)} \else \citet{el-hamamsy_competent_2023} \fi.}

The student-learning data was complemented by data on teachers' perception of CS and the CS-PD acquired in January 2021, and data regarding what teachers taught (which we refer to as adoption) between January and June 2021 (see Table \ref{tab:teacher_survey}). \added{The adoption data is based on the activities that the teachers were introduced to during their CS professional development program and is collected in the form of a number of periods per activity which we are then able to convert into boolean values and derive the amount of CS activities taught.}

Please note that the datasets includes missing data due to i) students not being present for either the full pre- and/or post- tests, ii) teachers not administering the test, or iii) teachers not answering the pre- and/or post- teacher survey.
As the analyses combine multiple datasets, a synthesis of the number of students and teachers for which the full responses are available with respect to the data subsets considered is provided in Table \ref{tab:study_1_participants}. 
Finally, while it would have been interesting to have a control group to be able to infer how learning compared between students who had access to CS courses and those who did not, \deleted{the department of education did not authorise} the administration of a performance assessment to students in non-CS-schools \added{was not authorised} due to ethical concerns. Nonetheless, given the variability in what the teachers taught, 4 grade 3 classes and 6 grade 4 classes did not receive any CS education and thus provide an interesting point of comparison. 
As the second data subset (test + adoption data) constitutes the core of the analysis, we provide more detailed demographics information \added{in appendix \ref{app:study1bdemographics} in } Table \ref{tab:study1_learning_adoption_demographics}. 
 \\

\begin{landscape}
\begin{table}[h]
\centering
\caption{\added{Teacher survey questions (7-Point Likert, excepted adoption and demographic questions). Cronbach's $\alpha$ is provided for each sub-scale. Please note that the sample size (n=67) was too small to validate the measurement model through Confirmatory Factor Analysis.}
}
\label{tab:teacher_survey}
\notsotiny
\begin{tabular}{p{1.5cm}P{2cm}p{3cm}p{4cm}p{6cm}p{3cm}}
\toprule
Dimension & Reference & Concept & Original question (when different from the final question) & Adaptation to the context of the curricular reform (translated from \ifdefined\Anonymous Anonymous language \else French\fi) & M$\pm$SD ([min, max]) \\ \midrule
\multirow{1}{2cm}{Professional development program perception (Cronbach's $\alpha=0.88$)} &  &  &  & Root: Overall, the CS PD program  \\
 & \multirow{1}{2cm}{\ifdefined\Anonymous \added{ Anonymous authors - details removed for peer review - a, 202X)}\else \citet{el-hamamsy_computer_2021}\fi} & Interest &  & was rich and interesting & $1.58\pm1.11$ ($[-2.0, 3.0]$) \\
 &  & Difficulty &  & had an adapted difficulty & $1.2\pm1.27$ ($[-2.0, 3.0]$)  \\
 &  & Equilibrium &  & had a good equilibrium between theory and practice & $1.48\pm1.16$ ($[-1.0, 3.0]$)  \\
 & & Workshops &  Not applicable & I enjoyed the workshops &  $2.05\pm1.06$ ($[-2.0, 3.0]$)  \\
 &  & Trainers & Not applicable & I appreciated the trainers & $2.26\pm0.79$ ($[0.0, 3.0]$) \\
 &  & Exchanges & Not applicable & I enjoyed the exchanges with the other participants & $1.8\pm1.07$ ($[-1.0, 3.0]$)  \\
 & \citet{danaher_comparison_1996} & PD recommendation & Not applicable & I would recommend the CS PD program to other teachers & $1.41\pm1.29$ ($[-3.0, 3.0]$)  \\
 & \citet{danaher_comparison_1996} & Pedagogical activity recommendation & Not applicable & I would recommend the pedagogical activities seen in CS training to other teachers & $1.7\pm1.18$ ($[-2.5, 3.0]$)  \\ \midrule

\multirow{1}{2cm}{CS perception (Cronbach's $\alpha=0.93$)} &  &  &  & Root: Today I believe that CS  \\
 & \citet{roche_acceptation_2019} & CS school mission &  & is part of the school's mission & $0.66\pm1.41$ ($[-3.0, 3.0]$) \\
 & \citet{roche_acceptation_2019} & CS transversal utility &  & is useful for learning in other subjects & $0.92\pm1.22$ ($[-2.5, 3.0]$)  \\
 &  & CS global-utility &  & is generally useful & $0.94\pm1.14$ ($[-2.5, 3.0]$)  \\
 & \citet{deci_self-determination_1989} & CS potential for advancement & Not applicable & enables students to have a better chance of professional integration & $0.4\pm1.23$ ($[-2.0, 3.0]$)  \\
 & \citet{deci_self-determination_1989} & CS potential for advancement & Not applicable & enable students to become digital actors rather than consumers & $0.69\pm1.2$ ($[-2.0, 3.0]$)  \\
 & \citet{deci_self-determination_1989} & CS potential for advancement & Not applicable & allows students to express their creativity & $-0.07\pm1.42$ ($[-3.0, 3.0]$)  \\
 &  & CS global-non-utility &  & is not really worth it & $-1.25\pm1.24$ ($[-3.0, 2.0]$)   \\
 &  & CS non-utility &  & will not necessarily bring much to the students & $-0.86\pm1.24$ ($[-3.0, 2.0]$) \\ \midrule

\multirow{1}{2cm}{Autonomous motivation to teach CS (Cronbach's $\alpha_{all}=0.68$, $\alpha=0.76$ without Introjected regulation 1) based on the Situational Motivational \citep{angot_dynamique_2013} and Motivation at work \citep{gagne_motivation_2010} scales} &  &  &  & Root:  The reason I have implemented/plan to implement CS activities with my students is that  \\
 & \citet{angot_dynamique_2013} & Intrinsic motivation 1 & I find this activity really pleasant. & Teaching CS is fun. & $0.65\pm1.28$ ($[-3.0, 3.0]$)  \\
 & \citet{angot_dynamique_2013} & Identified regulation 1 & I think this activity is important to me. & I believe that teaching CS is important for students. & $0.84\pm1.26$ ($[-2.0, 3.0]$)  \\
& \citet{angot_dynamique_2013} & External regulation 1 & I feel I am supposed to do it. & I feel I am supposed to do this. & $0.75\pm1.61$ ($[-3.0, 3.0]$) \\
 & \citet{angot_dynamique_2013} & Intrinsic motivation 2 & I find this activity interesting & I find that teaching CS interesting. & $0.87\pm1.18$ ($[-3.0, 3.0]$)  \\
 & \citet{angot_dynamique_2013} & Identified regulation 2 & I find that doing this activity is good for me. & I find that teaching CS is useful for my students & $0.9\pm1.2$ ($[-2.5, 3.0]$)  \\
 & \citet{angot_dynamique_2013} & External regulation 2 & It's something I have to do. & I feel that this is something I have to do. & $0.7\pm1.37$ ($[-3.0, 3.0]$)  \\
 & \citet{gagne_motivation_2010} & Introjected regulation 1 & My work is my life and I don’t want to fail & I want to show others that I can do it & $-1.35\pm1.62$ ($[-3.0, 2.0]$) \\
 & \citet{gagne_motivation_2010} & Introjected regulation 2 & My reputation depends on it & My reputation depends on it & $-1.54\pm1.54$ ($[-3.0, 3.0]$) \\  \\  \\  \\  \midrule
 
Adoption &\ifdefined\Anonymous \added{ (Anonymous authors - details removed for peer review - d, 202X)} \else \citep{el-hamamsy_tacs_2022} \fi & Number of activities & & Which activities did you teach? \added{[List of all the CS pedagogical activities the teachers were trained to introduce in the CS-PD program they participated in in 2018-2019 for which they had access to all the requires material and pedagogical resources]} & $2.3\pm1.88$ ($[0.0, 7.0]$)  \\
& \ifdefined\Anonymous \added{ (Anonymous authors - details removed for peer review - d, 202X)} \else \citep{el-hamamsy_tacs_2022} \fi & Periods & & How many periods did you teach per activity? \added{[List of the activities that the teachers selected in the previous question]} & $9.15\pm10.36$ ($[0.0, 45.0]$)  \\ \midrule

Demographics &  & Age &  & Age & $39.44\pm11.51$ ($[23.0, 62.0]$)  \\
 &  & Teaching experience &  & Years of teaching experience & $15.24\pm10.91$ ($[0.5, 38.0]$)  \\
 &  & ICT experience &  & Years of experience with informatics & $11.78\pm8.15$ ($[0.0, 30.0]$)  \\
 &  & Teaching Digital Education experience &  & Years of experience in teaching digital education & $2.49\pm2.22$ ($[0.0, 18.0]$)  \\ 
 &  & Perceived ICT competence & & When it comes to ICT, I consider myself to be (1=a non user, 2=a novice, 3=an intermediate, 4=somebody who is used to it, 5=an expert) & $3.32\pm0.75$ ($[1.0, 5.0]$)  \\ 
 &  & Relative ICT competence &  & I tend to use digital technologies (1=after most of my colleagues, 2=as the same time as most of my colleagues, 3=before most of my colleagues, I am an early adopter, 4=before anybody else, I am an innovator) & $2.17\pm0.69$ ($[1.0, 4.0]$)  \\ \bottomrule
\end{tabular}
\end{table}
\end{landscape}

\subsubsection{Analysis Methodology}

The student learning data is analysed in three stages. 
First, the January and June test data ($n=1319$) is analysed using multiple ANOVA with Benjamini-Hochberg p-value correction to reduce the false discovery rate (study 1a). The \deleted{significant} results are reported \added{as significant (i.e. $p<0.05$)} only if the minimum effect size \added{(Cohen's D}\footnote{Cohen's D is a means of quantifying the difference between the means of two samples ($\mu_1$, $\mu_2$) all the while accounting for their standard deviations ($sd_1$ and $sd_2$). Cohen's D is therefore computed as the difference between the two sample's means divided by the pooled standard deviation ($s_p$). Therefore Cohen's $D=\frac{\mu_1-\mu_2}{s_p}$ where $s_p=\sqrt{\frac{sd_1^2 + sd_2^2}{2}}$. The rule of thumb to interpret Cohen's D is as follows: if around $0.2$ the effect is considered small, if around 0.5 the effect is considered medium and if around $0.8$ or above the effect is considered large.}) required to achieve a statistical power of $0.8$ is reached with $\alpha=0.05$. 
Dunn's post-hoc test is then applied for multiple comparisons when significant. \added{When comparing responses between groups of students (according to the dependent variables) the delta between the average scores on the cCTt's scale is provided ($\Delta$), in addition to the F-value, degrees of freedom, corresponding p-value and effect size using Cohen's D.}
The ANOVA considers the students' scores as the dependent variable, and the interaction between the following independent variables: time (pre-test or post-test), grade (3 or 4) and gender (boy or girl \added{as indicated on the school's records}\footnote{\added{Please note that we never asked students to relate their gender throughout the data collection process to avoid biasing students' responses and performance as a result of stereotype threat. Indeed, as we could not guarantee that all students would participate in all the data collections which were conducted over multiple sessions, and therefore could not solely rely on collecting the gender information at the end of the final data collection, we relied on the gender information obtained from the school's records. This information is provided by students' parents to the schools and therefore most likely aligns with the students' sex, without a guarantee that this corresponds to up-to-date information regarding the way students identify themselves. Furthermore this gender information was provided by the schools in a binary format. Although we acknowledge that gender relates to a person's identity, differs from biological sex \citep{risman_gender_2018}, and is increasingly recognised as being non binary, this was not yet fully the case in the country where the study was conducted at the level of formal primary education and at the time of the data collection. 
Indeed, at the time of the data collections, gender at the level of primary school and formal education more broadly was mainly considered as a binary construct. Nonetheless, most international studies find that the proportion of people who identify as transgender is generally inferior to 1.5\% (e.g. 0.6\% of the population aged 13 or older in the US, \citealp{herman_how_2022}; between 0.5\% and 1.3\% for children, adolescents and adults according to \citealt{zucker_epidemiology_2017}'s international review). The potential discrepancy between the gender information on the school's records and students' gender identity represents therefore at most a 1.5\% error which is below the level of significance which would affect the validity of the findings with a confidence level $\alpha=0.05$. Therefore, in order to align with the current practice in the STEM education community which often employ the term gender and gender biases when actually gathering and analysing binary or biological sex data (e.g. \citealp{jensen_undergraduate_2023, sung_short-term_2023, malespina_gender_2023}), we maintain the term gender, gender-biases and gender-gaps when referring to our data and our analyses.}}). 
Second, the dataset that introduces the adoption data ($n=989$), i.e., what the teachers taught between the pre- and post-test, is analysed through hierarchical linear modelling which nests students in classes and classes in schools (study 1b). Finally, to determine whether teacher-level variables (see Table \ref{tab:teacher_survey}) influence student learning, the third dataset (study 1c) that includes teacher perception is analysed through a correlation analysis with averaged class-level student scores ($n=67$), prior to a hierarchical linear modelling at the student-level ($n=752$). \added{The hierarchical linear modelling done in these two stages was conducted in R (version 4.2.1, \citealp{r_core_team_r_2019}) with nlme (version 3.1-157, \citealp{nlme1, nlme2}) and sjstats (version 0.18.2, \citealp{sjstats}).}

\begin{table*}[t]
  \centering
  \caption{\added{Study 1 on Student Learning - Number of Complete Observations According to the Data Considered: January \& June student test data, teacher adoption data (June), teacher perception data (January)}}
  \footnotesize
  \begin{tabular}{lcccc}
  \toprule
  & \multicolumn{3}{c}{\textbf{Number of students per grade}}  \\
  \textbf{Data subset} & Grade & Boys  & Girls &  Total  \\
  \midrule
  
  0. January or June test data & Grade 3 &  $357$ & $313$ &  $670$  \\
   ($n_{classes}=83$) & Grade 4 &   $363$ &       $351$ &  $714$  \\
   & Total &  $720$ &   $664$ &   $1384$  \\ \midrule
   
  1. January \& June test data &  Grade 3 & $332$ & $297$ &  $629$  \\
  ($n_{classes}=74$) - study 1a & Grade 4 &      $353$ &       $337$ &  $690$  \\
  & Total &  $685$ &   $634$ &   $1319$  \\ \midrule
  
  2. January \& June test data \& Adoption data & Grade 3 & $256$ & $224$ &  $480$  \\
  ($n_{classes}=55$) - study 1b & Grade 4 &      $265$ &       $244$ &  $509$  \\
   & Total &   $521$ &   $468$ &    $989$  \\ \midrule
   
  3. January \& June test data \& Adoption data \& Perception data  & Grade 3 &  $207$ & $186$ &  $393$  \\
  ($n_{classes}=43$) - study 1c & Grade 4  &      $207$ &       $173$ &  $380$  \\
  & Total &  $414$ &   $359$ &    $773$  \\
  \midrule

  4. January \& June test data \& Perception data & Grade 3 & $257$ & $238$ &  $495$  \\
  ($n_{classes}=55$) - study 1c & Grade 4 &      $262$ &       $232$ &  $494$  \\
  & Total &   $519$ &   $470$ &    $989$  \\
  \bottomrule
  \end{tabular}
  \label{tab:study_1_participants}
\end{table*}

\pagebreak

\subsection{Results - The impact of teaching CS on student learning}

\subsubsection{Student learning and the influence of gender and when the test was taken (study 1a)}

The ANOVA indicates that all independent variables and their interactions significantly influence the test score (\added{see appendix \ref{app:study1a}} Table \ref{tab:study1_anova_sig} \added{for a synthesis of the effects}) and the following trends emerge. 

\paragraph{Are all students progressing?}

\newcommand*{\ShowStatsInText}{}

\begin{figure*}[!h]
    \centering
    \includegraphics[width=1\textwidth]{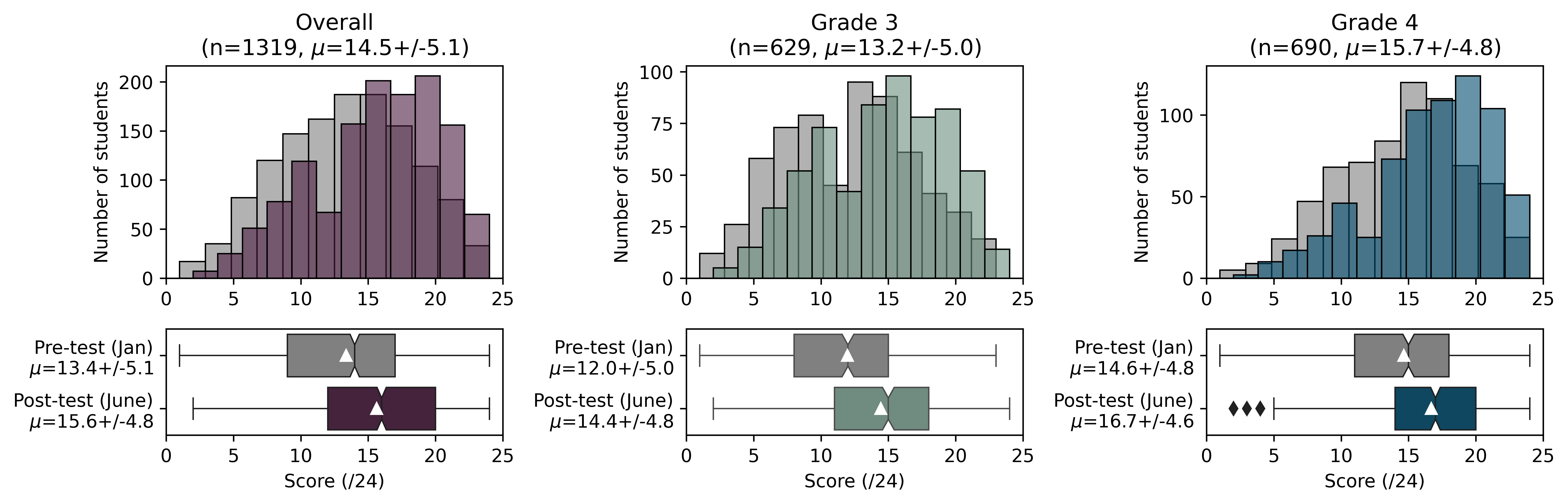}
    \caption{\added{Student performance distribution according to grade and whether in the pre- or post-test}}
    \label{fig:study1_grade_pre_post}
\end{figure*}

As Fig. \ref{fig:study1_grade_pre_post} \deleted{and Table \ref{tab:study1_anova_sig}} shows, grade 4 students perform better than grade 3 students with a medium effect size overall \ifdefined\ShowStatsInText (grade 4$>$3 $\Delta=2.468$, $p<0.0001$, \added{Cohen's} $D=0.502$)\fi
, in the pre-test
\ifdefined\ShowStatsInText  (pre-test grade 4 $>$ 3, $\Delta=2.686$, $p=0.0$, \added{Cohen's} $D=0.549$) \fi
and in the post-test\ifdefined\ShowStatsInText (post-test grade 4 $>$ 3 $\Delta=2.249$, $p=0.0$, \added{Cohen's} $D=0.482$)\fi
. 
Students also performed better on the post-test overall\ifdefined\ShowStatsInText  (post-test $>$ pre-test, $\Delta=+2.256$, $p<0.0001$, \added{Cohen's} $D=0.457$) \fi 
with students in grades 3 and grade 4 improving by a medium effect size\ifdefined\ShowStatsInText (grade 4 post $>$ pre, $\Delta=2.048$, $p=0.0$, \added{Cohen's} $D=0.436$; grade 3 post $>$ pre, $\Delta=2.485$, $p=0.0$, \added{Cohen's} $D=0.51$)\fi
. Interestingly, the grade 3 students' performance in the post-test (June) was equivalent to the grade 4 students' performance on the pre-test (January), although only 6 months separated the assessments \ifdefined\ShowStatsInText   (grade 4 pre-test $\sim$ grade 3 post-test, $\Delta=0.201$, $p=0.4444$, \added{Cohen's} $D=0.042$)\fi. 

\paragraph{Are there gender biases and are these closing?}

\begin{figure*}[h]
    \centering
    \includegraphics[width=1\textwidth]{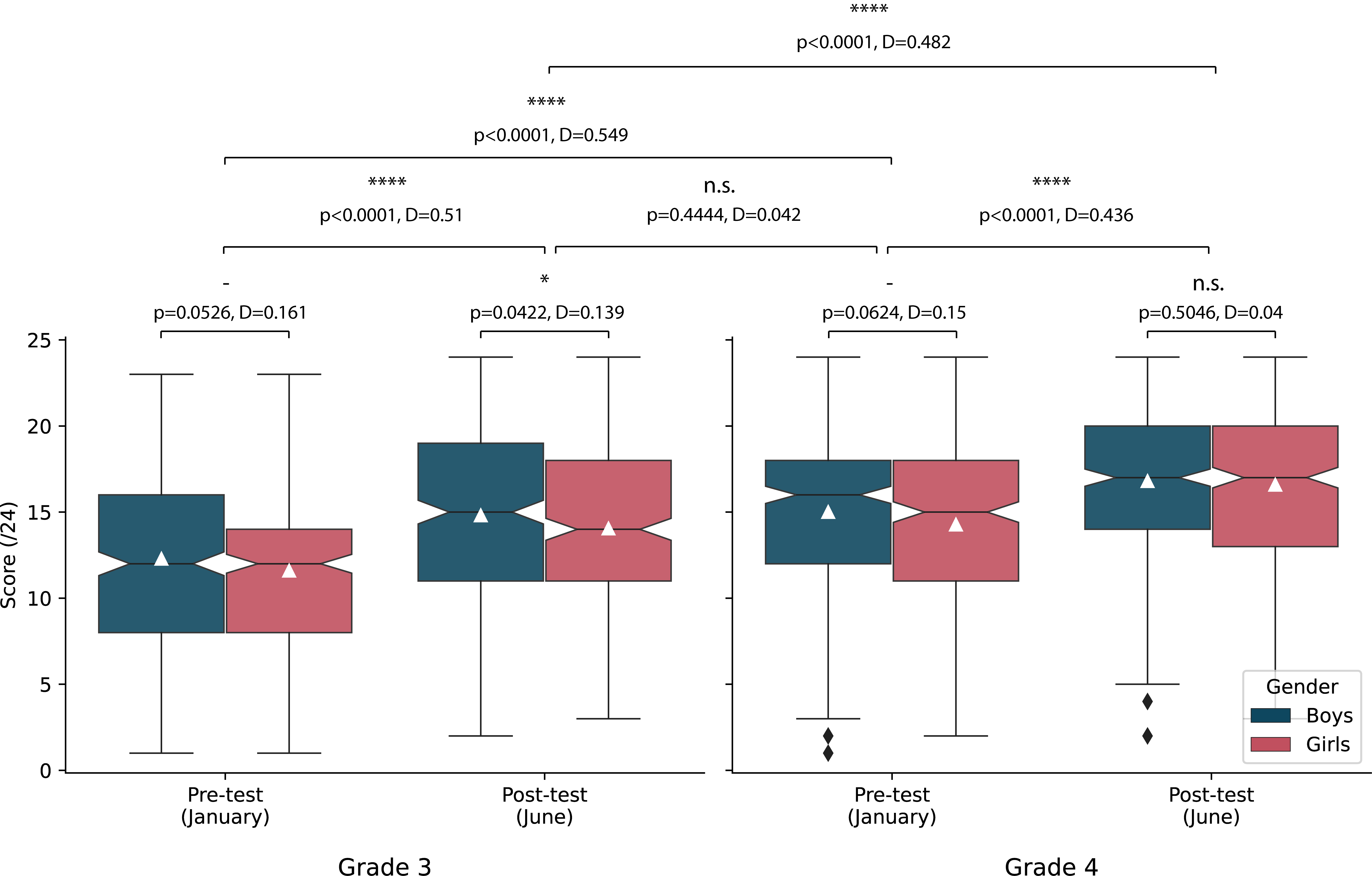}
    \caption{Student performance distribution according to grade, gender and whether in the pre- or post-test.  }\label{fig:study1_grade_gender_prepost}
\end{figure*}

The results that account for the students' gender alone show there is a significant main effect of students' gender on their performance. In particular, boys have significantly higher scores than girls overall with a small effect size \ifdefined\ShowStatsInText  (boys $>$ girls, $\Delta=0.551pts$, $p=0.0015$, \added{Cohen's} $D=0.109$)\fi
.
Considering the two-way interaction effects, we observe the following tendencies. Over all students, the gender gap is significant in the pre-test \ifdefined\ShowStatsInText  (January boys $>$ girls, $\Delta=0.664pts$, $p=0.0079$, \added{Cohen's} $D=0.131$)\fi
but decreases and is no longer significant by the post-test \ifdefined\ShowStatsInText   (June boys $\sim$ girls, $\Delta=0.438pts$, $p=0.0744$, \added{Cohen's} $D=0.091$)\fi
. 
Considering the two way interactions, these gender differences are significant in grade 3 \ifdefined\ShowStatsInText (grade 3 boys $>$ girls, $\Delta=0.725pts$, $p=0.004$, \added{Cohen's} $D=0.145$)\fi
, but not in grade 4 \ifdefined\ShowStatsInText (grade 4 boys $\sim$ girls, $\Delta=0.469pts$, $p=0.0604$, \added{Cohen's} $D=0.098$)\fi
. 
The three-way interaction between these variables thus helps shed some light on the trends observed (see Fig. \ref{fig:study1_grade_gender_prepost}) to draw conclusions:

\begin{itemize}
    \item In grade 3 there is a small marginally significant gap in the pre-test \ifdefined\ShowStatsInText  (grade 3 pre-test boys $\sim$ girls, $\Delta=0.764pts$, $p=0.0526$, \added{Cohen's} $D=0.161$) \fi
    and a small significant gap in post-test\ifdefined\ShowStatsInText  (grade 3 post-test boys $>$ girls, $\Delta=0.687pts$, $p=0.0422$, \added{Cohen's} $D=0.139$)\fi
    , with the effect sizes indicating that the gap is getting smaller, but has not yet closed. 
    \item In grade 4 there are small marginally significant differences in the pre-test \ifdefined\ShowStatsInText  (grade 4 pre-test boys $\sim$ girls, $\Delta=0.727pts$, $p=0.0624$, \added{Cohen's} $D=0.151$) \fi 
    and no significant differences observed in the post-test \ifdefined\ShowStatsInText  (grade 4 post-test boys $\sim$ girls, $\Delta=0.211pts$, $p=0.5046$, \added{Cohen's} $D=0.046$)\fi
    , indicating that the gender gap has closed.
\end{itemize}

\begin{figure*}[h]
    \centering
    \includegraphics[width=1\textwidth]{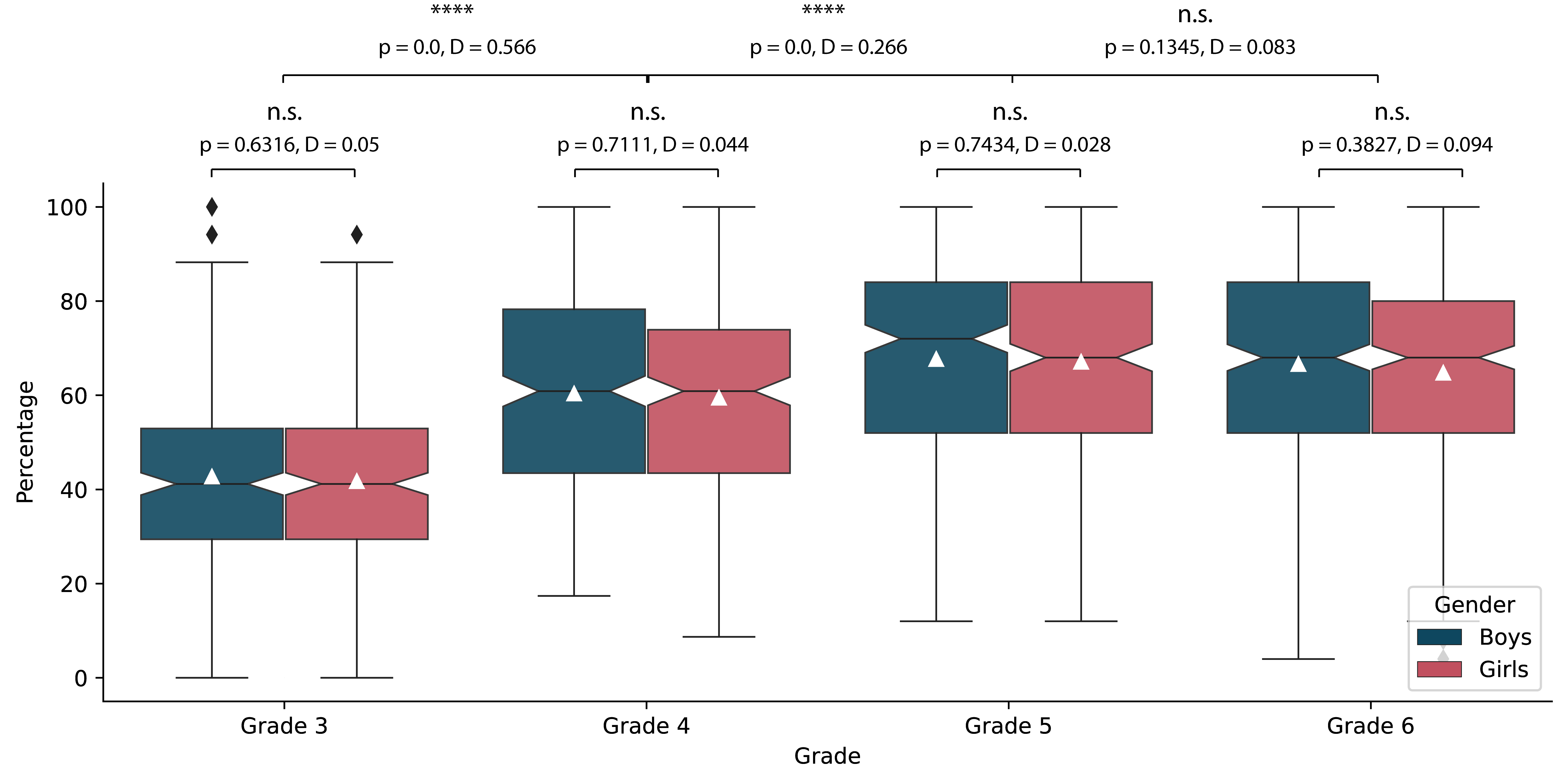}
    \caption{Student performance distribution according to grade and gender using data from the second study (n=2226, November 2021). All grade-differences are significant, excepted the one between grades 5 and 6 
    while the gender-differences per grade are non-significant.
    }
    \label{fig:study2_learning_grade_gender}
\end{figure*}

To complement these findings we consider the student learning data from study 2 (see section \ref{sec:study2}) that was conducted in November 2021 (5 months after the post-test of study 1) in the same schools and includes students from grades 3-6 (7-11). This is a particularly interesting cohort of students because students in grades 3 and 4 in study 2 are the first group of students to have had access to CS education starting first grade. 
Analysing the student performance data confirms that students continue to progress in terms of CT-concepts when moving on to grades 5 and 6 (see Fig. \ref{fig:study2_learning_grade_gender}). Indeed, the differences between grades 3 and 4 are significant ($\Delta =2.87pts$, $p < 0.0001$, \added{Cohen's} $D=0.566$), as well as those between grades 4 and 5 ($\Delta =1.35pts$, $p < 0.0001$, \added{Cohen's} $D=0.266$), although there is no significant difference between students in grades 5 and 6 ($\Delta =0.423pts$, $p=0.1345$, \added{Cohen's} $D=0.083$). This is consistent with the fact that students increase in maturity faster when they are younger. As such, students in grades 3 and 4 differ more significantly in terms of their cognitive abilities than students in grades 5-6 \citep{hartshorne_when_2015}. 

Evaluating the difference between boys' and girls' scores per grade indicates that the results are non-significant across grades (see Fig. \ref{fig:study2_learning_grade_gender}). As these students were in their 3rd or 4th year of CS education, this would appear to corroborate the previous findings: students who have had early and prolonged access to CS education are less likely to exhibit CS-performance gender-gaps. 

\FloatBarrier

\subsubsection{Student learning and the influence of the CS-education received (study 1b)}

\begin{figure*}[h]
    \centering
    \includegraphics[width=0.32\textwidth]{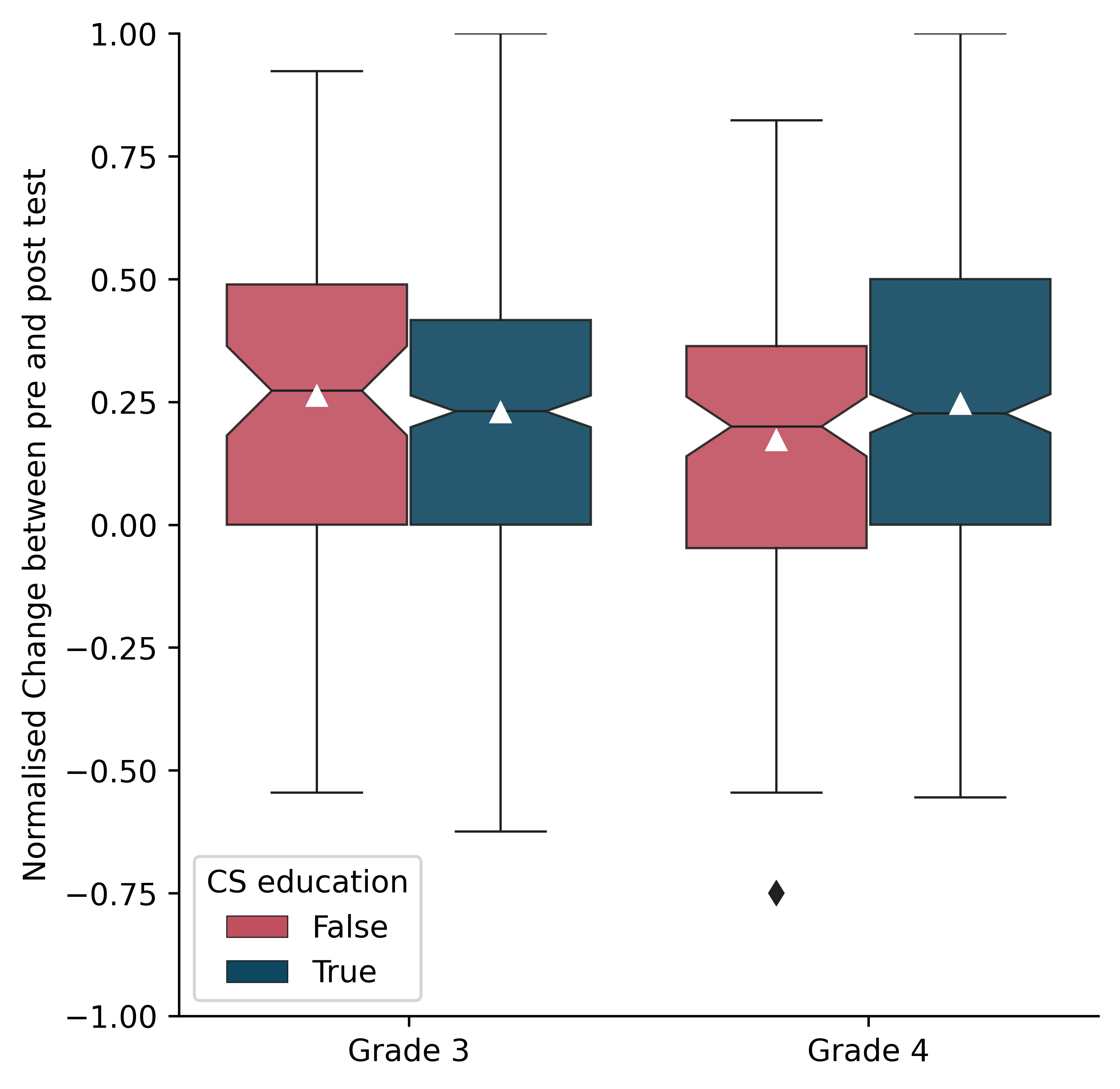}
    \includegraphics[width=0.64\textwidth]{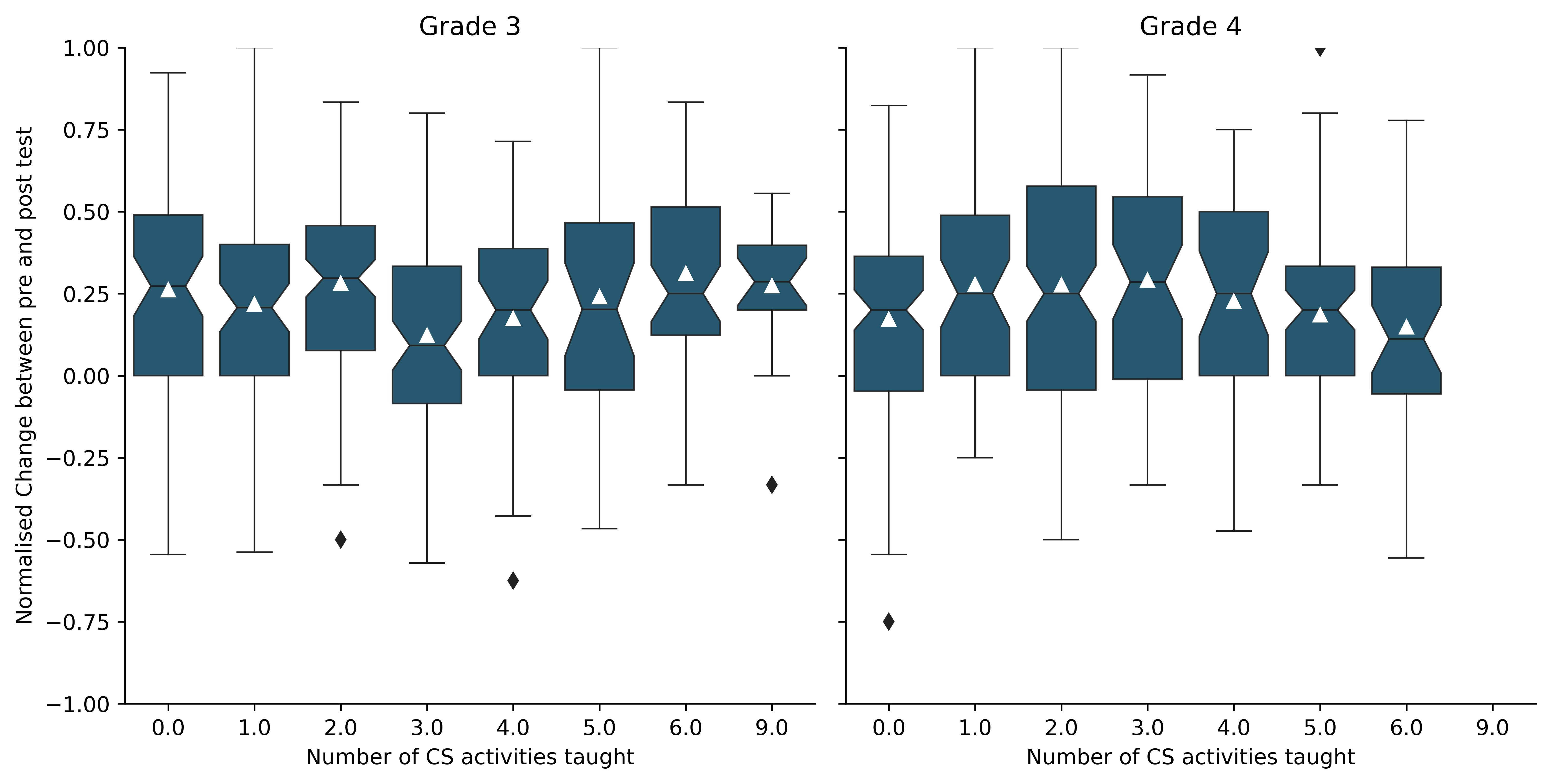}
    \caption{Student normalised change distribution according to grade, access to CS-education (left) and the number of CS-activities taught (centre for grade 3, right for grade 4). A two-way ANOVA between the grade and what was taught does not identify any significant differences between groups in terms of access to CS education ($F(2)=1.05$, $p=0.35$). A one-way ANOVA per grade did not identify any significant differences according to the number of activities taught (grade 3 $F_3(1)=0.13$, $p_3=0.72$; grade 4 $F_4(1)=0.89$, $p_4=0.35$)}
    \label{fig:study1_normalised_change_vs_adoption}
\end{figure*}

To understand how teaching the CS-pedagogical content \added{from the curriculum} may have influenced student learning, we consider the data from 989 students for whom the pre- and post- tests, and teacher adoption data (i.e. what the teachers taught, see section \ref{sec:study1_participants_data}) are available. We implemented \deleted{the} multiple hierarchical linear models while nesting \added{students in classes and} classes within schools to account for the different ways of considering student learning and adoption\footnote{
The hierarchical linear models considered the following: 
\begin{itemize}
    \item Dependent variables: the delta between the post-test and pre-test scores or the normalised change (a symmetrical version of the learning gain, \citealp{coletta_why_2020})
    \item Independent variables: the interaction between pre-test score, grade (3 or 4), and different adoption metrics (number of activities, or amount of CS-education time) 
    \item \added{Random effects: classes within schools. Please note that random effects are not the main focus of the analysis but still need to be included in the hierarchical linear model in order to account for their influence on the dependent variables. We therefore do not estimate the impact of each school or class on the outcome but rather control for them in order to avoid drawing erroneous conclusions.}
\end{itemize}
}.
These models consistently indicated that there was no direct link of adoption on students' post-test scores. For instance the model considering how the delta between the post and pre-tests is influenced by the students' grade, gender and the number of CS activities taught \added{estimates a non-significant} effect of the number of CS activities taught on the progress students made with $b=0.122$, $df=45$, $t=0.442$, and $p=0.661$ (see Table \ref{tab:student_learning_HR} \added{in appendix \ref{app:study1bHR}}).
Only the pre-test score significantly predicts the progress made in the post-test, with students performing lower at the pre-test progressing more. While the lack of a significant influence of CS activities taught on learning may appear surprising, visualising the trends between teaching and not teaching CS pedagogical content, as well as according to the number of activities taught, confirms the lack of an evident trend (see Fig. \ref{fig:study1_normalised_change_vs_adoption}).

\hspace{10pt}
\subsubsection{Student learning and the influence of teacher demographics, perception and the CS-PD received (study 1c)}

\added{Given the link between access to CS education and performance, and the lack of a direct link between what the teachers taught and student learning, it would appear that there are additional factors at play when affecting learning. Therefore} in a final phase, the teachers' aggregate i) perception of the PD program, ii) perception of CS, iii) autonomous motivation to teach CS\footnote{The Autonomous Motivation (AM) score is computed using the Relative Autonomy Index \citep{grolnick1989parent} by combining the sub-scales for intrinsic motivation (IM), identified regulation (IdR), introjected regulation (InR) and external regulation (ER) and aggregating them as explained by \citet{howard_review_2020}. That is to say: $AM=(2\times IM +1\times IdR-1\times InR-2\times ER)/6$} 
and the iv) demographic data collected at the same time as the pre-test was put in relation to the results of student learning. First, the students' results were averaged per class and correlated with the teacher-level variables. As the perception data is on a 7-Point Likert scale and non-normally distributed, Spearman's rank correlation was used. All the correlations with class performance were non-significant (demographics, prior experience, CS perception), with the exception of the training evaluation (Spearman's rho=0.33, p=0.007). 

As adoption was found to be not significantly related to student learning (study 1b), we compared two hierarchical linear models at the student-level, one with and one without adoption variables, with both including student-level, teacher perception-level and teacher demographic-level variables. An analysis of variance between the two models indicates that the difference is non-significant (p=0.768) and that the more parsimonious model should be preferred. This also has the advantage of containing a larger set of complete data (1027 observations versus 752). 
The resulting hierarchical linear model at the student-level (see Table \ref{tab:student_learning_teacher_HR}) confirms the trend observed in the correlation analysis, and indicates that the following dependent variables predict the delta between the pre and post test scores, with no influence of teacher demographic variables (including teaching and ICT experience): 

\begin{itemize}
    \item The pre-test score predicts the delta negatively ($p<0.0001$, $\beta=-0.35$), i.e. students performing lower at the pre-test progressed more.
    \item The average PD program evaluation score predicts the delta positively ($p=0.0053$, $\beta=1.02$), i.e. students of teachers who positively viewed the CS-PD progressed more.
\end{itemize}

\begin{table*}[h]
    \centering
    \caption{Hierarchical linear model for student learning with respect to student-, and teacher-level variables (dependent variable: Delta between pre- and post- test scores, $n=1027$ students in $57$ classes in $6$ schools). Significant variables are highlighted in bold. $R^2=0.279$, $AIC=5386$, $BIC=5474$, $RMSE=3.04$. \added{Random effects $\sigma^2=9.72$, $\tau_{class}=0.57$, $\tau_{school}=1.37$}}

    \footnotesize
    \begin{tabular}{p{2cm}lcccP{1.5cm}ccc}
    \toprule
    Variables & & Estimate & 95\% CI & Std.Error & Degrees of Freedom & t-value &  p-value  \\ \midrule
    & \textbf{(Intercept)} & $10.48$ & \added{$[5.42, 15.54]$} & $2.579$ & $968$ & $4.06$ & $0.0001$  \\  \midrule
    Student-level & $\textbf{Pre-test score} $ & $ -0.35$ & \added{[-0.40, -0.31]} &  $0.023$ & $968$ & $-15.54$ & $\textbf{0.0000}$  \\
    & Gender (girl)  & $ 0.17$ & \added{$[-0.22, 0.55]$} & $0.198$ & $968$ & $0.84$ & $0.4021$ \\
    & Grade (3)  & $ 0.35$ & \added{$[-0.34, 0.53]$} & $0.823$ & $39$ & $0.42$ & $0.6746$  \\
    & Grade (4)  & $ 1.15$ & \added{$[-0.13, 0.73]$} & $0.814$ & $39$ & $1.41$ & $0.1663$  \\ \midrule
    \multirow{1}{2cm}{Teacher-perception} & \textbf{CS-PD program evaluation} & $ \textbf{1.02}$ & \added{$[0.07, 0.38]$}  & $0.344$ & $39$ & $2.96$ & $\textbf{0.0053}$  \\
    & CS utility perception & $ 0.15$ & \added{$[-0.84, 1.06]$} & $0.628$ & $39$ & $0.24$ & $0.8122$  \\
    & CS non-utility perception & $ 0.27$ & \added{$[-0.72, 1.18]$} & $0.600$ & $39$ & $0.45$ & $0.6561$  \\
    & CS autonomous motivation & $ -0.21$ & \added{$[-0.97, 0.56]$} & $0.380$ & $39$ & $-0.54$ & $0.5896$  \\ \midrule
    \multirow{1}{2cm}{Teacher demographics} & Age & $ -0.14$ & \added{$[-0.30, 0.01]$} & $0.078$ & $39$ & $-1.83$ & $0.0743$  \\
    & Experience with informatics & $ 0.01$ & \added{$[-0.06, 0.09]$} & $0.036$ & $39$ & $0.41$ & $0.6828$  \\
    & Teaching experience & $ 0.11$ & \added{$[-0.05, 0.27]$} & $0.079$ & $39$ & $1.42$ & $0.1640$  \\
    & Digital education teaching experience & $ -0.10$ & \added{$[-0.31, 0.11]$} & $0.104$ & $39$ & $-0.94$ & $0.3552$  \\
    & Perceived ICT competence & $ -0.57$ & \added{$[-1.34,
0.19]$} & $0.376$ & $39$ & $-1.53$ & $0.1345$  \\
    & Perceived relative ICT competence & $0.20$ & \added{$[-0.65, 1.05]$} & $0.421$ & $39$ & $0.47$ & $0.6387$
      \\ \bottomrule
    \end{tabular}

    Please note that (i) the classes had an average of $18\pm2$ students per class (minimum $14$, maximum $22$); (ii) the schools had an average of $8\pm5$ classes (i.e. $8$, $8$, $1$, $17$, $6$, $7$, $8$ classes) who participated in the three data collections required for this analysis. These numbers are coherent with the relative sizes of the schools, with the exception of the third where the majority of teachers chose not to participate in the data collection.
    
    \label{tab:student_learning_teacher_HR}
\end{table*}

\pagebreak
\subsection{Synthesis and limitations of study 1}

The students progress in terms of CT-concepts over time with grade 3 students achieving a year's worth of CT-development in a 6-month window (study 1a, \textit{positive impact}).  However, the results of the hierarchical linear modelling indicate that there is no direct effect of what was taught with the progress students made (study 1b, \textit{no impact and therefore negative for equity}). The only factors that appear to influence learning are: i) the students' scores in the pre-test, with students who have lower scores progressing more thus contributing to \textit{performance-equity}; ii) the teachers' perception of the PD program (study 1c, \textit{positive impact}). There is additionally no influence of teachers' demographics on what the students have learnt, indicating that the PD program helped prepare teachers to teach CS content, irrespective of their prior teaching experience and ICT experience. This contributes once more to \textit{equity} by ensuring that all students have access to quality CS education, irrespective of the teachers' background (structural barriers). Finally, the findings indicate the existence of gender gaps (study 1a, likely due to social barriers) but that these get smaller the longer students are in contact with CS education (positive for \textit{gender-equity}). 

There are however limitations due to the lack of a true control group that has never had access to CS education. Indeed, the students in the present study were not compared to students who had not done any CS education between the pre- and post-tests, or since the start of their schooling. The fact that students with lower pre-test scores progress more may also be due to the existence of a ``ceiling effect'' for already higher performing students (either cognitively, with respect to what the cCTt measures, or what is attainable with the pedagogical content taught). In terms of teacher and class data, while the teachers were asked what they taught and for how long, this does not indicate their mastery of the content, the implementation fidelity (i.e. to what extent they put emphasis on the CS concepts in these activities) or whether they taught other activities that were not part of the PD program that may be linked to CS education or grid based concepts which are also part of the maths curriculum. Finally, the assessment:
\begin{itemize}
    \item focuses on CT-concepts, although there are other elements of CT that may be positively affected by access to CS education which are not measured \added{(in addition to other dimensions of the CS curricular reform including those pertaining to machines and networks, data and information and the impact of CS on society)};
    \item \added{is used in both the pre- and post-test due to the fact that (i) at the time of the studies there existed no valid and reliable assessment of CT-concepts in primary school for these grades; (ii) no validated assessment proposes isomorphic variants which have been proven to have the exact same difficulty and can therefore be reliably employed in the comparison of pre-post test design. To the best of our knowledge this remains true today as only \citet{parker_pair_2022} has begun investigating how to create an isomorphic version of their instrument (the ACES test) and analysed what types of changes to the questions could truly be considered isomorphic in this context. This is important because ``seemingly superficial changes in an item’s context can cause students to recruit different knowledge and cognitive processes when solving a problem'' \citep{parker_pair_2022}}.
\end{itemize}

\FloatBarrier

\section{Study 2 - Student Perception, the link with what teachers from the CS-schools implemented, and correlations with performance}
\label{sec:study2}

\subsection{Methodology}

\subsubsection{Participants and Data Collection}

This study extends the first by evaluating students' mastery of CT-concepts and their perception of the discipline. The data collection was conducted in November 2021 and involved all students from grades 3-6 in the 7 CS-schools involved in the first study (see Table \ref{tab:study2_participants}). The students first responded to a perception survey, before being administered the cCTt (which was shown to be adapted for grades 5-6 in \citet{el-hamamsy_competent_2023} to assess their mastery of CT-concepts.  \\

\begin{table*}[h]
    \centering
    \caption{Number of students participating in the first perception survey and the third test (study 2, November 2021) and their intersection with the teacher adoption survey}
    \label{tab:study2_participants}
    \footnotesize
    \begin{tabular}{llccccc}
    \toprule
    Subset & Gender &  \multicolumn{4}{c}{Grade} & Total  \\
    & & $3$ &    $4$ &    $5$ &    $6$ &   \\
    \midrule
     & Boys &  $263$ &  $307$ &  $314$ &  $334$ &   $1218$   \\
    Perception ($n_{classes}=142$) & Girls &  $265$ &  $286$ &  $311$ &  $328$ &   $1190$  \\
     & All &  $528$ &  $593$ &  $625$ &  $662$ &   $2408$   \\ \midrule
     
     & Boys &  $196$ &  $230$  &  $285$  &  $351$  &   $1062$   \\
    Perception \& adoption  ($n_{classes}=114$) & Girls &  $201$ &  $220$  &  $287$  &  $346$  &   $1054$   \\
     & All &  $397$ &  $450$  &  $572$  &  $697$  & $2116$   \\\midrule
     
     & Boys &  $240$ &  $282$  &  $289$  &  $317$  &   $1128$    \\
    Test ($n_{classes}=140$) & Girls &  $243$  &  $252$ &  $296$  &  $307$  &   $1098$   \\
    & All &  $483$ &  $534$  &  $585$  &  $624$  &   $2226$   \\\midrule
    
    & Boys &  $265$ &  $311$  &  $317$  &  $337$  &   $1230$   \\
    Test or Perception ($n_{classes}=142$) & Girls &  $272$ &  $287$  &  $314$  &  $330$  &   $1203$   \\
    & All &  $537$ &  $598$  &  $631$  &  $667$  &   $2433$ \\ \midrule
    
     & Boys &  $209$  &  $262$  &  $270$  &  $295$  &   $1036$   \\
    Test \& Perception ($n_{classes}=139$) & Girls &  $214$ &  $239$  &  $281$  &  $294$  &   $1028$   \\
     & All &  $423$ &  $501$  &  $551$  &  $589$  &   $2064$   \\\midrule
     
     & Boys &  $166$  &  $198$  &  $285$  &  $267$  &    $916$   \\
     Test \& Perception \& adoption ($n_{classes}=105$) & Girls &  $175$  &  $182$  &  $287$  &  $252$  &    $896$   \\
     & All &  $341$  &  $380$  &  $572$  &  $519$  & $1812$ \\
    \bottomrule
    \end{tabular}
\end{table*}

\begin{table*}[h]
\centering
\caption{\added{Student perception survey items translated from \ifdefined\Anonymous Anonymous language \else French\fi. 
Cronbach's $\alpha_{CS, Robotcs, Tablets}=0.67$ for the 9 items consisting of 3 sub-scales using the 5-Point Analog Visual Scale (5PT-AVS) and is considered to be between acceptable and good \citep{george_spss_2003}, and to have between moderate and high reliability \citep{hinton_spss_2004}.}}
\footnotesize
\label{tab:study23_student_survey_items}
\begin{tabular}{llp{7.5cm}l}
\toprule
\textbf{Dimension} & \textbf{Concept} & \textbf{Question} & \textbf{Format}  \\ \midrule
Computer Science & Interest & I like informatics & 5PT-AVS  \\
 & Self-efficacy & I am capable of learning informatics & 5PT-AVS  \\
 & Utility & We can do a lot of things with informatics & 5PT-AVS  \\
 & Role Models & When I think of someone who does informatics I think of (you can chose multiple answers): & Checkboxes  \\ 
 & & \hspace{20pt} \textit{The teacher} \\ 
 & & \hspace{20pt} \textit{My mother} \\
 & & \hspace{20pt} \textit{My father} \\ 
 & & \hspace{20pt} \textit{A sibling or friend} \\ 
 & & \hspace{20pt} \textit{Somebody else} \\ 
 & & \hspace{20pt} \textit{Nobody}  \\
 \midrule
Robotics & Interest & I like robots & 5PT-AVS  \\
 & Self-efficacy & I am capable of using robots & 5PT-AVS  \\
 & Utility & We can do a lot of things with robots & 5PT-AVS  \\
 & Usage & When I am at school or at home I use or play with the following robots (you can chose multiple answers): & Checkboxes  \\
 & & \hspace{20pt} \textit{Thymio} \\ 
 & & \hspace{20pt} \textit{Bluebot} \\ 
 & & \hspace{20pt} \textit{Lego Robots (WeDo, Spike, Prime, Mindstorm or Technic)} \\
 & & \hspace{20pt} \textit{Cubetto} \\ 
 & & \hspace{20pt} \textit{mBot} \\ 
 & & \hspace{20pt} \textit{Ozobot} \\ 
 & & \hspace{20pt} \textit{Other robots} \\ \midrule
Tablets & Interest & I like tablets & 5PT-AVS  \\
 & Self-efficacy & I am capable of using tablets & 5PT-AVS  \\
 & Utility & We can do a lot of things with tablets & 5PT-AVS  \\
 & Usage & When I am at school or at home I use a tablet or computer to (you can chose multiple answers) & Checkboxes \\ 
 & & \hspace{20pt} \textit{Take photos and videos} \\ 
 & & \hspace{20pt} \textit{Call, text, watch videos or listen to music} \\ 
 & & \hspace{20pt} \textit{Play games} \\ 
 & & \hspace{20pt} \textit{Read} \\ 
 & & \hspace{20pt} \textit{Program (e.g., Scratch)} \\ 
 & & \hspace{20pt} \textit{Draw, create interactive albums or music} \\ 
 & & \hspace{20pt} \textit{Nothing}  \\  
\midrule
General & \added{School-related} self-efficacy & I am capable of doing well at school & 5PT-AVS  \\ \bottomrule
\end{tabular}
\footnotesize Please note that the survey also included items pertaining to the usage of robots and tablets but that these are not investigated in the present article. 
\end{table*}

The perception survey (see Table \ref{tab:study23_student_survey_items}) targeted three dimensions.
The first is the students' perception of \emph{Computer Science}, including who they perceive as doing CS, called "informatics" in the region, a scalable alternative to the draw-a-computer-scientist test \citep{pantic_drawing_2018}. 
Students were asked whether they perceived certain role models (e.g., influencers such as parents and teachers, \citealp{wang_diversity_2017}), someone else, or nobody, as doing CS. One hypothesis is that students who have access to CS-education are more likely to perceive their teachers as role models. As primary school teachers are mainly women, they can be considered female role models, an element that is key to engaging girls in the field \citep{cheryan_why_2017, kong_study_2018}. Another hypothesis is that perceiving people "close to them" as doing CS (i.e., related to the idea that CS is becoming ubiquitous and accessible to all), will contribute to improved perception of CS overall. 
The second dimension is how students perceive \emph{robots}, as robotics is a means of teaching CS \ifdefined\Anonymous \added{(Anonymous authors - details removed for peer review - c, 202X)} \else \citep{el-hamamsy_symbiotic_2021}\fi, and CS and engineering tend to be subject to stronger stereotypes than science and maths among young students \citep{master_programming_2017}. 
The third dimension is how students perceive \emph{tablets and other digital devices} which are also employed as means of teaching CS (and ICT) in the curricular reform.  \\

For each of these dimensions (CS, robotics, tablets), the emphasis is placed on three factors that are ``different but related aspects of motivation'' \citep{master_programming_2017} and can be considered as predictors of academic achievement in general \citep{bandura_perceived_1993, olivier_student_2019, howard_student_2021}, educational choices, and career decisions \citep{blotnicky_study_2018, wang_role_2020, mason_development_2020}, in addition to being the most prominent in surveys evaluating students' (at all levels of education) perception of CS, coding or STEM \citep{mason_development_2020}: 

\begin{itemize}
    \item Interest, i.e. "how much the individual likes or is interested in the activity" \citep{mason_development_2020}, which is a key component of intrinsic motivation in self-determination theory \citep{ryan_intrinsic_2020} and expectancy-value theory \citep{eccles_expectancy-value_2020}. Several studies have found that boys tend to be more interested in CS than girls, as in the case of most STEM-related disciplines \citep{mason_development_2020}, but that interest increases after access to CS-related experience, in particular for girls, which contributes to closing the interest gender gap \citep{master_programming_2017}. 
    \item Self-efficacy \citep{bandura_perceived_1993, kong_study_2018}, i.e. ``a person's belief that they can complete a particular task or fulfil a particular role within a specific domain'' \citep{mason_development_2020}. Similarly to interest, self-efficacy has been found to be higher for boys than girls in STEM-related domains, and to increase with computing experience, in some cases even contributing to closing the gender gap \citep{mason_development_2020}, whether related to programming \citep{gunbatar_gender_2018}, or robotics \citep{master_programming_2017}. Please note that as domain-specific self-efficacy may also be related to general self-efficacy we also consider a school-related self-efficacy variable in the survey \added{(which refers to how well students believe they are able to perform in school in general)}. 
    \item Perceived utility \citep{wigfield_expectancyvalue_2000, eccles_expectancy-value_2020}, a component of expectancy-value theory referring ``to how a task fits into an individual's future plans'' and is considered to ``directly [influence] a person's achievement-related choices, and is influenced by a person's experiences, perceptions, goals, and self-schemata'' \citep{mason_development_2020, wigfield_expectancyvalue_2000}.
\end{itemize}

\begin{figure*}[h]
    \centering
    \includegraphics[width=0.8\textwidth]{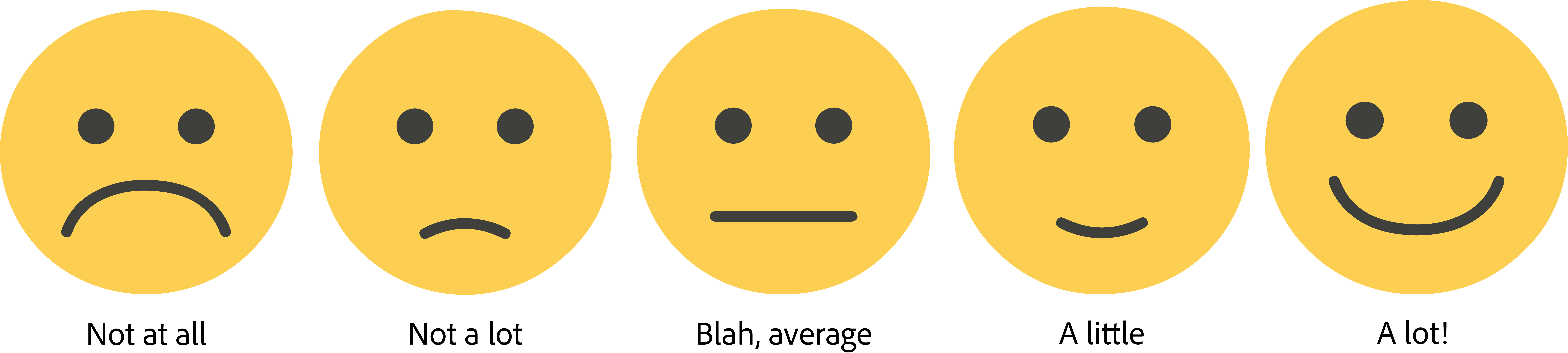}
    \caption{Analog Visual Scale employed for the student survey's Likert questions. The labels in \ifdefined\Anonymous Anonymous language \else French\fi (original survey language) were established with teachers and validated in a pilot run with two classrooms.}
    \label{fig:AVS_scale}
\end{figure*}

Given that the same survey was administered from grades 3 to 6 in conjunction with the test, the survey needed to be short to account for their age and attention span (see Table \ref{tab:study23_student_survey_items}). Cronbach's $\alpha$ measurement of internal consistency of scales is provided for all Likert-type questions employing an analog-visual scale (see Fig. \ref{fig:AVS_scale}). This is complemented by a Confirmatory Factor Analysis to confirm the adequacy of the complete measurement model (see section \ref{sec:study2_results}). Finally, the student survey was complemented by a teacher adoption-survey that asked each teacher the amount of time spent teaching each of the CS, ICT and robotics activities proposed in the PD-program.

Please note that the survey was initially intended as a pre-post administration to be put in relation with what the students did in between (as in study 1, see section \ref{sec:study1_methodology}). However, the positively skewed results (see section \ref{sec:study2_results}) indicated that the students' perception of the discipline was possibly impacted by the CS-education received in prior years. It was thus essential to compare with students who had not yet received any CS-education. Unlike administering an assessment of CT-concepts to students who had not received any CS-education, administering a perception survey to a control group  was accepted by the department of education (see study 3 in section \ref{sec:study3_methodology}). 


\subsubsection{Analysis Methodology}
\label{sec:study2_analysis_method}

\begin{figure*}[!h]
    \centering
    \includegraphics[width=\textwidth]{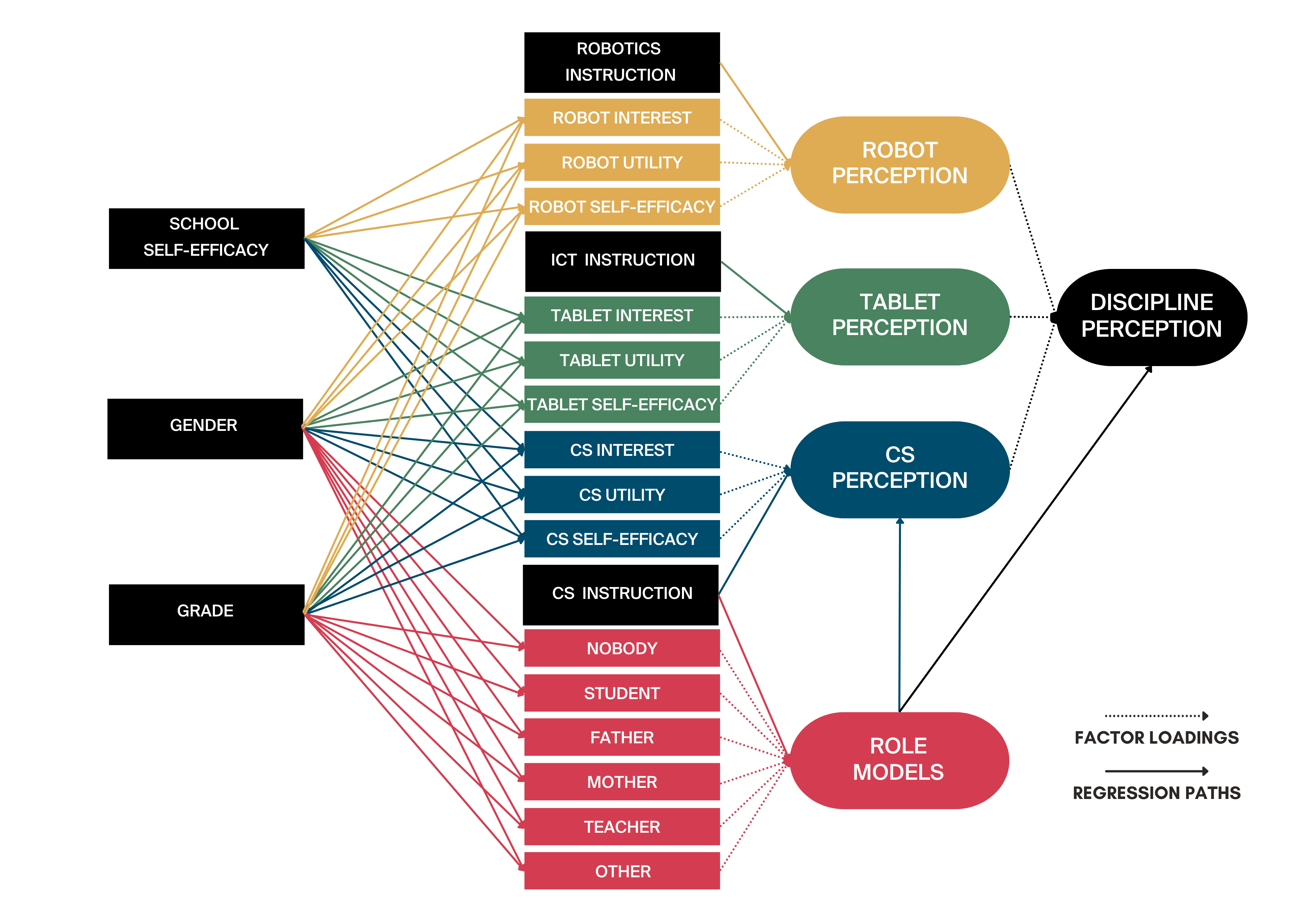}
    \caption{Study 2 Structural Equation Model for the Perception Survey}
    \label{fig:study2_SEMPerceptionmodel}
\end{figure*}

\begin{figure*}[!h]
    \centering
    \includegraphics[width=\textwidth]{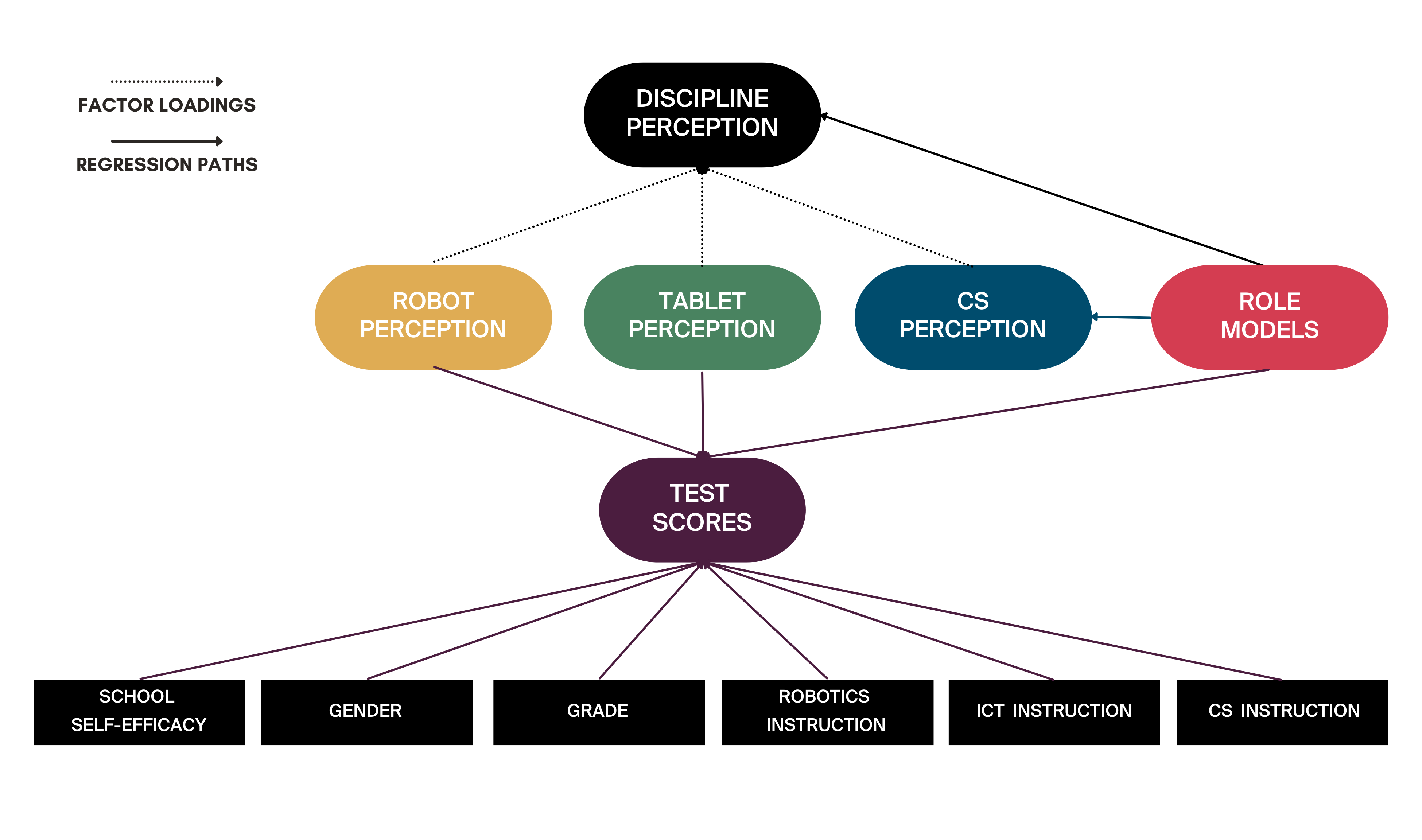}
    \caption{Study 2 Structural Equation Model for the link between Perception and Performance (as measured with the cCTt which targets CT-concepts which align with a subset of the CS concepts in the curriculum, i.e. sequences, loops, if-else statements, while statements). Please note that this model includes all paths from the model in Fig. \ref{fig:study2_SEMPerceptionmodel} but has been simplified for visualisation purposes.}
    \label{fig:study2_SEMPerceptionLearningmodel}
\end{figure*}

The analysis is conducted in three stages: 
\begin{enumerate}
    \item a descriptive analysis
    \item Structural Equation Modelling (SEM) to assess the impact of student demographic variables (gender, grade, general school-related self-efficacy), class-level variables (with respect to CS-, robotics- and ICT-related education received since the start of the year) on students' perception of the discipline (see Fig. \ref{fig:study2_SEMPerceptionmodel})
    \item Introducing student performance variables into the previous SEM to see how perception of the discipline may influence performance (see Fig. \ref{fig:study2_SEMPerceptionLearningmodel}).  
\end{enumerate}

To assess the model's goodness of fit, both the measurement model (CFA) and the structural model (SEM) must be validated. \citet{hu_cutoff_1999} recommend employing multiple complementary fit indices. Therefore, we employed local and global fit indices, namely the ratio between the $\chi^2$ statistic and the degrees of freedom, the comparative fit index (CFI), the Tucker-Lewis index (TLI), the root mean square error of approximation (RMSEA) and the standardised root mean square residual (SRMR). While $\chi^2$ statistic should be non significant \citep{alavi_chi-square_2020, prudon_confirmatory_2015}, this is rarely the case, which is why numerous researchers have recommended employing the ratio between the $\chi^2$ statistic and the degrees of freedom (df). The value of $\chi^2/df$ should be inferior to 5 for acceptable fit, and inferior to 3 for good fit \citep{kyriazos_applied_2018}. The CFI and TLI should be above 0.9 for acceptable fit and above 0.95 for good fit \citep{xia_rmsea_2019, byrne1994structural, schumacker2004beginner}. The RMSEA on the other hand should be below $.08$ for acceptable fit and below $.06$ for good fit \citep{xia_rmsea_2019, hu_cutoff_1999, chen_empirical_2008}. Finally, the SRMR should be below 0.08 \citep{xia_rmsea_2019, hu_cutoff_1999}. \\

As the data is not normally distributed and is positively skewed, in addition to including binary variables, the CFA and SEM analyses were conducted using robust diagonally weighted least square estimators. 
The modelling was conducted in R (version 4.2.1, \citealp{r_core_team_r_2019}) with lavaan (version 0.6-11, \citealp{lavaan}), \deleted{nlme (version 3.1-157),} semTools (version 0.5-6, \citealp{semTools}), semTable (version 1.8, \citealp{semtable}), psych (version 2.2.5, \citealp{psych}), and semPlot (version 1.1.5, \citealp{semPlot}).

\pagebreak
\subsection{Results - Perception, the influence of what was taught since the start of the year on perception, and the link with performance (study 2)}
\label{sec:study2_results}

\begin{figure*}[!h]
    \centering
    \includegraphics[width=\textwidth]{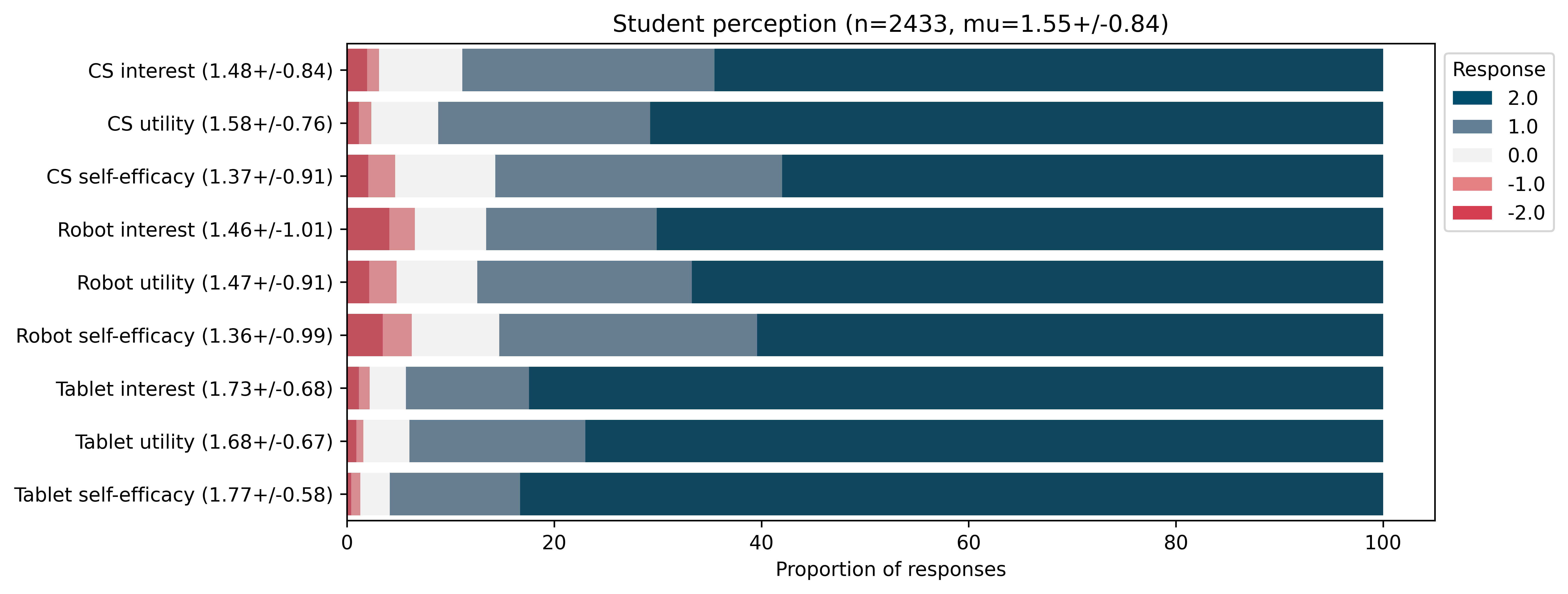}
    \caption{\added{Students' perception in schools that had been teaching CS for three years (n=2433)}}
    \label{fig:nov_2021_student_perception}
\end{figure*}

Students' perception of CS, robots and tablets is highly positive and nearly saturates ($M=1.55\pm0.84$ on the -2 to +2 scale, see Fig \ref{fig:nov_2021_student_perception}). An ANOVA however indicates that there are small significant gender differences. 

\begin{figure*}[h]
    \centering
    \includegraphics[width=0.9\textwidth]{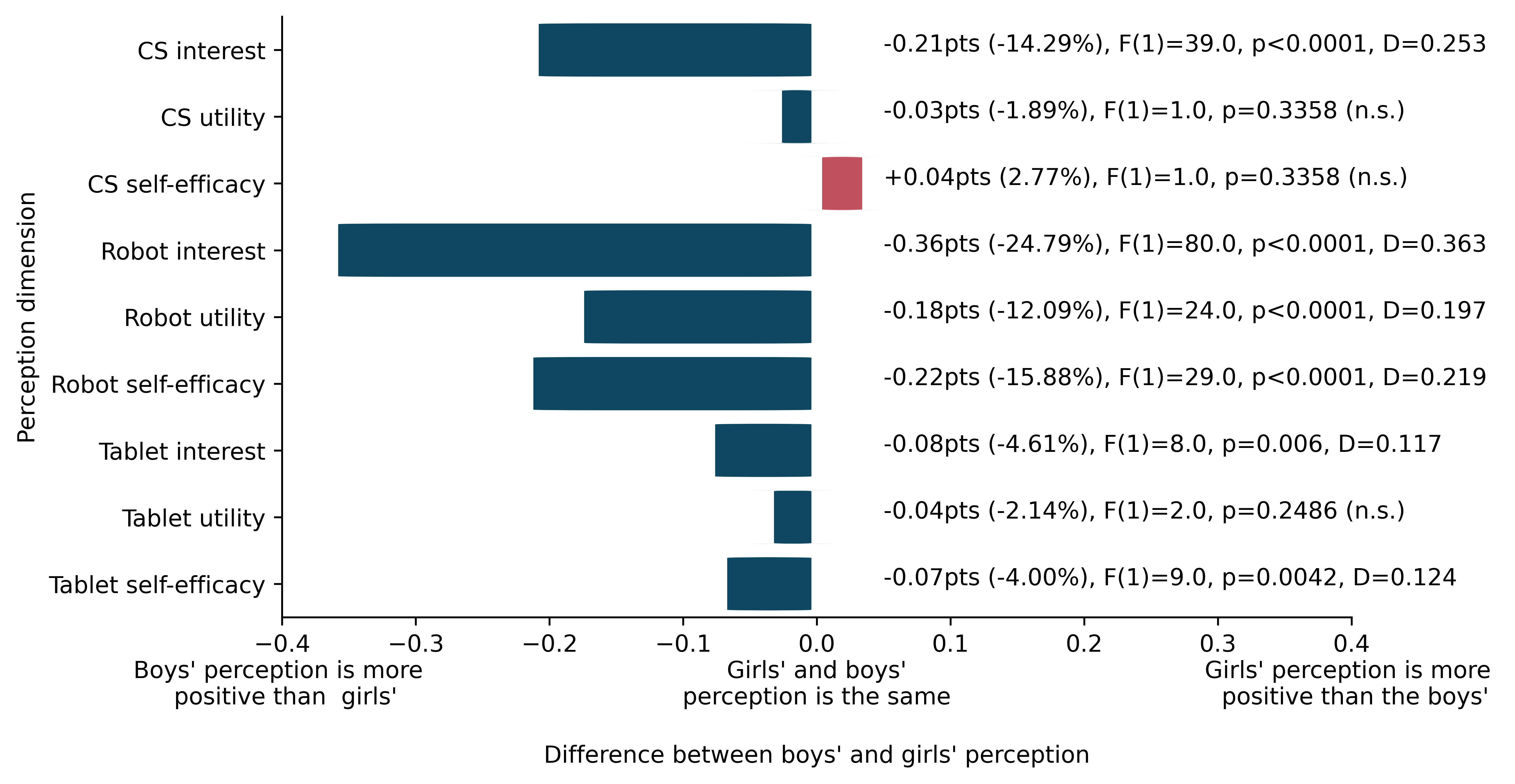}
    \caption{Delta between Boys' and Girls' perception on the 5 point Analog Visual Scale (scores between -2 and +2) in schools that had been teaching CS for three years  (n=2433)}
    \label{fig:nov_2021_student_perception_gender}
\end{figure*}

As Fig. \ref{fig:nov_2021_student_perception_gender} shows, boys: 
\begin{itemize}
    \item are more interested in CS ($p<0.0001$, Cohen's $D=0.253$)
    \item are more interested in tablets ($p=0.006$, Cohen's $D=0.117$)
    \item have higher tablet self-efficacy ($p=0.0042$, Cohen's $D=0.124$)
    \item perceive robots more favourably on all criteria ($p<0.0001$, Cohen's $D=[0.197, 0.363]$)
\end{itemize}

\begin{figure*}[h]
    \centering
    \includegraphics[width=0.8\textwidth]{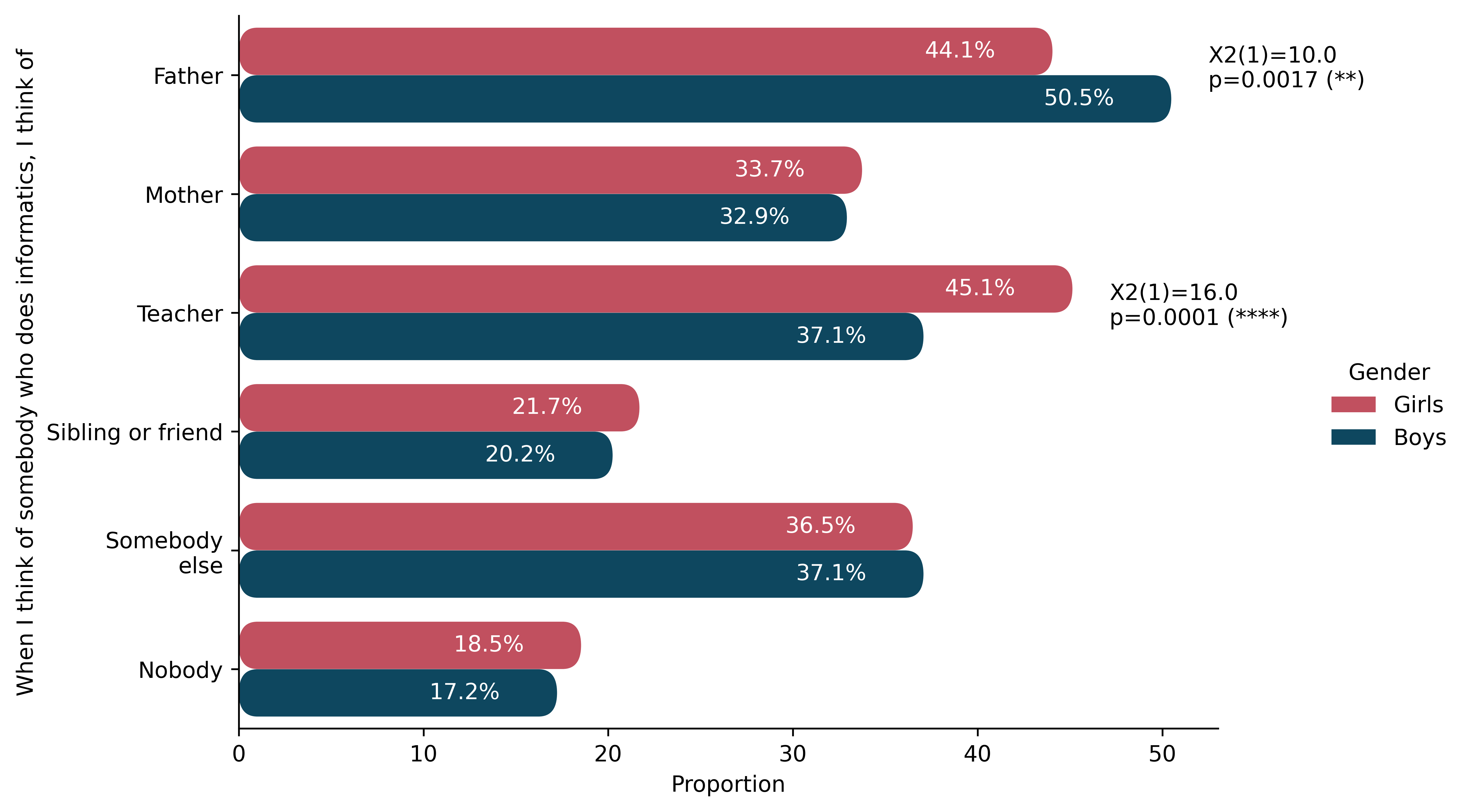}
    \caption{Students' perception of who does Computer Science in schools that had been teaching CS for three years (n=2433)}
    \label{fig:nov_2021_student_perception_who}
\end{figure*}

Gender biases are also found in terms of who is perceived by the students as doing CS ($\chi^2(5)=15.7$, $p=0.008$, see Fig. \ref{fig:nov_2021_student_perception_who}). In particular, boys consider that their father does CS more often than girls ($\chi^2(1)=10$, $p=0.0017$), while girls perceive that their teacher does CS more often than boys ($\chi^2(1)=16$, $p=0.0001$). 

To gain better insight into how the student-factors interact (demographic variables, perception of CS, tablets and robots, CS role models), and are influenced by what teachers taught, we employed SEM (n=2116, November 2021) \added{using Robust Diagonally Weighted Least Squares estimator (WLSMVS)}. 

\begin{itemize}
    \item \added{First, a} CFA indicates that the measurement model \added{does not have an adequate fit with the modification indices indicating that the issue is due to the ``other'' option in the CS role model question (Bartlett's test of sphericity $\chi^2(105)=3660$, $p<.001$, $KMO=0.70$, model fit $\chi^2(84)=373$, $p<0.001$, $\chi^2/df=4.44$, $CFI=0.888$, $TLI=0.859$, $RMSEA=0.040$, $RMSEA$ $0.90ci=[0.036$,$0.045]$, $SRMR=0.039$)} 
    \item \added{Second, a CFA conducted after removing the ``other'' option from the CS role model question indicates that the measurement model} has an adequate fit \deleted{after removing the ``other'' option from the CS role model question (see Table \ref{tab:CFA_SEM_fit})} \added{(Bartlett's test of sphericity $\chi^2(91)=3421$, $p<.001$, $KMO=0.72$, model fit $\chi^2(71)=216$, $p<0.001$, $\chi^2/df=3.05$, $CFI=0.939$, $TLI=0.922$, $RMSEA=0.031$, $RMSEA$ $0.90ci=[0.026$,$0.036]$, $SRMR=0.033$)}. 
    \item Finally, employing SEM on the model in Fig \ref{fig:study2_SEMPerceptionmodel} (see section \ref{sec:study2_analysis_method}) meets the fit requirements \deleted{(see Table XX)} \added{(Bartlett's test of sphericity $\chi^2(190)=7300$, $p<.001$, $KMO=0.72$, model fit $\chi^2(113)=260$, $p<0.001$, $\chi^2/df=2.30$, $CFI=0.941$, $TLI=0.908$, $RMSEA=0.025$, $RMSEA$ $0.90ci=[0.021,0.029]$, $SRMR=0.026$)}.
\end{itemize}

\begin{figure*}[!b]
    \centering
    \includegraphics[width=\textwidth]{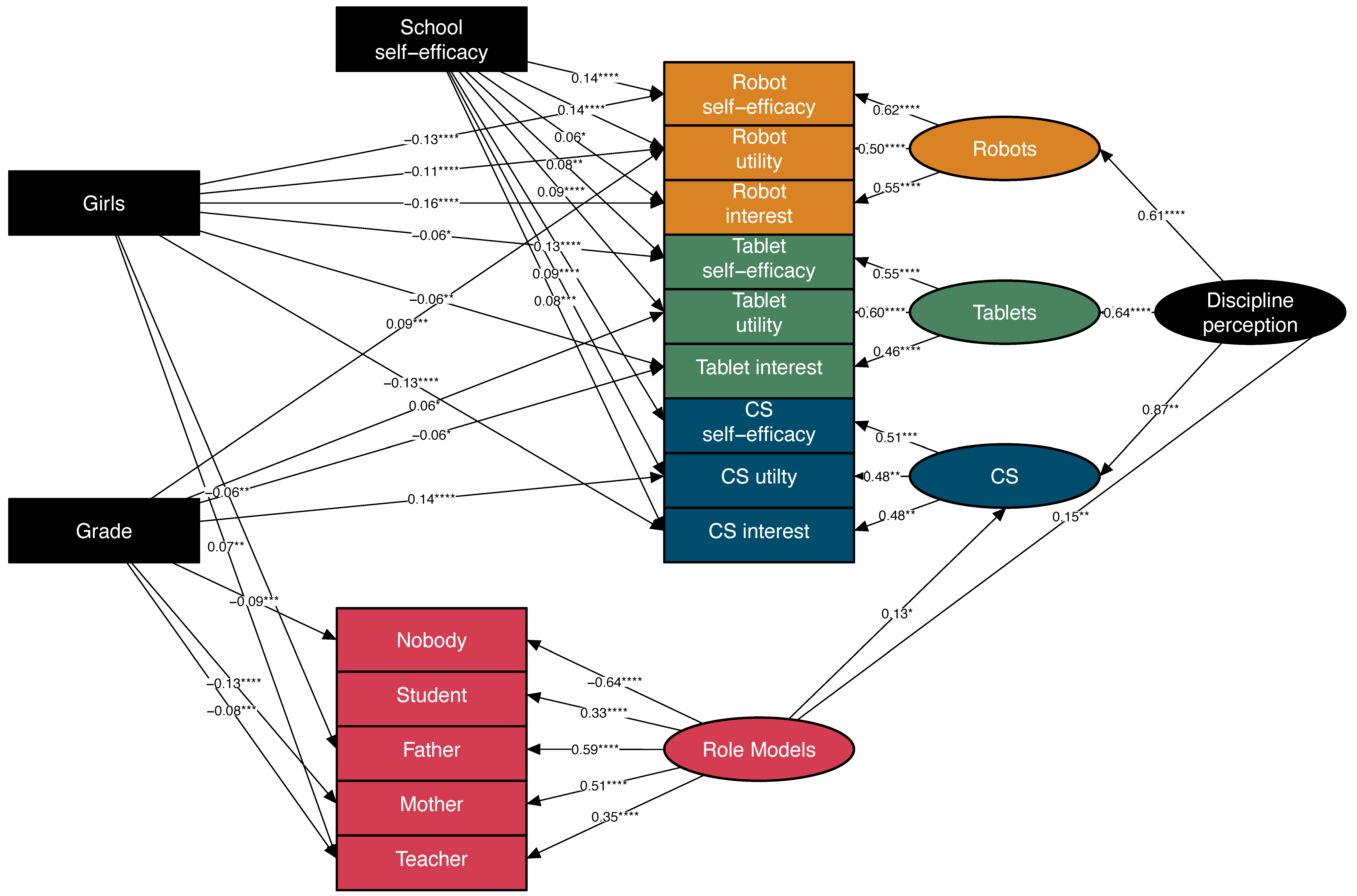}
    \caption{SEM perception structural model (n=2116, November 2021) path diagram with standardised variables for the measurement model that meets the requirements for adequate fit displaying only significant links in the model. Please note that all standardised factor loadings are above 0.3.}
    \label{fig:Nov_SEM_path_diagram}
\end{figure*}

Fig. \ref{fig:Nov_SEM_path_diagram} shows the significant paths and factors in the model (see Table \ref{tab:study2_SEMfit} \added{in appendix \ref{app:study2_SEM}} for all links) and indicates that: 

\begin{itemize}
    \item Perceiving an influencer or somebody close (e.g., teacher - $\beta=0.17 $, $p<0.001 $; parent - $\beta_{dad}=0.3, \beta_{mom}=0.23$, $p<0.001 $; or peer - $\beta=0.13 $, $<0.001 $) as doing CS, positively contributes to the perception of role models, while perceiving nobody has a negative influence $\beta=-0.024 $, $p<0.001 $). The role model latent factor then impacts the perception of CS ($\beta=0.3 $, $p=0.016 $) and of the discipline overall, i.e. the second order latent factor in the SEM, $\beta=0.15 $, $p=0.003 $).
    \item Higher school-related self-efficacy positively correlates with the perception of the discipline on all the Likert scale CS, robot and tablet related criteria, with the exception of interest in tablets\added{.}
    \item Girls tend to have a more negative perception of the discipline with respect to robots overall, tablets and CS interest, and tablets self-efficacy. They also perceive the father less often ($\beta=-0.06$, $p=0.005 $) and the teacher more often as doing CS ($\beta=0.06$, $p=0.003 $). 
    \item Older students are more likely to consider CS ($\beta=0.09 $, $p<0.001 $), tablets ($\beta=0.03 $, $p=0.020 $) and robots ($\beta=0.07 $, $p=0.000 $) useful; while being less interested in tablets ($\beta=-0.04 $, $p=0.014 $). They are also less likely to perceive their teacher ($\beta=-0.04 $, $p<0.001 $), mother ($\beta=-0.06$, $p<0.001$), and nobody as doing CS ($\beta=-0.03 $, $p<0.001 $). 
    \item The amount of CS education received since the start of the year does not significantly influence student perception on any dimensions ($p>0.05$). 
\end{itemize}

The lack of influence between teachers' adoption of CS pedagogical content and perception appears conjointly with a lack of influence between perception and performance. Indeed, the SEM that includes students' scores (n=1583, see Fig \ref{fig:study2_SEMPerceptionLearningmodel} in section \ref{sec:study2_analysis_method}) to see how performance is influenced by perception and demographics indicates that there is no significant link (see Table \ref{tab:study2_SEM_learning_reg}). The only variables that significantly influence the score are the grade (older students have higher scores) and their general self-efficacy (students that are more confident in their capacity to succeed in school have higher scores).  

\begin{table*}[h]
    \centering
    \caption{SEM Perception and Background to Performance Structural Model (n=1583, November 2021) Unstandardised Regression Parameters ($\chi^2(124)=221.462$, $p<0.001$, $chi^2/df=1.79$, $CFI=0.951$, $TLI=0.923$, $RMSEA=0.022$, $90$\%$ci=[0.017, 0.027]$, $SRMR=0.026$). Please note that on the smaller sample CS utility did not correlate highly with interest and self-efficacy and had to be removed from the model. For the full table see Table \ref{tab:study2_SEM_learning_perc_full} in appendix \ref{app:perc_learning_SEM}.}
    \label{tab:study2_SEM_learning_reg}
    \begin{tabular}{lrccccc}
        & \multicolumn{5}{c}{Model}\\ \hline
        & \multicolumn{1}{c}{Estimate}& \multicolumn{1}{c}{Std. Err.}& \multicolumn{1}{c}{Z}& \multicolumn{1}{c}{p}& \multicolumn{1}{c}{$R^2$}\\\hline
        
         \multicolumn{1}{l}{\underline{Percentage (/$100$)}} & & & & & $0.136$  \\
         CS perception & $-0.29$  & $1.31$ & $-0.22$ & $.824$ &  $0.707$ \\
         Tablets perception & $-0.04$ & $0.99$ & $-0.04$ & $.969$ &  $0.372$ \\
         Robots perception & $0.96$ & $1.11$ &  $0.86$ & $.387$ &  $0.411$ \\
         \textbf{General self-efficacy} & $1.54$ & $0.69$ &  $2.22$ &  \textbf{$.027$} &  \\
         Gender ($0=$boys, $1=$girls) & $-1.57$ & $1.06$ & $-1.47$ & $.141$ &   \\
        \textbf{Grade} & $7.53$ & $0.53$ &  $14.19$ &  \textbf{$.000$} &  \\
         Number of CS education periods SI & $0.09$ &  $0.13$ & $0.70$ & $.483$ &   \\
         Number of ICT education periods & $-0.02$ & $0.08$ & $-0.20$ & $.842$ &   \\
         Number of Robotics education periods & $-0.29$ & $0.42$ & $-0.68$ & $.496$ &   \\ \bottomrule
        \end{tabular}
\end{table*}

\subsection{Synthesis and limitations of study 2}

The students have a positive perception of the discipline, and the tools employed to teach it in schools with access to CS education. Although the results are nearly saturation, the structural equation models help identify that: 

\begin{itemize}
    \item gender influences the way the discipline is perceived, as girls have a more negative perception of the discipline then boys (in particular where robotics is concerned) which is aligned with stereotypes in these fields (social barriers)
    \item having a role model close to the students as doing CS positively influences the perception of CS and the overall discipline, but those perceived as doing CS differs according to gender since girls perceive the teacher more often, and boys the father more often as doing CS 
    \item there is no influence of the CS education received from the start of the year on perception.
    \item there is no link between students' perception of the discipline and their performance on the assessment (positive for equity)
    \item student general school-related self-efficacy positively correlates with the perception of the discipline and with students' performance on the test. 
\end{itemize}

There are however several limitations to this study, mainly that (i) the students were at least in their third year of CS education by the time the study was conducted, (ii) their perception was positively saturated and (iii) there was no control group. It would have been interesting to have access to a pre-test prior to their first CS lecture and to compare the evolution of perception over time. Where the link between perception and what the teachers taught is concerned, as for study 1, some of the findings may be biased by the fact that teachers may be teaching CS content that was not included in the PD program and are not accounted for in the analyses. 

\added{In terms of the perception survey itself, while the CFA analysis indicates that the perception survey is a short and valid instrument that can be employed to measure grades 3-6 students' perception of the discipline and the tools used to teach it, this is not without its limitations. Indeed, the survey measures interest, utility and self-efficacy concepts with only one item for each dimension (CS, robotics, tablets). Ideally, for each concept and dimension, there would be at least 3-4 items (for interest, utility, and self-efficacy) in order to improve the reliability of the instrument. This owes to our requirement of being able to administer the CS perception survey to grades 3-6 students before the cCTt (and not after to avoid having their performance bias their perception), without taking too much in-class time for both (i.e. the perception survey had to be short and take less than 20 minutes overall with grade 3 students). Nonetheless, researchers have investigated the reliability of single-item items and have shown that it is possible to have reliable measures with only single items (see \citealt{hoeppner_comparative_2011}).}

\FloatBarrier

\section{Study 3 - Student Perception between CS-schools and schools where teachers were not yet trained to teach Computer Science }
\label{sec:study3_methodology}

\subsection{Methodology}
\subsubsection{Participants and Data Collection}

To extend study 2, the perception survey (see Table \ref{tab:study23_student_survey_items}) was administered to all students in grades 3-6 ($n=1644$) from 3 schools with access to CS education (which we refer to as CS-schools, $n\sim=831$) and 2 \added{similar} schools without access to CS education (\added
{which we refer to as non CS-schools, $n\sim=813$, i.e. schools where, at the time of the study, the teachers were neither trained to introduce the new discipline into their practice nor had access to the material resources, infrastructure or support they require to teach the discipline - an element which was confirmed by an accompanying teacher survey}). \replaced{All 5 schools were}{that were} selected to be representative of the demographics of the region (see Table \ref{tab:study3_participants}). 
The objective was to compare the students' perception of the discipline between the two conditions (CS-schools and non CS-schools) as students in CS-schools had been in contact with the discipline for multiple years and perception was positively saturated in study 2.

\begin{table*}[h]
    \centering
    \caption{Number of participants in the second student perception survey (study 3, May 2021)}
    \label{tab:study3_participants}
    \footnotesize
    \begin{tabular}{cccccccc}
    \toprule
    CS education & Gender & \multicolumn{4}{c}{Grade} & Total  \\
                     &  &   $3$   & $4$  &  $5$ &   $6$ &  \\ \midrule                             
    False &  Boys       &        $84$ &  $91$ & $116$ & $122$ & $413$  \\
          &  Girls       &       $83$  & $98$ & $121$  & $98$ & $400$  \\
          &  Total       &       $167$ & $189$ & $237$ & $220$ & $813$  \\ \midrule
    True  &  Boys      &         $102$  & $91$  & $96$ & $135$ & $424$  \\
          &  Girls      &        $93$  & $99$  & $92$ & $123$ & $407$  \\
          &  Total      &        $195$ & $190$ & $188$ & $258$ & $831$  \\ \midrule
    Total &  Boys      &         $186$ & $182$ & $212$ & $257$ & $837$  \\
          &   Girls      &        $176$ & $197$ & $213$ & $221$ & $807$  \\
          &   Total        &        $362$ & $379$ & $425$ & $478$ & $1644$  \\ \bottomrule
    \end{tabular}
\end{table*}

\subsubsection{Analysis Methodology}

The comparison between both groups is established using Structural Equation Modelling by constraining the models to have equal factor loadings, and allowing the regression parameters to vary between the two groups (gender, grade, general self-efficacy). 
By comparing the intercepts of the two SEMs, it is possible to establish the effect of having received several years of CS-education on perception. By comparing the regression parameters, it is possible to establish whether there are interaction effects between the student variables (e.g., gender) and access to CS-education, and thus determine if gender-related gaps are indeed closing with the introduction of the novel curriculum.

\subsection{Results - Perception and the influence of having access to CS education on perception (study 3)}

The SEM to compare the groups (CS-schools vs. non CS-schools) constrained the loadings and thresholds, while leaving the intercepts and regression parameters free to vary between groups\footnote{The selection of model constraints was achieved by successively comparing through ANOVA the following SEMs: 1) without groupings, 2) groupings without constraints, 3) constrained loadings and thresholds, 4) constrained loadings, thresholds, regression parameters, 5) constrained loadings, thresholds, intercepts. }. 
We thus compare the intercepts and regression coefficients between the groups (for the full SEM see Table \ref{tab:study3_SEM_groups} in appendix \ref{app:study3_group_comp_SEM}).  \\

\begin{figure*}[h]
    \centering
    \includegraphics[width=\textwidth]{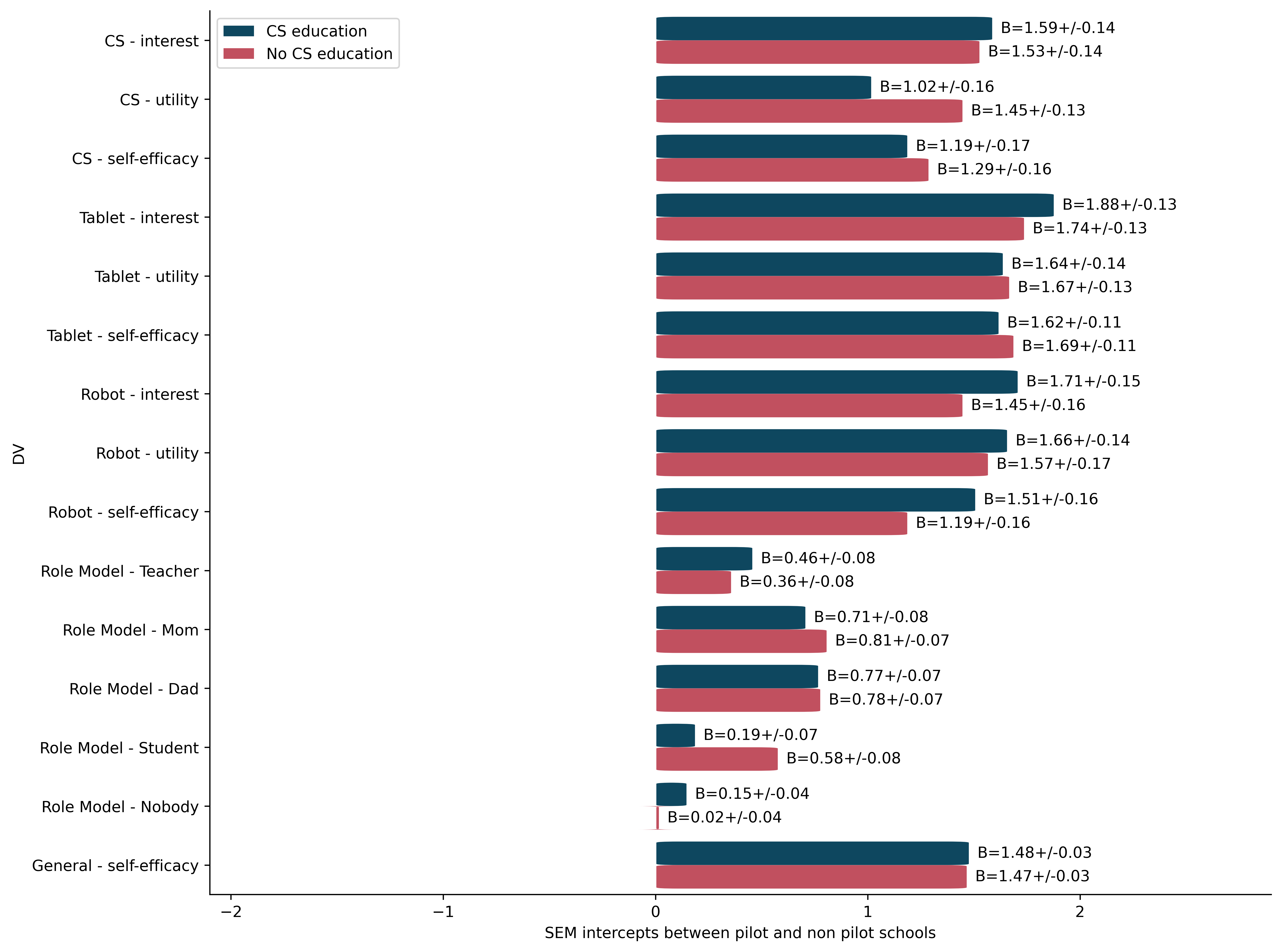}
    \caption{Study 3 - Comparison of the SEM intercepts between schools that had access to CS-education and schools that did not.}
    \label{fig:Study3_intercepts_comp}
\end{figure*}

The intercepts for both groups indicate that responses positively saturates for both groups for nearly all CS, robotic and tablet perception items are shown in Fig. \ref{fig:Study3_intercepts_comp}. Nonetheless, students in CS-schools appear more interested generally, and evaluate the robotics generally more favourably. However, CS and tablet utility and self-efficacy are lower for students in CS-schools. Students in CS-schools perceive the teacher more often as doing CS, which is coherent with the fact that their teachers over the past few years have been teaching CS pedagogical content. On the other hand, students in CS-schools perceive their mothers and other students less often as doing CS, possibly indicating that the students have a better awareness of what it means to "do" CS \citep{pantic_drawing_2018}, and that it is not only related to using a computer or tablet. 

\begin{figure*}[h]
    \centering
    \includegraphics[width=\textwidth]{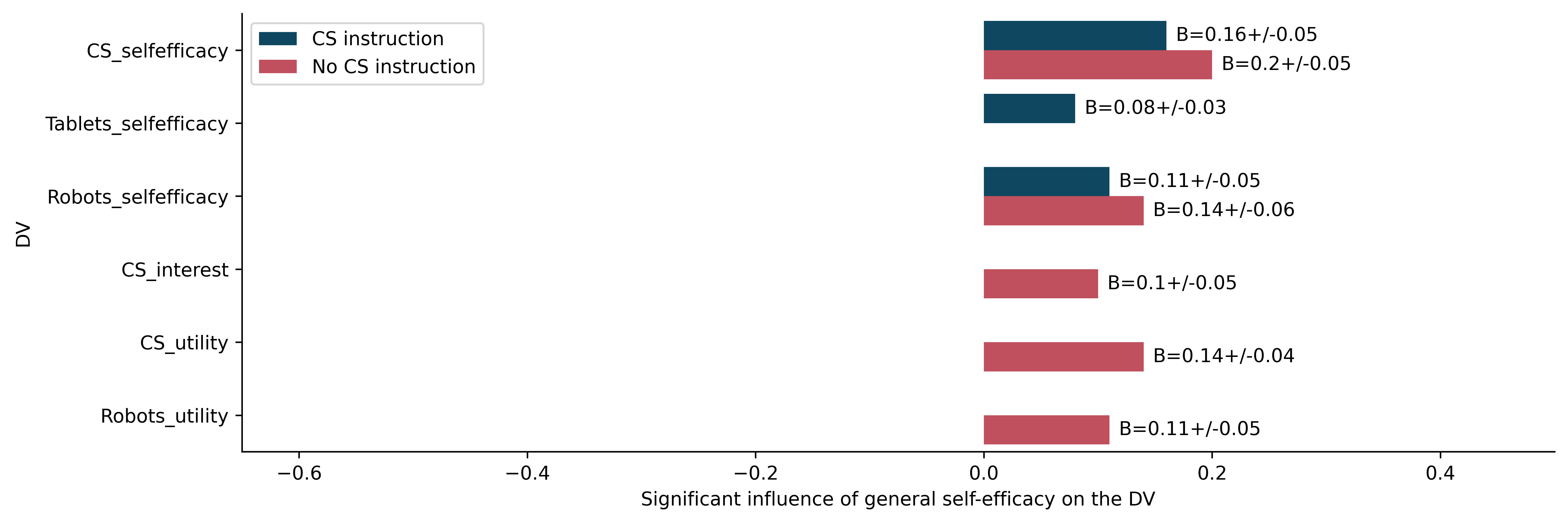}
    \caption{Study 3 - Comparison of the SEM regression coefficients for general self-efficacy between schools that had access to CS-education and schools that did not. Please note that only significant regressors are shown.}
    \label{fig:Study3_selfefficacy_reg_comp}
\end{figure*}

\begin{figure*}[h]
    \centering
    \includegraphics[width=\textwidth]{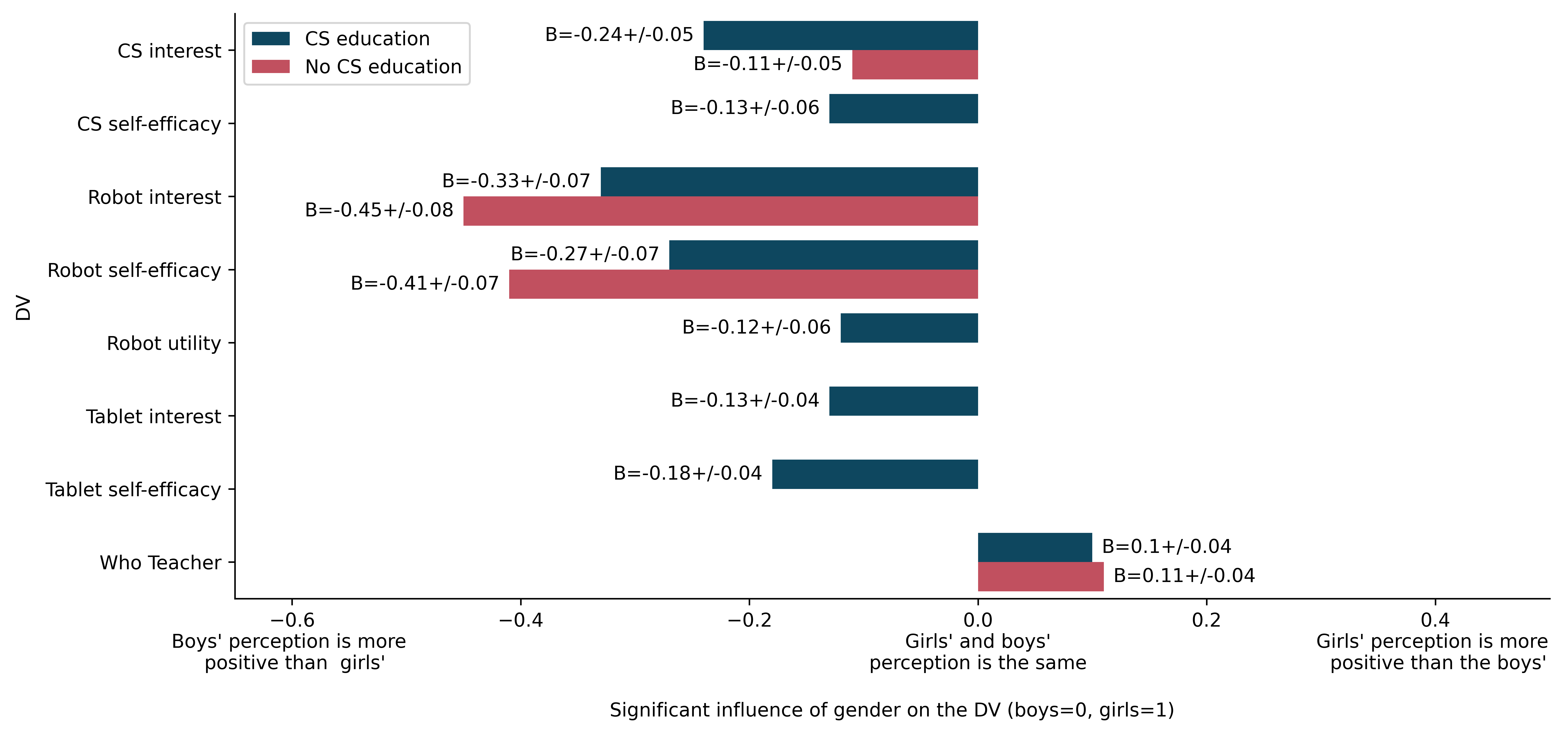}
    \caption{Study 3 - Comparison of the SEM regression coefficients for gender between schools that had access to CS-education and schools that did not. Please note that only significant regressors are shown.}
    \label{fig:Study3_gender_reg_comp}
\end{figure*}

The significant impact of general self-efficacy and gender on student perception are shown in Fig. \ref{fig:Study3_selfefficacy_reg_comp}, and \ref{fig:Study3_gender_reg_comp}.

Fig. \ref{fig:Study3_selfefficacy_reg_comp} shows that general school-related self-efficacy positively influences CS self-efficacy ($b_{CS}=0.16$, $p_{CS}=0.001$, $b_{no-CS}=0.2$, $p_{no-CS}<0.001$) and robotics self-efficacy ($b_{CS}=0.11$, $p_{CS}=0.033$, $b_{no-CS}=0.14$, $p_{no-CS}=0.016$) of all students. This reveals that students who consider themselves less capable of doing well in schools also think that they are less able to do CS and robotics, although the influence is less pronounced when students have received CS-education. Access to CS education may thus contribute to a wider range of students considering that they are capable of doing CS and robotics. On the other hand, for tablets, while there is no significant influence of school-related self-efficacy in non CS-schools ($p_{no-CS}=0.054$), it is present in CS-schools ($b_{CS}=0.08$, $p_{CS}=0.016$) which may indicate that students realise the range of possibilities (beyond merely passive activities) and that this may require more competencies to be able to make use of. 
Nonetheless, general self-efficacy does not influence interest or perceived utility in CS-schools ($p_{CS}>0.05$), contrary to non CS-schools for CS interest ($b_{no-CS}=0.1$, $p_{no-CS}=0.036$), CS utility ($b_{no-CS}=0.14$, $p_{no-CS}=0.001$), and Robotics' utility ($b_{no-CS}=0.11$, $p_{no-CS}=0.044$). It would thus appear that access to CS-education helps reduce these biases.

Where gender is concerned (see Fig \ref{fig:Study3_gender_reg_comp}), all gender gaps identified as significant confirm the stereotypes that boys perceive the discipline more favourably than girls. Some gender gaps are only present in CS-schools (CS \& tablet interest and self-efficacy, robots utility) suggesting that access to CS-education increases these gaps. There are nonetheless some gaps that are smaller in CS-schools, all the while remaining present in both types of schools: robotics interest and self-efficacy, as well as perceiving the teacher as doing CS in CS-schools. Only the CS-interest gap is present in both schools and stronger in CS-schools.

\subsection{Synthesis and limitations of study 3}
  
Students' perception of the discipline is highly positive and affected by gender biases (social barriers) in both schools with CS education and schools without. However, access to CS education leads to:
\begin{itemize}
    \item positive impacts through: increased interest in CS and the associated tools, a more positive perception of robotics on all dimensions, teachers being more often perceived as doing CS
    \item negative impacts through: lower self-efficacy with respect to CS and tablets
    \item positive outcomes for equity through: a closing gender gap for robotics interest and self-efficacy (gender-equity), a lesser influence of general self-efficacy on several perception dimensions (CS interest, utility, self-efficacy; robotics utility and self-efficacy) 
    \item negative outcomes for equity through: an increasing gender gap for CS and tablets self-efficacy (gender-equity), a higher influence of general self-efficacy on tablets' self-efficacy
\end{itemize}

As in the case of studies 1 and 2, this study has its limitations. Firstly, the sample is relatively small to do a comparison between groups (even when constraining parameters to be equal). As such, the minimum effect size \added{(Cohen's D)} that can be detected is smaller than in the case of study 2. This analysis would therefore benefit from a replication at a larger scale. As mentioned for study 2, there is also no view on how the perception evolves over time within these groups, and at the point where students gain access to CS education the first time. Therefore, it would be interesting to have access to a sample of students just before they began having access to CS \added{education}\deleted{instruction} and then follow up over time, and compare with a group that has no access to CS \added{education}\deleted{instruction}. This type of analysis has temporal constraints and must be planned for at the start of the reforms and prior to deployment to all schools if the objective is to be able to compare for an extended period of time. 

\FloatBarrier

\section{Discussion}

\begin{figure*}[!h]
    \centering
    \includegraphics[width=0.95\textwidth]{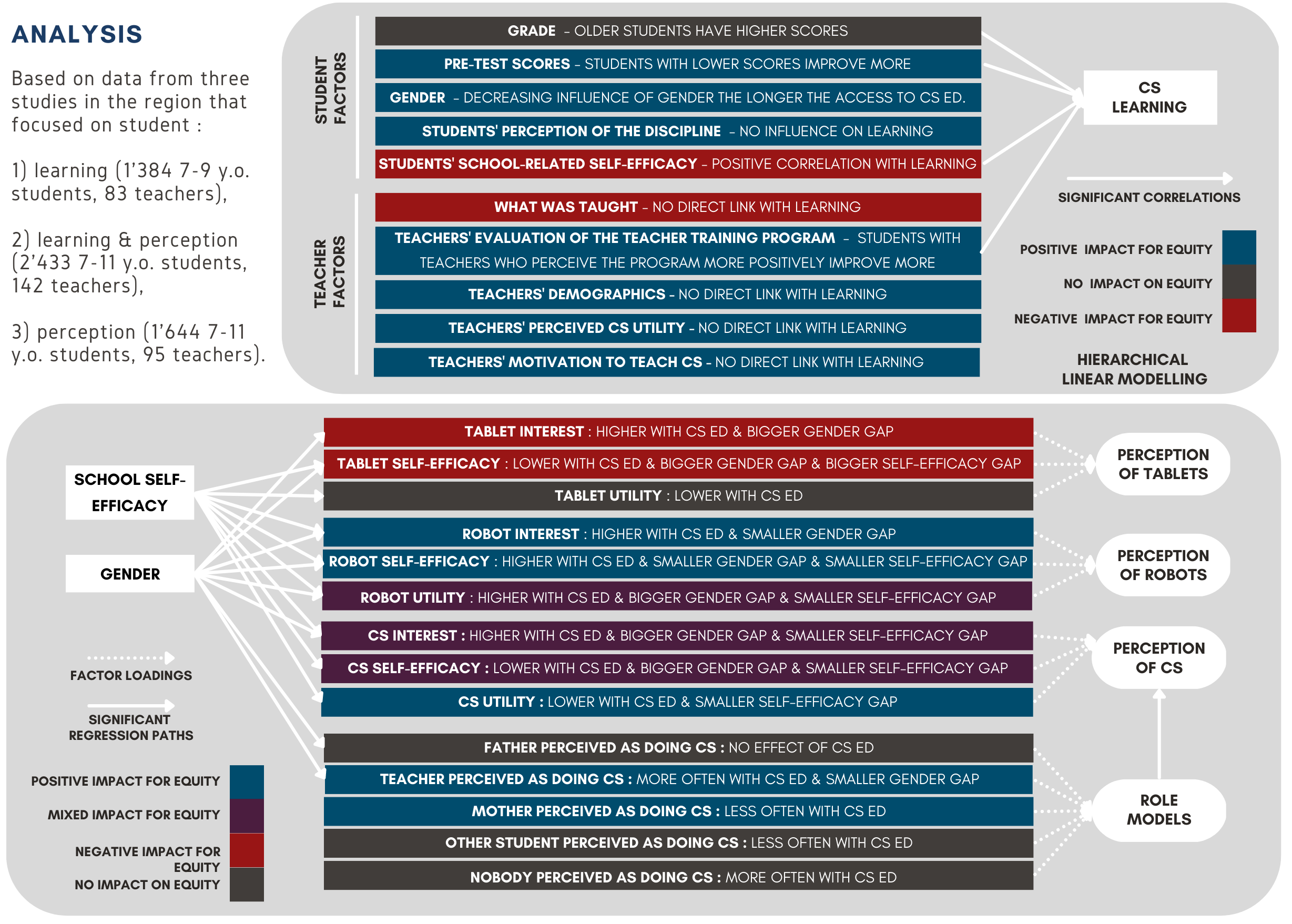}
    \caption{\added{Visual Synthesis of the study's findings and how these relate to impact and equity. Each factor considered is indicated in a rectangle in bold. 
    For the student learning results that are based on hierarchical linear models, the identified effect of said factor on the outcome variable is indicated in plain text in the same rectangle. 
    For the student perception results that are based on structural equation modelling, the impact of general school-related self-efficacy and gender on a given factor are indicated in the factor's rectangle in plain text.  
    In both cases, the impact of the measured effect (or lack thereof) on equity is colour coded (blue for a positive impact, red for a negative impact, purple for a mixed impact and black for an absence of impact). \added{Please note that we only indicate significant links / effects (i.e. $p>0.05$) which does not reflect on the strength of the effect detected (for that please refer to the results section and see the effect sizes and regression coefficients).}}}
    \label{fig:synthesis}
\end{figure*}

This article investigates whether a large-scale mandatory primary school CS curricular reform and accompanying PD program has an impact, and contributes to achieving equity goals, in terms of learning and perception. As indicated in the introduction, achieving equity goals requires addressing structural (i.e. access related) and social (i.e. stereotype related) barriers that lead to under-representation in the field by influencing performance and perception early on. While equity in terms of access is ensured by the fact that the reform is being deployed to all teachers in the region, two main questions drive the study: (RQ1) How does teaching CS pedagogical content impact student learning? And how does it impact learning-related gender- and performance-equity? (RQ2) How does teaching CS pedagogical content impact students' perception of CS? And how does it impact perception-related self-efficacy- and gender-equity? We provide a visual synthesis of the findings in Fig. \ref{fig:synthesis} based on the learning and perception drawn from 3 studies conducted over two years and involving respectively $n_1=1384$, $n_2=2433$ and $n_3=1644$ grade 3-6 students (ages 7-11) and their $n_1=83$, $n_2=142$ and $n_3=95$ teachers. The findings are further discussed in the following subsections. 

\subsection{Impact of the curricular reform on student learning, and learning-related performance- and gender-equity (RQ1)}

\subsubsection{Student learning impact}

The findings of studies 1 and 2 indicate that the students progress in terms of CT-concepts \added{(sequences, loops, if-else statements, while statements)} over time. In particular, we observe that grade 3 students achieved a year's worth of CT-development in the 6 months that separated the pre- and post-tests (positive impact). Indeed, the grade 3 students' post test scores were equivalent to the grade 4 students' pre-test scores. 
However, there is no direct link between learning and the amount of CS education received (absence of impact). There is, on the other hand, a positive influence of the teachers' perception of the PD program (positive impact). While this likely acts as a mediating variable for teachers' assimilation of the underlying CS-concepts and their appropriation of the pedagogical content, it does indicate the need to find means of motivating teachers to introduce CS into their practices \ifdefined\Anonymous \added{ (Anonymous authors - details removed for peer review - d, 202X)} \else \citep{el-hamamsy_tacs_2022} \fi and to ensure that they see the utility of doing so \ifdefined\Anonymous \added{ (Anonymous authors - details removed for peer review - e, 202X)} \else \citep{el-hamamsy_sustainability_2022} \fi. 

The lack of a direct link between what the teachers taught (i.e. adoption) and learning could be due to two main factors and their interaction: the adequacy of the content with respect to the targeted concepts, and the teachers' appropriation of the CS pedagogical content. We have synthesised the corresponding hypotheses in Table \ref{tab:student_learning_hypotheses} depending on whether either or both of these factors are indeed at play in the present context. As a reminder: the teachers were trained to introduce the specific CS-pedagogical activities which were designed by experts in CS and pedagogy from multiple institutions. Therefore, considering conjointly these elements, and the link between student learning and the teachers' perception of the PD-program, it appears likely that the second hypothesis is true. More specifically: the lack of direct link with adoption could be partially or entirely due to teacher-level factors (their mastery of the concepts, and how they are teaching the pedagogical activities), although we may not presently rule out the other hypotheses. 

\begin{table*}[h]
\centering
\caption{Hypotheses related to the absence of direct links between CS-education and student learning}
\label{tab:student_learning_hypotheses}
\footnotesize
\begin{tabular}{p{2cm}cp{5cm}p{7cm}}
\toprule
 &  & \multicolumn{2}{c}{\textbf{The teachers' appropriation of the CS-pedagogical content is aligned with the curricular objectives}}  \\
 
 &  & {True} & {False}  \\ \midrule
 
\multirow{2}{2cm}{\textbf{The CS-pedagogical content is adequate with respect to the targeted concepts}} 

& True & H1: The students have reached the limit of their cognitive abilities and are not capable of progressing more, irrespective of the additional content and CS education received 
& H3: The teacher, while teaching the CS-pedagogical activities is not teaching the CS-concepts well (H3.1) either because they do not have sufficient mastery the concepts themselves; or (H3.2) because they do not put the emphasis on the CS concepts while teaching and focus on other facets, such as disciplinary links (e.g., maths or verbalisation), coherently with the differences between intended, enacted and attained curricula that are present generally \citep{van_den_akker_curriculum_2003} and in the context of CS \citep{falkner_international_2019}. The PD program should be revised.  \\  \\
& False & H2: The CS-pedagogical content is either (H2.1) not developmentally appropriate \citep{ottenbreit-leftwich_computational_2022, bers_state_2022}, or (H2.2) does not go sufficiently in depth for students to progress beyond what they are acquiring without the CS-education, and should be revised. & H2 + H3  \\ \bottomrule
\end{tabular}
\end{table*}

To better understand the impact of teaching CS on learning, it would be important to investigate the various hypotheses by considering teacher assessments \added{to gain insight into their mastery of the concepts}, classroom observations \added{to gain insight into teachers' implementation fidelity}, and comparing with students in non CS-schools. Such an approach would not only make it possible to assess each of the pedagogical activities individually, but would also give the opportunity to provide guidelines regarding how best to teach the pedagogical content to promote learning. 
Doing so however requires getting past certain barriers in the field, whether in terms of teacher reticence towards classroom observation and evaluation \citep{hickmott_assess_2018}, or in terms of acceptability for policy makers (e.g. access to a control group for performance assessments). However, it is only by gaining such insight that it will be possible to adapt the CS curricular reform so that they are successfully implemented and sustained in teachers' practices. This could include adapting the CS pedagogical content, PD program, or even considering a different strategy to introduce CS into formal K-12 education. The latter could involve having specialised teachers, or introducing CS transversally to support other disciplines thus contributing to "build computational litteracies in all students"  \citep{peel_algorithmic_2022}, all the while accounting for time struggles \citep{ottenbreit-leftwich_computational_2022} which according to \citet{fofang_mutually_2020} would provide pedagogical and equity benefits (but may also run the risk of decreasing the impact of the curricular reform, \citealp{suessenbach_informatik_2022}). We therefore argue that a complete assessment of CS and CT curricula would benefit greatly from expanding to other dimensions of CT (e.g. CT-processes, \citealp{brennan_new_2012}), and evaluating the impact that CS-pedagogical content may have on learning in other disciplines
\ifdefined\Anonymous  (\citealp{ottenbreit-leftwich_computational_2022}, \added{Anonymous authors - details removed for peer review -b, 202X)}\else \citep{ottenbreit-leftwich_computational_2022, el-hamamsy_case_2022}\fi, transversal and 21st century skills \ifdefined\Anonymous  (\citealp{barr_computational_2011, gretter_computational_2016, nouri_development_2020}, \added{Anonymous authors - details removed for peer review -b, 202X)}\else \citep{barr_computational_2011, gretter_computational_2016, nouri_development_2020, el-hamamsy_case_2022} \fi. 

\subsubsection{Student learning equity}

The findings of study 1 indicate that students performing lower at the pre-test progress more in the 6 months before the post-test. This indicates that the performance gap is closing and contributing to \textit{performance-equity} and is consistent with \citet{vygotsky_mind_1978}'s concept of the Zone of Proximal Development (ZPD). The ZPD is determined by the learning activity and its relation to what students are capable of doing along and with a specific instruction. Therefore it would appear that the content is adapted to all students because: 

\begin{itemize}
\item students with low scores on the pre-test progress more, indicating that the pedagogical content is within their ZPD
\item students with high scores on the pre-test may already master the concepts and therefore not progress more with the instruction provided
\end{itemize}

Provided the additional lack of influence of teacher-demographics (including teaching and ICT experience), their perceived utility of CS and their autonomous motivation to teach CS, on student learning, this would appear to indicate that the PD-program contributes to fostering student learning, and \textit{learning equity} more generally, by reducing the impact of teachers' perception of CS and socio-demographics \added{(including their prior teaching experience and age which have been found to impact student achievement in various contexts, \citealp{croninger_teacher_2007, kini_does_2016, ladd_returns_2017, burroughs_review_2019})}.

The findings of study 1 also indicate that a marginally significant gender gap exists in grades 3-4 (likely due to stereotypes and social barriers), and that it appears to be closing over time (positive for gender-equity). This is corroborated by the data from study 2 (from the following academic year) where students who have had more access to CS education do not exhibit gender gaps. These findings therefore confirm the importance of providing prior CS experience to address performance-related gender gaps. As the study did not include grade 1-2 students, it would appear relevant to follow up on the cohort of students over multiple years (and from the start of their schooling) to see how these differences appear and evolve over time.  \\

The findings would therefore appear to corroborate that the CS-curricular reform contributes to achieving \emph{learning equity goals}. This would align with the findings of a recent independent study conducted in Germany to evaluate the impact of the introduction of "informatics" into the curriculum throughout the country. In a longitudinal study, \citet{suessenbach_informatik_2022} found that i) lower secondary schools students' ICT competence increased with access to informatics education, ii) the gap between students with low and high socio-economic backgrounds decreased, iii) gender gaps were closing with girls catching up with boys' performance, and iv) the impact was stronger in the case of informatics as its own discipline rather than having informatics transversally integrated into other subjects. 

\FloatBarrier

\subsection{Impact of the curricular reform on student perception and perception-related self-efficacy- and gender-equity (RQ2)}

\subsubsection{Student perception impact}

Students' perception of the discipline and the tools employed to teach it is globally positive in primary school, whether in CS schools or not (studies 2, 3), as the results are positively saturated. Nonetheless, students' overall perception of the discipline is influenced by access to CS education. Indeed, access to CS education contributes to increased interest in CS and the associated tools, with a more positive perception of robotics overall (positive impact). Perceived utility and self-efficacy towards CS and tablets is however lower (negative impact). The latter may be indicative of a better understanding of what CS is, and the extent of the applications that are possible with tablets, contributing to more realistic expectations \citep{pantic_drawing_2018} which are presently lacking in the region \ifdefined\Anonymous \added{ (Anonymous authors - details removed for peer review - g, 202X)} \else \citep{el-hamamsy_C3PP_2022}\fi. 
As the results remain globally positive, these results appear promising for both CS and robotics as interest in the former, and interest, self-efficacy and perceived utility in the latter are key motivational factors that influence academic performance and career choices. Future studies should therefore i) continue to monitor how these factors evolve and how they relate to students' decision or not to pursue studies in these fields once they are able to decide what they want to study and what type of career to pursue (which in the present educational system, begins at the end of 8th grade) and ii) investigate using qualitative methodologies why certain trends are observed.   \\

\subsubsection{\added{Student perception equity with respect to the effect of gender}
}

Gender gaps are present already in grades 3-6 with boys having a more positive perception of the discipline than girls on nearly all criteria, coherently with \citet{master_gender_2021} and \citet{sullivan_girls_2016}, despite access to CS-education from grade 1 (study 2). Robotics in particular appears to be subject to the largest gender gaps (study 2, 3). 
Nonetheless, the perception of CS-role models, and in particular influencers \citep{wang_diversity_2017} such as teachers and parents being perceived as doing CS has a positive influence on the perception of the discipline, but is subject to gender biases (study 2). As access to CS-education contributes to more students perceiving their teachers as doing CS (study 3), and these teachers are mainly women in primary school, the introduction of CS for all in schools may contribute to addressing social perceptions and counter the creation of early gender gaps evoked by \citet{wang_diversity_2017}.  
The model selection further indicates that the influence of gender on perception varies with access to CS education (impact on gender-equity). While for interest and self-efficacy the gender gap appears to be closing for Robotics in CS-schools (positive for gender-equity), the gap is increasing for CS and Tablets (negative for gender equity). 
Different trends along these dimensions indicate that the CS pedagogical activities might need to be adjusted to ``provid[e] students with early experiences that signal equally to both girls and boys that they belong and can succeed'' \citep{cheryan_why_2017} (e.g., by adopting more collaborative settings, \citealp{sullivan_girls_2016}, or introducing social aspects). 
The findings further indicate that introducing robotics as a means of teaching CS may also contribute to broadening participation in STEM by reducing robotics-related gender biases. This complements a prior study that found that employing robots to teach CS benefited both CS and robotics at the teacher and PD-level \ifdefined\Anonymous \added{(Anonymous authors - details removed for peer review - c, 202X)} \else \citep{el-hamamsy_symbiotic_2021}\fi. As such, robotics to teach CS benefits both the PD-, teacher-, and student-levels. Robotics and STEM more broadly could therefore take advantage of ongoing CS-curricular reforms worldwide to broaden participation and engage more students in these fields.

\subsubsection{\added{Student perception equity with respect to the effect of school-related self-efficacy}}

\emph{The influence of general self-efficacy varies between CS and non CS schools} (study 3). On the one hand, in CS-schools, there is a positive influence of general self-efficacy on Tablet related self-efficacy, which is not the case for non CS-schools. Once again, this may be due to students having a better awareness of what it means to "do" CS and more creative activities with Tablets \citep{pantic_drawing_2018}. Indeed, the tablet is a ubiquitous tool in the region which students easily have access to. Their traditional usage of this tool differs significantly from the type of activities that are proposed in the curricular reform which tend to be active and creative. The learning objectives of these tasks push students to adjust their "imagination" around this tool \citep{flichy_place_2001}. Concretely, we believe that the students are forced to reconsider the affordances and the potential of this tool, thus reevaluating their own competencies beyond the more traditional use involving playing games, texting, taking photos and watching videos. 

On the other hand, CS perception (interest, self-efficacy and utility), and robots perception (self-efficacy, and utility) are positively influenced by general self-efficacy in both CS and non-CS schools, but to a lesser degree when students have received CS education. Contrary to tablets, CS and robotics are novel, with students having little to no access in schools \added{(or at home)} where the CS education curricular reform has not yet occurred. Therefore, we believe that the positive influence of general self-efficacy on domain-specific self-efficacy is consistent with \citet{bandura_social_1986}'s sociocognitive theory on auto-evaluation: people's belief in their efficacy to do a task is developed through vicarious experience, i.e. by comparing themselves to others. However, it is also built through mastery experiences: by experiencing CS and robotics related activities, the students are more influenced by their own CS- and robotics-specific experiences, and less by their overall assessment of their capacity to succeed in school. Therefore, given the influence of self-efficacy on students' choices and career decisions in the long term, such experiences may ultimately contribute to broadening participation in the field to a wider range of students, and namely to not only those who believe they are good in school.  \\

\subsubsection{\added{Student perception equity with respect to the link between performance and perception}}

Similarly, there is \emph{no evident link between student performance and perception of the discipline} (study 2, positive for equity), such as those found in other studies in middle school \citep{rachmatullah_toward_2022, hinckle_relationship_2020}. However, student performance is related to students' general self-efficacy, with students who consider that they are better at school performing better on the test. This would suggest that there may be a link between students' performance on CT-concepts and other disciplines, irrespective of how students perceive the discipline. This may be indicative that perception is not yet biased by performance and inversely. Nonetheless, given the role that perception (and stereotypes) has been found to play on academic and career decisions (see section \ref{sec:introduction}), it is important to continue to monitor how students' perception evolves over time and establish at which point this may influence their sense of belonging and career decisions.  \\

The trends observed confirm not only the importance of introducing the discipline in formal education for all, but also the complex interactions that this introduction may have on students' perception. The latter indeed may not necessarily contribute to closing all perception-related gaps but may also exacerbate others. Therefore, in addition to conducting the study with a larger sample to be able to detect smaller effect sizes, it would be important to complement the results of the study with qualitative data to gain better insight into how students perceive the discipline, how this differs, and why, between students with and without access to CS-education

\FloatBarrier

\section{Conclusion}

Early exposure to Computer Science (CS) \& Computational Thinking (CT) for all is important to broaden participation and promote equity in the field. This is contingent on addressing structural related barriers (lack of access) and social barriers (stereotypes) in order to reduce performance and perception gaps which affect sense of belonging and career decision. Addressing these barriers requires a system-wide implementation of Computer Science and Computational Thinking curricula for all students starting early foundational years. That is why numerous countries are introducing CS \& CT into their curricula starting primary school. The question is though, are these curricular reforms contributing to learning and reducing performance gaps? Curricular reforms and professional development programs are seldom evaluated at the student-level despite the importance of establishing their effectiveness in terms of student learning and perception. Therefore, in the present article, we evaluate the implementation of a regional CS-curricular reform in order to determine if the reform contributes to achieving equity goals. More specifically, we study how the implementation of the CS curriculum by teachers impacts and contributes to equity in terms of student learning (with respect to gender and performance gaps, RQ1) and perception (with respect to gender and self-efficacy gaps, RQ2). To answer these questions, the analysis employs hierarchical linear modelling and structural equation modelling using data from three studies involving respectively $n_1=1384$, $n_2=2433$ and $n_3=1644$ grade 3-6 students (ages 7-11) and their $n_1=83$, $n_2=142$ and $n_3=95$ teachers.  \\

In terms of student learning impact, the students are progressing over time. There is however no direct link between what the teachers taught (i.e. adopted) over an extended period of time and student learning. Although certain studies have suggested that perception may play a mediating role on performance, this is not the case in the present study. There is however a link between student learning and how teachers perceived the CS-PD program. Teacher perception may thus be acting as a mediating variable or be confounding with other dimensions such as teachers’ assimilation of Technological Pedagogical and Content Knowledge \citep{mishra_technological_2006} obtained during the PD, their appropriation of the content and the depth of the associated change in their practice, supporting the need to gain better insight into how the content is taught.
As there are known differences between intended, enacted and attained curricula \citep{van_den_akker_curriculum_2003}, the findings indicate the need to investigate not just whether, but how teaching the discipline, and individual pedagogical content, influences learning.
In terms of student learning equity, the findings indicating that i) the performance gap between lower and higher achieving students are closing and that ii) pre-existing gender gaps appear to be closing. Whether in terms of impact or equity, it would be important to expand to other dimensions of learning that may be influenced, whether to have a more complete evaluation of CT (by including practices and perspectives, \citealp{brennan_new_2012}), or by looking more generally into the impact on learning in other disciplines, or in terms of transversal competences.  \\

Where student perception is concerned, in terms of impact, the results are relatively straightforward: students in both CS and non CS schools perceive CS and the tools involved with teaching CS positively. Interest in the discipline and perception of robotics is nonetheless more positive in schools with access to CS-education which may contribute to broadening participation in the field. The findings in terms of equity indicate that there are gender gaps which indicate that boys have a better perception of the discipline than girls. However, whether in schools with access to CS education or not, the perception of role models close to them as doing CS contributing to student's positive perception of the discipline. As teachers are mainly women in primary school, introducing CS as a discipline taught by all teachers contributes to teachers being more often perceived as doing CS, and may ultimately contribute to gender-equity. Comparing students in schools with and without access to CS education indicates that there are differences in how the discipline is perceived in both types of schools and that there are interaction effects with gender: ii) initially smaller gender gaps are widening (e.g. CS \& tablet interest and self-efficacy, robots utility) while initially higher gender gaps are closing (e.g. robotics interest and self-efficacy, perceiving teachers as doing CS in CS-schools) with access to CS-education. Teaching CS thus has a complex influence on perception which requires investigating more deeply why students perceive the discipline the way they do and how it is influenced by access to CS-education. Monitoring this perception over time is also critical in order to understand how it evolves over time and influences long term career decisions.  \\

\added{Answering the overarching question ``how does the curricular reform impact student-level outcomes and equity in the field?'' is therefore not as straightforward as it seems. On the one hand, introducing CS for all in the curriculum and being taught CS has a positive impact and affects equity by:}
\begin{itemize}
    \item \added{Promoting student learning and contributing to performance-equity by reducing (i) differences between initially high and low performing students, (ii) the performance gender gap, and (iii) the impact of teacher demographics on student learning.}
    \item \added{Contributing to perception gender-equity by reducing the largest gender-related perception gaps (namely those pertaining to robotics).}
\end{itemize}

\added{On the other hand, the curricular reform does not automatically lead to improvements on all fronts. The impact is neither direct, as shown by the student learning results which lack a direct link between what was taught and learning; nor straightforward, as shown by the fact that there is an interaction effect between gender and access to CS education, with initially smaller (or not initially present) gender gaps increasing. \\
The findings of the study therefore demonstrate that the following elements are important to achieve equity and broadening participation in the field:} 
\begin{itemize}
    \item \added{Introducing CS for all students starting the first years of formal education.}
    \item \added{Preparing the teachers to teach CS, removing the influence of teacher demographic and teacher motivational factors on student learning.}
    \item \added{Having activities that signal to girls and boys equally and that are in students' Zone of Proximal Development in order to help all achieve the desired learning objectives.}
    \item \added{Investigating the impact of CS curricular reform and PD program implementations at the student level, and including teacher-level insight, all the while considering that the complex dynamics that may be involved in CS-education}.
\end{itemize}

\ifdefined\IJSTEMEduc
\else
\input{4-final-statements.tex}
\fi

\begin{backmatter}


\section*{Data availability}

The data will be publicly available on Zenodo upon publication (doi:10.5281/zenodo.7489244, \citealp{el-hamamsy_students_dataset}).

\section*{Ethics}
The researchers were granted ethical approval to conduct the study by the head of the Department of Education and by the Human Research Ethics Committee of \ifthenelse{\boolean{ISanonymous}}{ the Anonymous university (details removed for peer review), HREC protocol XXX-2019 (details removed for peer review)}{EPFL (project HREC 033-2019)}.

\section*{Conflicts of interest}
The authors declare that the research was conducted without any conflicts of interest.

\section*{Funding}
This work was funded by the \ifthenelse{\boolean{ISanonymous}}{Anonymous Funding Agency (details removed for peer review)}{the NCCR Robotics, a National Centre of Competence in Research, funded by the Swiss National Science Foundation (grant number 51NF40\_185543)}.

\section*{Authors' contributions}

\begin{itemize}
    \item L.E.-H.: Conceptualisation, methodology, investigation, data curation, analysis, visualisation, validation, writing - original draft preparation, writing - review \& editing
    \item B.B.: Conceptualisation, methodology, writing - review \& editing
    \item C.A.: methodology, validation, writing - review \& editing
    \item M.C.: conceptualisation, writing - review \& editing
    \item S.A.: methodology, writing - review \& editing
    \item J.D.Z.: conceptualisation, writing - review \& editing
    \item F.M.: conceptualisation, supervision, writing - review \& editing, funding acquisition
\end{itemize}

\section*{Acknowledgements}
We would like to thank all the participants and the members of the different institutions \ifdefined\Anonymous for supporting the Anonymous project \else (Department of Education - DEF, the University of Teacher Education – HEP Vaud, the teams from the two universities - EPFL and Unil) for supporting the EduNum \fi  project led by the ministry of education of the \ifdefined\Anonymous Anonymous region \else Canton Vaud\fi. \ifdefined\Anonymous \else We would also like to address a special thanks to Sylvie Bui who was an immense help in setting up the first data collections and to Emilie-Charlotte Monnier for going into the classrooms and pre-testing the student surveys\fi.

\FloatBarrier

\end{backmatter}

\bibliographystyle{apa_} 
\bibliography{0-bib}

\pagebreak
\appendix

\section{Appendix for Study 1}
\label{app:study1}

\subsection{Student demographics for study 1b}
\label{app:study1bdemographics}

\begin{table*}[!h]
  \centering
  \caption{\added{Study 1b - Student Learning demographics for the data subset which considers n=989 students with complete January \& June test data and Adoption data.}}
  \label{tab:study1_learning_adoption_demographics}
  
  \footnotesize
  \begin{tabular}{llccccc}
  \toprule
   & & \multicolumn{2}{c}{\textbf{Grade 3}} & \multicolumn{2}{c}{\textbf{Grade 4}} & \textbf{Total}  \\
  {} & Metric & No CS education classes & CS-education classes & No CS education & CS-education &  \\
  \midrule
  Students & count &  $71$ & $409$ &   $113$ & $396$ &  $989$  \\
  Classes & count   &   $4$ &  $24$ &   $6$ &  $21$ & $55$  \\
  Activities & M$\pm$SD  & $0.00\pm0.00$ &  $3.29\pm2.02$ & $0.00\pm0.00$ &  $3.12\pm1.59$ &  \\
   & range   & $[0;0]$ & $[1;9]$ & $[0;0]$ & $[1;6]$ &  \\
  Hours spent & M$\pm$SD   & $0.00\pm0.00$ &  $16.12\pm11.25$ & $0.00\pm0.00$ &   $12.54\pm6.80$ &  \\
   & range & $[0;0]$ &   $[1;48]$ & $[0;0]$ &   $[2;26]$ &  \\
  January cCTt Score &  M$\pm$SD &  $12.37\pm5.42$ &   $12.21\pm4.92$ &  $14.51\pm4.38$ &   $14.78\pm4.78$ &  \\
   &  range  &  $[1;23]$ &   $[1;23]$ &  $[3;23]$ &   $[1;24]$ &  \\
  June cCTt Score &   M$\pm$SD  &  $15.66\pm4.26$ &   $14.91\pm4.75$ &  $16.12\pm3.92$ &   $16.87\pm4.51$ &  \\
   &  range  &  $[3;23]$ &   $[3;24]$ &  $[2;23]$ &   $[4;24]$ &  \\
  \bottomrule
  \end{tabular}
  Please note that the grade 3 students in the "control" group (i.e. those who have not had access to CS education between the pre and the post tests) have slightly higher scores than those in the CS-education classes but this difference is not significant according to an ANOVA 
    neither in the pre-test ($\Delta=0.158pts$, $p=0.8502$, Cohen's $D=0.031$), nor in the post-test ($F(1)=0.77$, $p=0.38$, $\Delta=0.752pts$, Cohen's $D=0.167$).
\end{table*}

\FloatBarrier
\subsection{Hierarchical linear regression model for study 1b}
\label{app:study1bHR}

\begin{table*}[!h]
    \centering
    \caption{\added{Hierarchical linear model for student learning (dependent variable: Delta between pre- and post- test scores, n=989) with significant variables in bold. $R^2=0.285$, $RMSE=2.89$, $AIC=5132$, $BIC=5225$, Log-Likelihood$=-2547$. Abbreviations: NCS=Number of CS activities taught. Random effects $\sigma^2=8.85$, $\tau_{class}=0.00$, $\tau_{school}=1.50$ for $55$ classes in $7$ schools}}
    \footnotesize
    \begin{tabular}{lcccP{1.5cm}ccc}
    \toprule
    & Estimate & \added{95\% ci} & Std.Error & Degrees of Freedom & t-value &  p-value  \\ \midrule
    (Intercept)  &  $7.11$ & \added{$[5.08, 9.15]$} &  $1.04$  &  $922$  &  $6.86$  &  $p<0.0001$  \\
    Pre-test score  &  $-0.379$  &  \added{$[-0.52, -0.24]$} & $0.0722$  &  $922$  &  $-5.25$  &  \textbf{$p<0.0001$}  \\
    Gender (girls)  &  $0.697$ & \added{$[-1.62, 3.01]$}  &  $1.18$  &  $922$  &  $0.591$  &  $0.555$  \\
    Grade (4) & $1.15$ & \added{$[-1.97, 4.27]$}  &  $1.55$  &  $45$  &  $0.741$  &  $0.462$  \\
    NCS  &  $0.122$ & \added{$[-0.43, 0.68]$}  &  $0.275$  &  $45$  &  $0.442$  &  $0.661$  \\ \midrule
    Pre-test score:Gender (girls)  &  $-0.0221$ & \added{$[-0.20, 0.16]$}  &  $0.0906$  &  $922$  &  $-0.243$  &  $0.808$  \\
    Pre-test score:Grade (4) & $-0.0383$ & \added{$[-0.24,
0.16]$}  &  $0.101$  &  $922$  &  $-0.377$  &  $0.706$  \\
    Gender (girls):Grade (4) & $-0.880$ & \added{$[-4.52,
2.76]$}  &  $1.86$  &  $922$  &  $-0.474$  &  $0.636$  \\
    Pre-test score:NCS  &  $-0.00386$  & \added{$[-0.04, 0.04]$} &  $0.0198$  &  $922$  &  $-0.194$  &  $0.846$  \\
    Gender (girls):NCS  &  $-0.346$  & \added{$[-0.96,
0.26]$} &  $0.311$  &  $922$  &  $-1.11$  &  $0.267$  \\
    Grade 4:NCS  &  $-0.260$  & \added{$[-1.22, 0.70]$} &  $0.478$  &  $45$  &  $-0.544$  &  $0.589$  \\ \midrule
    Pre-test score:Gender (girls):Grade (4) & $0.0308$  & \added{$[-0.23, 0.29]$}  &  $0.131$  &  $922$  &  $0.235$  &  $0.814$  \\
    Pre-test score:Gender (girls):NCS  &  $0.0224$  & \added{$[-0.03, 0.07]$} &  $0.0255$  &  $922$  &  $0.876$  &  $0.381$  \\
    Pre-test score:Grade (4):NCS  &  $0.00979$  & \added{$[-0.05, 0.07]$} &  $0.0326$  &  $922$  &  $0.300$  &  $0.764$  \\
    Gender (girls):Grade (4):NCS  &  $0.195$  & \added{$[-0.91, 1.30]$} &  $0.562$  &  $922$  &  $0.347$  &  $0.729$  \\ \midrule
    Pre-test score:Gender (girls):Grade (4):NCS  &  $-0.0129$  & \added{$[-0.09, 0.07]$} &  $0.0412$  &  $922$  &  $-0.313$  &  $0.755$  \\ \bottomrule
    \end{tabular}
    \label{tab:student_learning_HR}
\end{table*}

\pagebreak
\subsection{Analysis of variance on the student learning data for study 1a}
\label{app:study1a}

\begin{table*}[!h]
  \centering
  \caption{\added{ANOVA of student learning data with Benjamini-Hochberg p-value correction and minimum effect size (Cohen's D) that can be detected with the sample}}
  \footnotesize
  \begin{tabular}{p{2cm}P{1.25cm}P{1cm}cP{1.25cm}P{1cm}cP{1.25cm}P{4.5cm}}
  \toprule
  Independent Variable &  Sum of squares &  Degrees of Freedom &  F &  Number of groups &  Residual Degrees of Freedom &   p &  Min Cohen's D & Significant difference (effect size and Dunn's post-hoc test for interaction effects)  \\
  \midrule
  time & $3357$ & $1$ &  $137.9$ & $2$ &  $2636$ &  $0.0000$ & $0.109$ & Post-test $>$ Pre-test  \\
  & & & & & & & &  
  $\Delta=+2.256pts$, $p<0.0001$, \added{Cohen's} $D=0.457$  \\\addlinespace[0.1cm]
 
  Grade   & $4007$ & $1$ &  $166.3$ & $2$ &  $2636$ &  $0.0000$ & $0.109$ & Grade 4$>$3 \\ 
  & & & & & & & & $\Delta=2.468pts$, \added{Cohen's} $D=0.502$  \\\addlinespace[0.1cm]
  
  Gender   &  $199$ & $1$ &  $7.8$ & $2$ &  $2636$ &  $0.0052$ & $0.109$ & Boys $>$ Girls  \\
  & & & & & & & &  $\Delta=0.551pts$, $p=0.0015$, \added{Cohen's} $D=0.109$  \\
  
  time:Grade & $7396$ & $3$ &  $108.0$ & $4$ &  $2634$ &  $0.0000$ & $0.129$ & See Fig. \ref{fig:study1_grade_gender_prepost}  \\
  time:Gender   & $3565$ & $3$ & $48.9$ & $4$ &  $2634$ &  $0.0000$ & $0.129$ & Pre-test Boys $>$ Girls  \\ 
  & & & & & & & & $\Delta=0.664pts$, $p=0.0079$, \added{Cohen's} $D=0.131$  \\ \addlinespace[0.1cm]
  & & & & & & & & Post-test Boys $\sim$ Girls  \\ 
  & & & & & & & & $\Delta=0.438pts$, $p=0.0744$, \added{Cohen's} $D=0.091$  \\ \addlinespace[0.1cm]
  
  Grade:Gender  & $4248$ & $3$ & $58.9$ & $4$ &  $2634$ &  $0.0000$ & $0.129$ & Grade 3 Boys $>$ Girls  \\ 
  & & & & & & & & $\Delta=0.725pts$, $p=0.004$, \added{Cohen's} $D=0.145$  \\ \addlinespace[0.1cm]
  & & & & & & & & Grade 4 Boys $\sim$ Girls  \\ 
  & & & & & & & & $\Delta=0.469pts$, $p=0.0604$, \added{Cohen's} $D=0.098$ \\ \addlinespace[0.1cm]
  time:Grade:Gender & $7660$ & $7$ & $48.1$ & $8$ & $2630$ &  $0.0000$ & $0.148$ & See Fig. \ref{fig:study1_grade_gender_prepost} \\
  \bottomrule
\end{tabular}
  \label{tab:study1_anova_sig}
\end{table*}

\pagebreak
\section{Appendix for Study 2}

\subsection{Structural Equation Model for the effect of student-related variables on their perception of the discipline (study 2)}
\label{app:study2_SEM}

\begin{table*}[!h]
\caption{\added{SEM Perception Structural Model (n=2116, November 2021) Unstandardised Factor Loadings and Regression Slopes}}
\label{tab:study2_SEMfit}
\scriptsize
\begin{tabular}{lrccccc}
\toprule
\textbf{Factor Loadings} & & {Estimate} & {Std. Err.} & {Z} & {p} & {$R^2$} \\\midrule
\underline{Discipline perception}

 & Tablets & $0.83$ & $0.11$ & $7.37$ & $.000$ & $0.41$ \\
 & Robots  & $0.76$ & $0.10$ & $7.85$ & $.000$ & $0.37$ \\
 & CS & $1.98$ & $0.70$ & $2.82$ & $.005$ & $0.81$ \\
 \underline{CS}
 & CS interest  & $0.17$ & $0.05$ & $3.25$ & $.001$ & $0.25$ \\
 & CS utility  & $0.16$ & $0.05$ & $3.19$ & $.001$ & $0.25$ \\
 & CS self-efficacy  & $0.20$ & $0.06$ & $3.31$ & $.001$ & $0.28$ \\
 \underline{Role Models}
 & Role Models - Teacher  & $0.17$ & $0.01$ & $15.60$ & $.000$ & $0.13$ \\
 & Role Models - Mom  & $0.23$ & $0.01$ & $25.37$ & $.000$ & $0.28$ \\
 & Role Models - Dad  & $0.30$ & $0.01$ & $32.03$ & $.000$ & $0.36$ \\
 & Role Models - Student & $0.13$ & $0.01$ & $15.47$ & $.000$ & $0.11$ \\
 & Role Models - Nobody  & $-0.24$ & $0.01$ & $-22.85$ & $.000$ & $0.41$ \\
 \underline{Tablets}
 & Tablets interest  & $0.25$ & $0.03$ & $7.97$ & $.000$ & $0.22$ \\
 & Tablets utility  & $0.30$ & $0.03$ & $9.29$ & $.000$ & $0.37$ \\
 & Tablets self-efficacy  & $0.24$ & $0.03$ & $8.96$ & $.000$ & $0.31$ \\
 \underline{Robots}
 & Robots interest  & $0.44$ & $0.04$ & $10.70$ & $.000$ & $0.33$ \\
 & Robots utility  & $0.36$ & $0.04$ & $9.36$ & $.000$ & $0.29$ \\
 & Robots self-efficacy  & $0.49$ & $0.04$ & $11.34$ & $.000$ & $0.42$ \\ \midrule
\textbf{Regression Slopes}  &  & {Estimate}  & {Std. Err.}  & {Z}  & {p}  &  \\
\midrule 
\underline{Discipline perception}
 & Role Models  & $0.15$ & $0.05$ & $2.94$ & $.003$ &  \\
 \underline{CS}
 & Role Models  & $0.30$ & $0.12$ & $2.41$ & $.016$ &  \\
 & Number of CS education Periods  & $-0.01$ & $0.01$ & $-0.62$ & $.536$ &  \\
 \underline{Role Models}
 & Number of CS education Periods  & $-0.01$ & $0.01$ & $-1.18$ & $.238$ &  \\
 & Number of ICT Instruction Periods  & $-0.00$ & $0.00$ & $-0.13$ & $.896$ &  \\
 & Number of Robotics Instruction Periods  & $0.03$ & $0.02$ & $1.22$ & $.224$ &  \\
 \underline{Tablets}
 & Number of ICT Instruction Periods  & $-0.00$ & $0.00$ & $-0.80$ & $.426$ &  \\
 \underline{Robots}
 & Number of Robotics Instruction Periods  & $0.00$ & $0.01$ & $0.06 $& $.949$ &  \\ 
 \underline{CS interest}
& Gender (boys=0 girls=1)  & $-0.20$ & $0.03$ & $-5.75$ & $.000$ &   \\
 &  Grade  & $-0.02$ & $0.02$ & $-1.12$ & $.262$ &   \\
 &  General self-efficacy  &   $0.09$ & $0.03$ &  $3.29$ & $.001$ &   \\
 \underline{CS utility}
 &  Gender (boys=0 girls=1)  & $-0.05$ & $0.03$ & $-1.60$ & $.110$ &   \\
 &  Grade  & $0.09$ & $ 0.02$ & $5.49$ & $.000$ &   \\
 &  General self-efficacy  &   $0.09$ & $0.02$ &  $3.99$ & $.000$ &   \\
 \underline{CS self-efficacy}
 &  Gender (boys=0 girls=1)  & $-0.01$ & $0.04$ & $-0.33$ & $.741$ &   \\
 &  Grade  & $0.01$ &  $0.02$ & $0.74$ & $.459$ &   \\
 &  General self-efficacy  &  $ 0.15$ & $0.03$ &  $5.19$ & $.000$ &   \\
 \underline{Role Models - Teacher}
 &  Gender (boys=0 girls=1)  & $0.06$ &  $0.02$ & $2.98$ & $.003$ &   \\
 &  Grade  & $-0.04$ & $0.01$ & $-3.78$ & $.000$ &   \\
 \underline{Role Models - Mom}
 &  Gender (boys=0 girls=1)  & $0.01$ &  $0.02$ & $0.35$ & $.723$ &   \\
& Grade & $-0.06$ & $0.01$ & $-5.84$ & $.000$ &   \\
 \underline{Role Models - Dad}
& Gender (boys=0 girls=1) & $-0.06$ & $0.02$ & $-2.84$ & $.005$ &   \\
& Grade & $-0.01$ & $0.01$ & $-1.07$ & $.286$ &   \\
 \underline{Role Models - Student}
& Gender (boys=0 girls=1) & $0.00$ & $0.02$ &  $0.04$ & $.971$ &   \\
& Grade & $-0.00$ & $0.01$ & $-0.56$ & $.572$ &   \\
 \underline{Role Models - Nobody}
& Gender (boys=0 girls=1) & $0.01$ & $0.02$ &  $0.43$ & $.667$ &   \\
& Grade & $-0.03$ & $0.01$ & $-3.70$ & $.000$ &   \\
 \underline{Tablets interest}
& Gender (boys=0 girls=1) & $-0.09$ & $0.03$ & $-2.70$ & $.007$ &   \\
& Grade & $-0.04$ & $0.02$ & $-2.45$ & $.014$ &   \\
& General self-efficacy & $-0.00$ & $0.02$ & $-0.07$ & $.946$ &   \\
 \underline{Tablets utility}
& Gender (boys=0 girls=1) & $-0.05$ & $0.03$ & $-1.66$ & $.097$ &   \\
& Grade & $0.03$ & $0.01$ &  $2.33$ & $.020$ &   \\
& General self-efficacy & $0.08$ & $ 0.02$ & $3.91$ & $.000$ &   \\
 \underline{Tablets self-efficacy}
& Gender (boys=0 girls=1) & $-0.07$ & $0.03$ & $-2.55$ & $.011$ &   \\
& Grade & $-0.01$ & $0.01$ & $-1.07$ & $.283$ &   \\
& General self-efficacy & $0.06$ &  $0.02$ & $3.29$ & $.001$ &   \\
 \underline{Robots interest}
& Gender (boys=0 girls=1) & $-0.32$ & $0.04$ & $-7.32$ & $.000$ &   \\
& Grade & $0.03$ & $0.02$ &  $1.44$ & $.151$ &   \\
& General self-efficacy & $0.09$ &  $0.03$ & $2.50$ & $.012$ &   \\
 \underline{Robots utility}
& Gender (boys=0 girls=1) & $-0.19$ & $0.04$ & $-4.95$ & $.000$ &   \\
& Grade & $0.07$ & $0.02$ &  $3.78$ & $.000$ &   \\
& General self-efficacy & $0.16$ &  $0.03$ & $4.98$ & $.000$ &   \\
 \underline{Robots self-efficacy}
& Gender (boys=0 girls=1) & $-0.25$ & $0.04$ & $-5.90$ & $.000$ &   \\
& Grade & $0.02$ & $0.02$ &  $0.71$ & $.477$ &   \\
& General self-efficacy & $0.19$ &  $0.03$ & $5.52$ & $.000$ &   \\
\bottomrule

\end{tabular}
\end{table*}

\FloatBarrier
\pagebreak
\subsection{Structural Equation Model for the Link between Perception and Learning (study 2)}
\label{app:perc_learning_SEM}

\begin{table*}[!h]
    \centering
    \caption{SEM Perception and Background to Learning Structural Model ($n=1583$, November 2021) Unstandardised Regression Parameters ($\chi^2(124)=221.462$, $p<0.001$, $chi^2/df=1.79$, $CFI=0.951$, $TLI=0.923$, $RMSEA=0.022$, $ci=[0.017, 0.027]$, $SRMR=0.026$). As CS utility did not correlate highly with interest and self-efficacy it was removed from this model.}
    \label{tab:study2_SEM_learning_perc_full}
    \notsotiny
    \begin{tabular}{rrrrrrr}
        \toprule
        & & \multicolumn{1}{c}{Estimate} &  \multicolumn{1}{c}{Std. Err.} &  \multicolumn{1}{c}{Z} &  \multicolumn{1}{c}{p} &  \multicolumn{1}{c}{$R^2$} \\\midrule
        & & \multicolumn{5}{c}{\underline{Factor Loadings}} \\ \multicolumn{1}{l}{\underline{Discipline perception}} 
        & Tablets & $0.76$  & $0.10$  & $7.70$  & $.000$ & $0.37$  \\
        & Robots & $0.82$ & $0.11$ & $7.22$ & $.000$ &  $0.41$  \\
        & CS & $1.51$ & $0.38$ & $3.93$ & $.000$ &  $0.71$  \\
         \multicolumn{1}{l}{\underline{CS}}
        & CS interest & $0.21$ & $0.05$ & $4.62$ & $.000$ &  $0.28$  \\
        & CS self-efficacy & $0.27$ & $0.06$ & $4.57$ & $.000$ &  $0.32$  \\
         \multicolumn{1}{l}{\underline{Role Model}}
        & Role Models - Teacher & $0.18$ & $0.01$ & $14.77$ & $.000$ &  $0.17$ \\
        & Role Models - Mom & $0.24$ & $0.01$ & $23.17$ & $.000$ & $ 0.30 $\\
        & Role Models - Dad &  $0.30$ & $0.01$ & $29.01$ & $.000$ &  $0.37$ \\
        & Role Models - Student & $0.14$ & $0.01$ &  $13.72$ & $.000$ &  $0.11 $\\
        & Role Models - Nobody & $-0.23$ & $0.01$ & $-19.99$ & $.000$ &  $0.39$ \\
         \multicolumn{1}{l}{\underline{Tablets}}
        & Tablets interest & $0.27$ & $0.04$ &  $7.73$ & $.000$ &  $0.24$ \\
        & Tablets utility & $0.36$ & $0.03$ &  $10.34 $ & $.000$ &  $0.44$ \\
        & Tablets self-efficacy & $0.27$ & $0.03$ &  $9.00$ & $.000$ &  $0.36$ \\
         \multicolumn{1}{l}{\underline{Robots}} 
        & Robots interest & $0.42$ & $0.05$ & $ 8.92$ & $.000$ & $ 0.34$ \\
        & Robots utility & $0.36$ & $0.04$ &  $8.38$ & $.000$ &  $0.31$ \\
        & Robots self-efficacy & $0.46$ & $0.05$ &  $9.50$ & $.000$ &  $0.41$ \\ \midrule
        & & \multicolumn{5}{c}{\underline{Regression Slopes}}  \\ 
        \multicolumn{1}{l}{\underline{Discipline perception}} 
        &  Role Models & $0.17$ & $0.06$ &  $2.94$ & $.003$ &   \\
         \multicolumn{1}{l}{\underline{CS}}
        &  Role Models & $0.13$ & $0.10$ &  $1.33$ & $.182$ &   \\
        & Number of CS education periods & $-0.01$ & $0.01$ & $-1.09$ & $.275$ &   \\
         \multicolumn{1}{l}{\underline{CS interest}}
        & Gender (boys=0, girls=1) & $-0.19$ & $0.04$ & $-4.85$ & $.000$ &   \\
        & Grade & $-0.02$ & $0.02$ & $-1.05$ & $.292$ &   \\
        & General self-efficacy & $0.09$ & $0.03$ &  $3.08$ & $.002$ &   \\
         \multicolumn{1}{l}{\underline{CS utility}}
        & Gender (boys=0, girls=1) & $-0.06$ & $0.04$ & $-1.69$ & $.091$ &   \\
        & Grade & $0.05$ & $0.02$ &  $2.71$ & $.007$ &   \\
        & General self-efficacy & $0.10$ & $0.03$ &  $3.76$ & $.000$ &   \\
         \multicolumn{1}{l}{\underline{CS self-efficacy}}
        & Gender (boys=0, girls=1) & $0.05$ & $0.05$ &  $1.10$ & $.270$ &   \\
        & Grade & $-0.01$ & $0.02$ & $-0.63$ & $.527$ &   \\
        & General self-efficacy & $0.14$ & $0.03$ &  $4.15$ & $.000$ &   \\
         \multicolumn{1}{l}{\underline{Role Models - Teacher}}
        & Gender (boys=0, girls=1) & $0.08$ & $0.02$ &  $3.30$ & $.001$ &   \\
        &  Grade & $-0.06$ & $0.01$ & $-4.79$ & $.000$ &   \\
         \multicolumn{1}{l}{\underline{Role Models - Mom}}
        & Gender (boys=0, girls=1) & $-0.00$ & $0.02$ & $-0.13$ & $.900$ &   \\
        & Grade & $-0.06$ & $0.01$ & $-5.55$ & $.000$ &   \\
         \multicolumn{1}{l}{\underline{Role Models - Dad}}
        & Gender (boys=0, girls=1) & $-0.07$ & $0.03$ & $-2.56$ & $.010$ &   \\
        & Grade & $-0.01$ & $0.01$ & $-1.16$ & $.247$ &   \\
         \multicolumn{1}{l}{\underline{Role Models - Student}}
        & Gender (boys=0, girls=1) & $0.01$ & $0.02$ &  $0.55$ & $.585$ &   \\
        & Grade & $-0.00$ & $0.01$ & $-0.23$ & $.821$ &   \\
         \multicolumn{1}{l}{\underline{Role Models - Nobody}}
        & Gender (boys=0, girls=1) & $0.01$ & $0.02$ &  $0.68$ & $.497$ &   \\
        & Grade & $-0.02$ & $0.01$ & $-2.36$ & $.018$ &   \\
         \multicolumn{1}{l}{\underline{Tablets interest}}
        & Gender (boys=0, girls=1) & $-0.07$ & $0.04$ & $-1.86$ & $.063$ &   \\
        & Grade & $-0.04$ & $0.02$ & $-2.08$ & $.037$ &   \\
        & General self-efficacy & $0.01$ & $0.03$ & $ 0.37$ & $.713$ &   \\
         \multicolumn{1}{l}{\underline{Tablets utility}}
        & Gender (boys=0, girls=1) & $-0.05$ & $0.04$ & $-1.45$ & $.148$ &   \\
        & Grade & $0.02$ & $0.02$ &  $1.10$ & $.270$ &   \\
        & General self-efficacy & $0.07$ & $0.02$ &  $2.97$ & $.003$ &   \\
         \multicolumn{1}{l}{\underline{Tablets self-efficacy}}
        & Gender (boys=0, girls=1) & $-0.06$ & $0.03$ & $-2.17$ & $.030$ &   \\
        & Grade & $-0.01$ & $0.02$ & $-0.63$ & $.527$ &   \\
        & General self-efficacy & $0.08$ & $0.02$ &  $3.58$ & $.000$ &   \\
         \multicolumn{1}{l}{\underline{Robots interest}}
        & Gender (boys=0, girls=1) & $-0.35$ & $0.05$ & $-6.83$ & $.000$ &   \\
        & Grade & $0.04$ & $0.03$ &  $1.35$ & $.176$ &   \\
        & General self-efficacy & $0.09$ & $0.04$ &  $2.15$ & $.031$ &   \\
         \multicolumn{1}{l}{\underline{Robots utility}}
        & Gender (boys=0, girls=1) & $-0.16$ & $0.04$ & $-3.67$ & $.000$ &   \\
        & Grade & $0.06$ & $0.02$ &  $2.44$ & $.015$ &   \\
        & General self-efficacy & $0.16$ & $0.04$ &  $4.20$ & $.000$ &   \\
         \multicolumn{1}{l}{\underline{Robots self-efficacy}}
        & Gender (boys=0, girls=1) & $-0.21$ & $0.05$ & $-4.39$ & $.000$ &   \\
        & Grade & $0.00$ & $0.03$ &  $0.04$ & $.970$ &   \\
        & General self-efficacy & $0.20$ & $0.04$ &  $5.31$ & $.000$ &   \\
         \multicolumn{1}{l}{\underline{Role Models}}
        & Number of CS education periods & $0.00$ &  $0.00$ & $0.13$ & $.896$ &   \\
         \multicolumn{1}{l}{\underline{Tablets}}
        & Number of ICT education periods & $-0.00$ & $0.00$ & $-0.52$ & $.601$ &   \\
         \multicolumn{1}{l}{\underline{Robots}}
        & Number of Robotics education periods & $-0.01$ & $0.02$ & $-0.79$ & $.429$ &   \\
          \multicolumn{1}{l}{\underline{Percentage ($/100$)}} & & & & & & $0.136$  \\
        & CS perception & $-0.29$ & $1.31$ & $-0.22$ & $.824$ &  $0.707$ \\
        & Tablets perception & $-0.04$ & $0.99$ & $-0.04$ & $.969$ &  $0.372$ \\
        & Robots perception & $0.96$ & $1.11$ &  $0.86$ & $.387$ &  $0.411$ \\
        & \textbf{General self-efficacy} & $1.54$ & $0.69$ &  $2.22$ &  \textbf{$.027$} &   \\
        & Gender (0=boys, 1=girls) & $-1.57$ & $1.06$ & $-1.47$ & $.141$ &   \\
        & \textbf{Grade} & $7.53$ & $0.53$ &  $14.19$ &  \textbf{$.000$} &  \\
        & Number of CS education periods SI & $0.09$ &  $0.13$ & $0.70$ & $.483$ &   \\
        & Number of ICT education periods & $-0.02$ & $0.08$ & $-0.20$ & $.842$ &   \\
        & Number of Robotics education periods & $-0.29$ & $0.42$ & $-0.68$ & $.496$ &   \\ \bottomrule
        \end{tabular}
\end{table*}

\FloatBarrier
\pagebreak
\section{Appendix for Study 3}

\subsection{Structural Equation Model between students in schools with and without access to CS education (study 3)}
\label{app:study3_group_comp_SEM}

\begin{table*}[!h]
\caption{SEM Perception Structural Model (n=1640, June 2022) Unstandardised Factor Loadings, Regression Slopes and Intercepts}
\label{tab:study3_SEM_groups}
\centering
\scriptsize
\hskip-0.5cm\begin{tabular}{lrccccc|ccccc}
\toprule
& & \multicolumn{5}{c|}{CS education} &  \multicolumn{5}{c}{No CS education} \\ \midrule
& &  \multicolumn{10}{c}{\textbf{Factor Loadings}} \\ 
& & {Est.} &  {Std. Err.} &  {Z} &  {p} &  {$R^2$} &  {Est.} &  {Std. Err.} &  {Z} &  {p} &  {$R^2$} \\\midrule
\underline{CS} & CS interest & $0.43$ & $0.04$ &  $10.38$ & $.000$ & $0.41$ &  $0.43$ & $0.04$ &  $10.38$ & $.000$ &  $0.34$ \\
 & CS utility & $0.35$ & $0.04$ &  $9.04$ & $.000$ & $0.24$ &  $0.35$ & $0.04$ &  $9.04$ & $.000$ &  $0.24$ \\
 & CS self-efficacy & $0.36$ & $0.04$ &  $8.96$ & $.000$ & $0.19$ &  $0.36$ & $0.04$ &  $8.96$ & $.000$ &  $0.23$ \\
 \underline{Tablets} & Tablets interest & $0.39$ & $0.04$ &  $10.62$ & $.000$ & $0.38$ &  $0.39$ & $0.04$ &  $10.62$ & $.000$ &  $0.35$ \\
 & Tablets utility & $0.37$ & $0.03$ &  $11.48$ & $.000$ & $0.32$ &  $0.37$ & $0.03$ &  $11.48$ & $.000$ &  $0.28$ \\
 & Tablets self-efficacy & $0.31$ & $0.03$ & $ 9.25$ & $.000$ & $0.28$ &  $0.31$ & $0.03$ &  $9.25$ & $.000$ &  $0.39$ \\
 \underline{Robots} & Robots interest & $0.55$ & $0.05$ &  $11.24$ & $.000$ & $0.38$ &  $0.55$ & $0.05$ &  $11.24$ & $.000$ &  $0.35$ \\
 & Robots utility & $0.48$ & $0.04$ &  $10.79$ & $.000$ & $0.34$ &  $0.48$ & $0.04$ &  $10.79$ & $.000$ &  $0.29$ \\
 & Robots self-efficacy & $0.59$ & $0.05$ &  $11.43$ & $.000$ & $0.39$ &  $0.59$ & $0.05$ &  $11.43$ & $.000$ &  $0.42$ \\
 \underline{Role Model} & Role Models - Teacher & $0.19$ & $0.02$ &  $12.67$ & $.000$ & $0.16$ &  $0.19$ & $0.02$ &  $12.67$ & $.000$ &  $0.14$ \\
 & Role Models - Mom & $0.24$ & $0.01$ &  $16.56$ & $.000$ & $0.26$ &  $0.24$ & $0.01$ &  $16.56$ & $.000$ &  $0.24$ \\
 & Role Models - Dad & $0.23$ & $0.02$ &  $14.73$ & $.000$ & $0.23$ &  $0.23$ & $0.02$ &  $14.73$ & $.000$ &  $0.21$ \\
 & Role Models - Student & $0.16$ & $0.01$ &  $12.38$ & $.000$ & $0.13$ & $0.16$ & $0.01$ &  $12.38$ & $.000$ &  $0.11$ \\
 & Role Models - Other & $0.16$ & $0.01$ &  $11.12$ & $.000$ & $0.15$ &  $0.16$ & $0.01$ &  $11.12$ & $.000$ &  $0.10$ \\
 & Role Models - Nobody & $-0.21$ & $0.01$ & $-13.91$ & $.000$ & $0.43$ & $-0.21$ & $0.01$ & $-13.91$ & $.000$ &  $0.40 $\\ \midrule
 & & \multicolumn{10}{c}{\textbf{Regression Slopes}} \\ 
 & & {Est} &  {Std. Err.} &  {Z} &  {p} &  {$R^2$} &  {Est} &  {Std. Err.} &  {Z} &  {p} &  {$R^2$} \\\hline
 \underline{CS interest} & Gender (boys=0, girls=1) & $-0.24$ & $0.05$ & $-5.02$ & $.000$ &  & $-0.11$ & $0.05$ & $-2.14$ & $.032$ &   \\
 & Grade & $0.02$ & $0.02$ &  $0.82$ & $.411$ &  & $-0.00$ & $0.03$ & $-0.04$ & $.971$ &   \\
 & General self-efficacy & $0.02$ & $0.04$ &  0.55 & $.583$ &  & $0.10$ & $0.05$ &  $2.09$ & $.036$ &   \\
 {\underline{CS utility}} & Gender (boys=0, girls=1) & $-0.03$ & $0.05$ & $-0.61$ & $.542$ &  & $-0.10$ & $0.05$ & $-1.75$ & $.080$ &   \\
 & Grade & $0.10$ & $0.03$ &  $3.77$ & $.000$ &  & $-0.00$ & $0.03$ & $-0.03$ & $.975$ &   \\
 & General self-efficacy & $0.07$ & $0.04$ &  $1.69$ & $.092$ &  & $0.14$ & $0.04$ &  $3.45$ & $.001$ &   \\
 \underline{CS self-efficacy} & Gender (boys=0, girls=1) & $-0.13$ & $0.06$ & $-2.15$ & $.031$ &  & $-0.08$ & $0.06$ & $-1.37$ & $.172$ &   \\
 & Grade & $0.02$ & $0.03$ &  $0.54$ & $.592$ &  & $0.01$ & $0.03$ &  $0.46$ & $.643$ &   \\
 & General self-efficacy & $0.16$ & $0.05$ &  $3.26$ & $.001$ &  & $0.20$ & $0.05$ &  $3.80$ & $.000$ &   \\
 \underline{Tablets interest} & Gender (boys=0, girls=1) & $-0.13$ & $0.04$ & $-2.93$ & $.003$ &  & $0.09$ & $0.05$ &  $1.83$ & $.067$ &   \\
 & Grade & $-0.03$ & $0.02$ & $-1.55$ & $.121$ &  & $-0.02$ & $0.02$ & $-0.90$ & $.367$ &   \\
 & General self-efficacy & $0.04$ & $0.04$ &  0.87 & $.386$ &  & $0.01$ & $0.05$ &  $0.22$ & $.824$ &   \\
 \underline{Tablets utility} & Gender (boys=0, girls=1) & $-0.03$ & $0.05$ & $-0.75$ & $.453$ &  & $-0.03$ & $0.06$ & $-0.51$ & $.611$ &   \\
 & Grade & $-0.01$ & $0.02$ & $-0.63$ & $.529$ &  & $-0.04$ & $0.03$ & $-1.42$ & $.157$ &   \\
 & General self-efficacy & $0.05$ & $0.03$ &  $1.52$ & $.130$ &  & $0.07$ & $0.04$ &  $1.66$ & $.098$ &   \\
 \underline{Tablets self-efficacy} & Gender (boys=0, girls=1) & $-0.18$ & $0.04$ & $-4.14$ & $.000$ &  & $-0.07$ & $0.04$ & $-1.71$ & $.087$ &   \\
 & Grade & $0.02$ & $0.02$ &  $0.88$ & $.377$ &  & $0.00$ & $0.02$ &  $0.18$ & $.858$ &   \\
 & General self-efficacy & $0.08$ & $0.03$ &  $2.41$ & $.016$ &  & $0.08$ & $0.04$ &  1.93 & $.054$ &   \\
 \underline{Robots interest} & Gender (boys=0, girls=1) & $-0.33$ & $0.07$ & $-4.89$ & $.000$ &  & $-0.45$ & $0.08$ & $-5.72$ & $.000$ &   \\
 & Grade & $-0.03$ & $0.03$ & $-1.18$ & $.238$ &  & $-0.06$ & $0.04$ & $-1.59$ & $.111$ &   \\
 & General self-efficacy & $0.02$ & $0.04$ &  $0.43$ & $.667$ &  & $0.09$ & $0.06$ &  $1.53$ & $.127$ &   \\
 \underline{Robots utility} & Gender (boys=0, girls=1) & $-0.12$ & $0.06$ & $-2.06$ & $.040$ &  & $-0.03$ & $0.07$ & $-0.46$ & $.643$ &   \\
 & Grade & $-0.03$ & $0.03$ & $-1.25$ & $.212$ &  & $-0.09$ & $0.04$ & $-2.61$ & $.009$ &   \\
 & General self-efficacy & $0.01$ & $0.04$ &  $0.37$ & $.712$ &  & $0.11$ & $0.05$ & $ 2.01$ & $.044$ &   \\
 \underline{Robots self-efficacy} & Gender (boys=0, girls=1) & $-0.27$ & $0.07$ & $-3.94$ & $.000$ &  & $-0.41$ & $0.07$ & $-5.72$ & $.000$ &   \\
 & Grade & $-0.05$ & $0.03$ & $-1.81$ & $.070$ &  & $-0.01$ & $0.04$ & $-0.31$ & $.753$ &   \\
 & General self-efficacy & $0.11$ & $0.05$ &  2.14 & $.033$ &  & $0.14$ & $0.06$ &  $2.42$ & $.016$ &   \\
 \underline{Role Models - Teacher} & Gender (boys=0, girls=1) & $0.10$ & $0.04$ &  $2.57$ & $.010$ &  & $0.11$ & $0.04$ &  $3.16$ & $.002$ &   \\
 & Grade & $-0.01$ & $0.02$ & $-0.64$ & $.522$ &  & $0.00$ & $0.02$ &  $0.27$ & $.789$ &   \\
 \underline{Role Models - Mom} & Gender (boys=0, girls=1) & $0.03$ & $0.04$ &  $0.74$ & $.458$ &  & $0.01$ & $0.04$ &  $0.21$ & $.833$ &   \\
 & Grade & $-0.06$ & $0.02$ & $-3.73$ & $.000$ &  & $-0.07$ & $0.02$ & $-4.62$ & $.000$ &   \\
 \underline{Role Models - Dad} & Gender (boys=0, girls=1) & $-0.07$ & $0.04$ & $-1.82$ & $.069$ &  & $-0.01$ & $0.03$ & $-0.19$ & $.849$ &   \\
 & Grade & $-0.04$ & $0.02$ & $-2.32$ & $.020$ &  & $-0.05$ & $0.02$ & $-2.98$ & $.003$ &   \\
 \underline{Role Models - Student} & Gender (boys=0, girls=1) & $-0.05$ & $0.03$ & $-1.38$ & $.169$ &  & $-0.07$ & $0.04$ & $-1.87$ & $.062$ &   \\
 & Grade & $0.03$ & $0.01$ &  $2.13$ & $.033$ &  & $-0.04$ & $0.02$ & $-2.74$ & $.006$ &   \\
 \underline{Role Models - Other} & Gender (boys=0, girls=1) & $0.05$ & $0.04$ &  $1.43$ & $.154$ &  & $-0.01$ & $0.03$ & $-0.19$ & $.848$ &   \\
 & Grade & $0.09$ & $0.01$ &  $5.98$ & $.000$ &  & $0.00$ & $0.02$ &  $0.09$ & $.929$ &   \\
 \underline{Role Models - Nobody} & Gender (boys=0, girls=1) & $-0.02$ & $0.02$ & $-1.00$ & $.318$ &  & $0.00$ & $0.02$ &  $0.04$ & $.969$ &   \\
 & Grade & $-0.01$ & $0.01$ & $-0.73$ & $.466$ &  & $0.03$ & $0.01$ &  $3.10$ & $.002$ &   \\ \bottomrule
\end{tabular}
\end{table*}

\addtocounter{table}{-1}
\begin{table*}[!h]
\caption{(Continued) SEM Perception Structural Model (n=1640, June 2022) Unstandardised Factor Loadings, Regression Slopes and Intercepts}
\centering
\scriptsize
\hskip-0.5cm\begin{tabular}{lrccccc|ccccc}
\toprule
& & \multicolumn{5}{c|}{CS education} &  \multicolumn{5}{c}{No CS education} \\ \hline
& & \multicolumn{10}{c}{\textbf{Intercepts}} \\
& & {Est} &  {Std. Err.} &  {Z} &  {p} &  {$R^2$} &  {Est} &  {Std. Err.} &  {Z} &  {p} &  {$R^2$} \\\midrule
 & CS interest & $1.59$ & $0.14$ &  $11.15$ & $.000$ &  & $1.53$ & $0.14$ &  $10.81$ & $.000$ &   \\
 & CS utility & $1.02$ & $0.16$ &  $6.27$ & $.000$ &  & $1.45$ & $0.13$ &  $10.87$ & $.000$ &   \\
 & CS self-efficacy & $1.19$ & $0.17$ &  $7.12$ & $.000$ &  & $1.29$ & $0.16$ &  $7.93$ & $.000$ &   \\
 & Tablets interest & $1.88$ & $0.13$ &  $14.00$ & $.000$ &  & $1.74$ & $0.13$ &  $13.28$ & $.000$ &   \\
 & Tablets utility & $1.64$ & $0.14$ &  $11.98$ & $.000$ &  & $1.67$ & $0.13$ &  $13.03$ & $.000$ &   \\
 & Tablets self-efficacy & $1.62$ & $0.11$ &  $14.19$ & $.000$ &  & $1.69$ & $0.11$ &  14.81 & $.000$ &   \\
 & Robots interest & $1.71$ & $0.15$ &  $11.19$ & $.000$ &  & $1.45$ & $0.16$ &  $9.14$ & $.000$ &   \\
 & Robots utility & $1.66$ & $0.14$ &  $11.81$ & $.000$ &  & $1.57$ & $0.17$ &  $9.42$ & $.000$ &   \\
 & Robots self-efficacy & $1.51$ & $0.16$ &  $9.27$ & $.000$ &  & $1.19$ & $0.16$ &  7.59 & $.000$ &   \\
 & Role Models - Teacher & $0.46$ & $0.08$ &  $6.10$ & $.000$ &  & $0.36$ & $0.08$ &  4.66 & $.000$ &   \\
 & Role Models - Mom & $0.71$ & $0.08$ &  $9.46$ & $.000$ &  & $0.81$ & $0.07$ &  $11.28$ & $.000$ &   \\
 & Role Models - Dad & $0.77$ & $0.07$ &  $10.50$ & $.000$ &  & $0.78$ & $0.07$ &  $11.26$ & $.000$ &   \\
 & Role Models - Student & $0.19$ & $0.07$ &  $2.71$ & $.007$ &  & $0.58$ & $0.08$ &  7.52 & $.000$ &   \\
 & Role Models - Other & $0.01$ & $0.07$ &  $0.18$ & $.855$ &  & $0.32$ & $0.08$ &  $4.22$ & $.000$ &   \\
 & Role Models - Nobody & $0.15$ & $0.04$ &  $3.43$ & $.001$ &  & $0.02$ & $0.04$ &  $0.57$ & $.566$ &   \\
 & Gender (boys=0, girls=1) & $0.49$ & $0.02$ &  $28.17$ & $.000$ &  & $0.49$ & $0.02$ &  $28.09$ & $.000$ &   \\
 & Grade & $4.61$ & $0.04$ &  $115.25$ & $.000$ &  & $4.63$ & $0.04$ &  $120.76$ & $.000$ &   \\
 & General self-efficacy & $1.48$ & $0.03$ &  $59.29$ & $.000$ &  & $1.47$ & $0.03$ &  $50.86$ & $.000$ &   \\ \bottomrule
\end{tabular}
\end{table*}

\end{document}